\documentclass[10pt,twocolumn]{article}
\usepackage[a4paper,top=0.70in,bottom=0.76in,left=0.68in,right=0.68in,columnsep=0.27in]{geometry}
\usepackage{bm}
\usepackage{enumitem}
\usepackage{placeins}
\IfFileExists{GeneralVersion/macros.tex}
  {\usepackage[T1]{fontenc}
\usepackage[english]{babel}
\usepackage{microtype}

\usepackage{newtxtext}
\usepackage{amsmath,amssymb,amsthm,amsbsy,amsfonts}
\usepackage{newtxmath}
\usepackage{tabularx}
\usepackage{mathtools}
\usepackage{latexsym}
\usepackage{esint}
\usepackage{physics}

\usepackage{array}
\usepackage{booktabs}
\usepackage{longtable}
\usepackage{makecell}

\usepackage{graphicx}
\usepackage{float}
\usepackage{placeins}
\usepackage{epstopdf}
\usepackage{caption}

\usepackage{xcolor}
\usepackage{tikz}
\usetikzlibrary{positioning,arrows.meta}
\usepackage{adjustbox}
\usepackage{multicol}

\usepackage{enumitem}
\usepackage{CJKutf8}

\usepackage{appendix}
\usepackage{titletoc}
\usepackage{titlesec}

\usepackage{listings}
\usepackage[most]{tcolorbox}
\tcbuselibrary{breakable,skins,theorems,listings}

\usepackage{xurl}

\usepackage[numbers,sort&compress]{natbib}
\setcitestyle{square,comma}
\usepackage[
  colorlinks=true,
  hypertexnames=false,
  linkcolor=blue,
  citecolor=blue,
  urlcolor=blue,
  filecolor=blue
]{hyperref}

\usepackage[capitalize,nameinlink]{cleveref}

\definecolor{myred}{RGB}{194,80,93}
\definecolor{myblue}{RGB}{82,119,194}
\hypersetup{
  linkcolor=myblue,
  citecolor=myblue,
  urlcolor=myblue,
  filecolor=myblue
}

\setlength{\parskip}{0pt}
\setlength{\parindent}{1em}
\setlength{\emergencystretch}{2em}
\setcounter{secnumdepth}{0}

\setlist[itemize]{leftmargin=1.5em, itemsep=0.15em, topsep=0.2em}

\setlength{\columnsep}{0.27in}
\setlength{\textfloatsep}{10pt plus 2pt minus 2pt}
\setlength{\floatsep}{8pt plus 2pt minus 2pt}
\setlength{\intextsep}{9pt plus 2pt minus 2pt}
\setlength{\abovecaptionskip}{4pt}
\setlength{\belowcaptionskip}{0pt}
\setlength{\footnotesep}{6.5pt}
\setlength{\skip\footins}{8pt plus 2pt minus 1pt}
\widowpenalty=10000
\clubpenalty=10000
\displaywidowpenalty=10000
\interfootnotelinepenalty=10000

\AtBeginDocument{%
  \setlength{\abovedisplayskip}{7pt plus 2pt minus 2pt}%
  \setlength{\belowdisplayskip}{7pt plus 2pt minus 2pt}%
  \setlength{\abovedisplayshortskip}{4pt plus 2pt minus 1pt}%
  \setlength{\belowdisplayshortskip}{5pt plus 2pt minus 1pt}%
}

\newcommand{\KGHeadingNoBreak}{\raggedright\hyphenpenalty=10000\exhyphenpenalty=10000}
\titleformat{\section}[block]
  {\normalfont\rmfamily\bfseries\fontsize{12.6}{14.6}\selectfont\KGHeadingNoBreak}
  {}{0pt}{}
\titlespacing*{\section}{0pt}{2.4ex plus 0.7ex minus 0.3ex}{0.85ex}

\titleformat{\subsection}[block]
  {\normalfont\rmfamily\bfseries\fontsize{10.5}{12.5}\selectfont\KGHeadingNoBreak}
  {}{0pt}{}
\titlespacing*{\subsection}{0pt}{2.0ex plus 0.5ex minus 0.25ex}{0.55ex}

\titleformat{\subsubsection}[block]
  {\normalfont\rmfamily\bfseries\normalsize\KGHeadingNoBreak}
  {}{0pt}{}
\titlespacing*{\subsubsection}{0pt}{1.6ex plus 0.4ex minus 0.2ex}{0.45ex}

\titleformat{\paragraph}[runin]
  {\normalfont\rmfamily\bfseries\normalsize}
  {}{0pt}{}[.]
\titlespacing*{\paragraph}{0pt}{1.25ex plus 0.3ex minus 0.2ex}{0.55em}

\DeclareCaptionLabelSeparator{kgbar}{\enspace\textbar\enspace}
\captionsetup{
  font={small,stretch=0.98},
  labelfont={bf},
  labelsep=kgbar,
  justification=raggedright,
  singlelinecheck=false
}
\captionsetup[figure]{name={Fig.}}
\renewcommand{\arraystretch}{1.08}

\newcommand{\MakeManuscriptTitle}[3]{%
  \begingroup
  \setlength{\parindent}{0pt}%
  \vspace*{0.15em}%
  {\rmfamily\bfseries\fontsize{17.2}{20.3}\selectfont\raggedright\hyphenpenalty=10000\exhyphenpenalty=10000 #1\par}%
  \vspace{0.82em}%
  {\fontsize{10.2}{12.2}\selectfont #2\par}%
  \vspace{0.35em}%
  {\fontsize{8.8}{10.6}\selectfont #3\par}%
  \vspace{0.82em}%
  \endgroup
}

\newcommand{\MakeCorrespondenceFootnotes}[2]{%
  \begingroup
  \renewcommand{\thefootnote}{\fnsymbol{footnote}}%
  \footnotetext[1]{\raggedright\footnotesize\href{mailto:#1}{#1}}%
  \footnotetext[2]{\raggedright\footnotesize\href{mailto:#2}{#2}}%
  \endgroup
}

\renewenvironment{abstract}{%
  \par\small\noindent\ignorespaces
}{%
  \par\vspace{0.8em}%
}

\newcommand{\BackMatterHeading}[1]{%
  \par\addvspace{1.7em}%
  \phantomsection
  \noindent{\rmfamily\bfseries\fontsize{10.5}{12.5}\selectfont #1\par}%
  \nobreak\vspace{0.35em}%
  \noindent\ignorespaces
}

\newcommand{\Eqref}[1]{%
  \hyperref[#1]{\textcolor{myred}{Eq.~(\ref*{#1})}}%
}

\newcommand{\Eqsref}[2]{%
  \hyperref[#1]{\textcolor{myred}{Eqs.~(\ref*{#1})}} and~%
  \hyperref[#2]{\textcolor{myred}{(\ref*{#2})}}%
}

\newcommand{\Figref}[1]{%
  \hyperref[#1]{\textcolor{myred}{Fig.~\ref*{#1}}}%
}

\newcommand{\Figpanelref}[2]{%
  \hyperref[#1]{\textcolor{myred}{Fig.~\ref*{#1}#2}}%
}

\newcommand{\Tabref}[1]{%
  \hyperref[#1]{\textcolor{myred}{Table~\ref*{#1}}}%
}

\newcommand{\SuppFigref}[1]{%
  \hyperref[#1]{\textcolor{myred}{Supplementary Fig.~\ref*{#1}}}%
}

\newcommand{\SuppFigpanelref}[2]{%
  \hyperref[#1]{\textcolor{myred}{Supplementary Fig.~\ref*{#1}#2}}%
}

\newcommand{\SuppTabref}[1]{%
  \hyperref[#1]{\textcolor{myred}{Supplementary Table~\ref*{#1}}}%
}

\newcommand{\SuppEqref}[1]{%
  \hyperref[#1]{\textcolor{myred}{Supplementary Eq.~(\ref*{#1})}}%
}

\newcommand{\SuppNoteref}[1]{%
  \hyperref[#1]{\textcolor{myred}{Supplementary Note~\ref*{#1}}}%
}

\newcommand{\SuppSubsecref}[1]{%
  \hyperref[#1]{\textcolor{myred}{Supplementary Note~\ref*{#1}}}%
}

\newcommand{\SuppSubsecsref}[2]{%
  \hyperref[#1]{\textcolor{myred}{Supplementary Notes~\ref*{#1}}} and~%
  \hyperref[#2]{\textcolor{myred}{\ref*{#2}}}%
}

\newcommand{\BeginSupplementaryInformation}[5]{%
  \clearpage
  \onecolumn
  \FloatBarrier
  \setcounter{page}{1}%
  \setcounter{section}{0}%
  \setcounter{subsection}{0}%
  \setcounter{subsubsection}{0}%
  \setcounter{figure}{0}%
  \setcounter{table}{0}%
  \setcounter{equation}{0}%
  \setcounter{footnote}{0}%
  \setcounter{secnumdepth}{1}%
  \renewcommand{\thefigure}{\arabic{figure}}%
  \renewcommand{\thetable}{\arabic{table}}%
  \renewcommand{\theequation}{S\arabic{equation}}%
  \captionsetup[figure]{name={Supplementary Fig.}}%
  \captionsetup[table]{name={Supplementary Table}}%

  \titleformat{\section}[block]
    {\normalfont\rmfamily\bfseries\fontsize{13.0}{15.2}\selectfont\KGHeadingNoBreak}
    {Supplementary Note~\thesection\enspace\textbar\enspace}{0pt}{}%
  \titlespacing*{\section}{0pt}{2.6ex plus 0.7ex minus 0.3ex}{0.9ex}%
  \titleformat{\subsection}[block]
    {\normalfont\rmfamily\bfseries\fontsize{10.8}{12.8}\selectfont\KGHeadingNoBreak}
    {}{0pt}{}%
  \titlespacing*{\subsection}{0pt}{2.0ex plus 0.5ex minus 0.2ex}{0.55ex}%
  \titleformat{\subsubsection}[block]
    {\normalfont\rmfamily\bfseries\normalsize\KGHeadingNoBreak}
    {}{0pt}{}%
  \titlespacing*{\subsubsection}{0pt}{1.5ex plus 0.4ex minus 0.2ex}{0.4ex}%
  \titleformat{\paragraph}[runin]
    {\normalfont\rmfamily\bfseries\normalsize}
    {}{0pt}{}[.]%
  \titlespacing*{\paragraph}{0pt}{1.2ex plus 0.3ex minus 0.2ex}{0.55em}%

  \begin{center}
    \vspace*{1.2em}
    \begin{minipage}{0.92\textwidth}
      \centering
      {\rmfamily\bfseries\fontsize{16.2}{19.0}\selectfont #1\par}
      \vspace{0.75em}
      {\rmfamily\bfseries\fontsize{13.2}{15.5}\selectfont Supplementary Information\par}
      \vspace{1.7em}
      {\normalsize #2\par}
      \vspace{0.9em}
      {\small #3\par}
      \vspace{1.45em}
      {\small #4\par}
      \vspace{0.7em}
      {\small\itshape #5\par}
    \end{minipage}
  \end{center}
  \vfill
  \clearpage
}

\newcommand{\knotlinewidth}{1.5pt}

\tikzset{
  knot/.style={
    line width=\knotlinewidth,
    baseline=-.5ex
  },
  move/.style={
    line width=\knotlinewidth
  },
  rest/.style={
    line width=\knotlinewidth,
    dashed
  },
  vertex/.style={
    circle,
    fill=black,
    inner sep=1.6pt
  },
  overcross/.style={
    double,
    line width=1.5,
    white,
    double=#1,
    double distance=\knotlinewidth
  },
  overcross/.default=black
}

\newcommand{\CrossPic}[1][]{%
  \tikz[knot,#1]{%
    \draw[move,blue] (-.48,.48) -- (.48,-.48);
    \draw[overcross=red] (-.48,-.48) -- (.48,.48);
  }%
}

\newcommand{\SidePic}[1][]{%
  \tikz[knot,#1]{%
    \draw[red,looseness=1.4] (-.5,-.5) to[out=0,in=0] (-.5,.5);
    \draw[blue,looseness=1.4] (.5,.5) to[out=180,in=180] (.5,-.5);
  }%
}

\newcommand{\TopPic}[1][]{%
  \tikz[knot,rotate=90,#1]{%
    \draw[red,looseness=1.4] (-.5,-.5) to[out=0,in=0] (-.5,.5);
    \draw[blue,looseness=1.4] (.5,.5) to[out=180,in=180] (.5,-.5);
  }%
}

\newcommand{\VertexPic}[1][]{%
  \tikz[knot,#1]{%
    \node[vertex] (v) at (0,0){};
    \draw[move,red]   (v) -- ++(45:0.55);
    \draw[move,green] (v) -- ++(135:0.55);
    \draw[move,black] (v) -- ++(225:0.55);
    \draw[move,blue]  (v) -- ++(315:0.55);
  }%
}

\newcommand{\Ypic}[1]{\vcenter{\hbox{\scalebox{0.55}{#1}}}}

\newcommand{\YCross}{\vcenter{\hbox{\scalebox{0.55}{\CrossPic}}}}
\newcommand{\YSide}{\Ypic{\SidePic}}
\newcommand{\YTop}{\Ypic{\TopPic}}
\newcommand{\YVertex}{\Ypic{\VertexPic}}

}
  {\usepackage[T1]{fontenc}
\usepackage[english]{babel}
\usepackage{microtype}

\usepackage{newtxtext}
\usepackage{amsmath,amssymb,amsthm,amsbsy,amsfonts}
\usepackage{newtxmath}
\usepackage{tabularx}
\usepackage{mathtools}
\usepackage{latexsym}
\usepackage{esint}
\usepackage{physics}

\usepackage{array}
\usepackage{booktabs}
\usepackage{longtable}
\usepackage{makecell}

\usepackage{graphicx}
\usepackage{float}
\usepackage{placeins}
\usepackage{epstopdf}
\usepackage{caption}

\usepackage{xcolor}
\usepackage{tikz}
\usetikzlibrary{positioning,arrows.meta}
\usepackage{adjustbox}
\usepackage{multicol}

\usepackage{enumitem}
\usepackage{CJKutf8}

\usepackage{appendix}
\usepackage{titletoc}
\usepackage{titlesec}

\usepackage{listings}
\usepackage[most]{tcolorbox}
\tcbuselibrary{breakable,skins,theorems,listings}

\usepackage{xurl}

\usepackage[numbers,sort&compress]{natbib}
\setcitestyle{square,comma}
\usepackage[
  colorlinks=true,
  hypertexnames=false,
  linkcolor=blue,
  citecolor=blue,
  urlcolor=blue,
  filecolor=blue
]{hyperref}

\usepackage[capitalize,nameinlink]{cleveref}

\definecolor{myred}{RGB}{194,80,93}
\definecolor{myblue}{RGB}{82,119,194}
\hypersetup{
  linkcolor=myblue,
  citecolor=myblue,
  urlcolor=myblue,
  filecolor=myblue
}

\setlength{\parskip}{0pt}
\setlength{\parindent}{1em}
\setlength{\emergencystretch}{2em}
\setcounter{secnumdepth}{0}

\setlist[itemize]{leftmargin=1.5em, itemsep=0.15em, topsep=0.2em}

\setlength{\columnsep}{0.27in}
\setlength{\textfloatsep}{10pt plus 2pt minus 2pt}
\setlength{\floatsep}{8pt plus 2pt minus 2pt}
\setlength{\intextsep}{9pt plus 2pt minus 2pt}
\setlength{\abovecaptionskip}{4pt}
\setlength{\belowcaptionskip}{0pt}
\setlength{\footnotesep}{6.5pt}
\setlength{\skip\footins}{8pt plus 2pt minus 1pt}
\widowpenalty=10000
\clubpenalty=10000
\displaywidowpenalty=10000
\interfootnotelinepenalty=10000

\AtBeginDocument{%
  \setlength{\abovedisplayskip}{7pt plus 2pt minus 2pt}%
  \setlength{\belowdisplayskip}{7pt plus 2pt minus 2pt}%
  \setlength{\abovedisplayshortskip}{4pt plus 2pt minus 1pt}%
  \setlength{\belowdisplayshortskip}{5pt plus 2pt minus 1pt}%
}

\newcommand{\KGHeadingNoBreak}{\raggedright\hyphenpenalty=10000\exhyphenpenalty=10000}
\titleformat{\section}[block]
  {\normalfont\rmfamily\bfseries\fontsize{12.6}{14.6}\selectfont\KGHeadingNoBreak}
  {}{0pt}{}
\titlespacing*{\section}{0pt}{2.4ex plus 0.7ex minus 0.3ex}{0.85ex}

\titleformat{\subsection}[block]
  {\normalfont\rmfamily\bfseries\fontsize{10.5}{12.5}\selectfont\KGHeadingNoBreak}
  {}{0pt}{}
\titlespacing*{\subsection}{0pt}{2.0ex plus 0.5ex minus 0.25ex}{0.55ex}

\titleformat{\subsubsection}[block]
  {\normalfont\rmfamily\bfseries\normalsize\KGHeadingNoBreak}
  {}{0pt}{}
\titlespacing*{\subsubsection}{0pt}{1.6ex plus 0.4ex minus 0.2ex}{0.45ex}

\titleformat{\paragraph}[runin]
  {\normalfont\rmfamily\bfseries\normalsize}
  {}{0pt}{}[.]
\titlespacing*{\paragraph}{0pt}{1.25ex plus 0.3ex minus 0.2ex}{0.55em}

\DeclareCaptionLabelSeparator{kgbar}{\enspace\textbar\enspace}
\captionsetup{
  font={small,stretch=0.98},
  labelfont={bf},
  labelsep=kgbar,
  justification=raggedright,
  singlelinecheck=false
}
\captionsetup[figure]{name={Fig.}}
\renewcommand{\arraystretch}{1.08}

\newcommand{\MakeManuscriptTitle}[3]{%
  \begingroup
  \setlength{\parindent}{0pt}%
  \vspace*{0.15em}%
  {\rmfamily\bfseries\fontsize{17.2}{20.3}\selectfont\raggedright\hyphenpenalty=10000\exhyphenpenalty=10000 #1\par}%
  \vspace{0.82em}%
  {\fontsize{10.2}{12.2}\selectfont #2\par}%
  \vspace{0.35em}%
  {\fontsize{8.8}{10.6}\selectfont #3\par}%
  \vspace{0.82em}%
  \endgroup
}

\newcommand{\MakeCorrespondenceFootnotes}[2]{%
  \begingroup
  \renewcommand{\thefootnote}{\fnsymbol{footnote}}%
  \footnotetext[1]{\hypertarget{corr-hakan}{}\raggedright\footnotesize\href{mailto:#1}{#1}}%
  \footnotetext[2]{\hypertarget{corr-lee}{}\raggedright\footnotesize\href{mailto:#2}{#2}}%
  \endgroup
}

\renewenvironment{abstract}{%
  \par\small\noindent\ignorespaces
}{%
  \par\vspace{0.8em}%
}

\newcommand{\BackMatterHeading}[1]{%
  \par\addvspace{1.7em}%
  \phantomsection
  \noindent{\rmfamily\bfseries\fontsize{10.5}{12.5}\selectfont #1\par}%
  \nobreak\vspace{0.35em}%
  \noindent\ignorespaces
}

\newcommand{\Eqref}[1]{%
  \hyperref[#1]{\textcolor{myred}{Eq.~(\ref*{#1})}}%
}

\newcommand{\Eqsref}[2]{%
  \hyperref[#1]{\textcolor{myred}{Eqs.~(\ref*{#1})}} and~%
  \hyperref[#2]{\textcolor{myred}{(\ref*{#2})}}%
}

\newcommand{\Figref}[1]{%
  \hyperref[#1]{\textcolor{myred}{Fig.~\ref*{#1}}}%
}

\newcommand{\Figpanelref}[2]{%
  \hyperref[#1]{\textcolor{myred}{Fig.~\ref*{#1}#2}}%
}

\newcommand{\Tabref}[1]{%
  \hyperref[#1]{\textcolor{myred}{Table~\ref*{#1}}}%
}

\newcommand{\SuppFigref}[1]{%
  \hyperref[#1]{\textcolor{myred}{Supplementary Fig.~\ref*{#1}}}%
}

\newcommand{\SuppFigpanelref}[2]{%
  \hyperref[#1]{\textcolor{myred}{Supplementary Fig.~\ref*{#1}#2}}%
}

\newcommand{\SuppTabref}[1]{%
  \hyperref[#1]{\textcolor{myred}{Supplementary Table~\ref*{#1}}}%
}

\newcommand{\SuppEqref}[1]{%
  \hyperref[#1]{\textcolor{myred}{Supplementary Eq.~(\ref*{#1})}}%
}

\newcommand{\SuppNoteref}[1]{%
  \hyperref[#1]{\textcolor{myred}{Supplementary Note~\ref*{#1}}}%
}

\newcommand{\SuppSubsecref}[1]{%
  \hyperref[#1]{\textcolor{myred}{Supplementary Note~\ref*{#1}}}%
}

\newcommand{\SuppSubsecsref}[2]{%
  \hyperref[#1]{\textcolor{myred}{Supplementary Notes~\ref*{#1}}} and~%
  \hyperref[#2]{\textcolor{myred}{\ref*{#2}}}%
}

\newcommand{\BeginSupplementaryInformation}[5]{%
  \clearpage
  \onecolumn
  \FloatBarrier
  \setcounter{page}{1}%
  \setcounter{section}{0}%
  \setcounter{subsection}{0}%
  \setcounter{subsubsection}{0}%
  \setcounter{figure}{0}%
  \setcounter{table}{0}%
  \setcounter{equation}{0}%
  \setcounter{footnote}{0}%
  \setcounter{secnumdepth}{1}%
  \renewcommand{\thefigure}{\arabic{figure}}%
  \renewcommand{\thetable}{\arabic{table}}%
  \renewcommand{\theequation}{S\arabic{equation}}%
  \captionsetup[figure]{name={Supplementary Fig.}}%
  \captionsetup[table]{name={Supplementary Table}}%

  \titleformat{\section}[block]
    {\normalfont\rmfamily\bfseries\fontsize{13.0}{15.2}\selectfont\KGHeadingNoBreak}
    {Supplementary Note~\thesection\enspace\textbar\enspace}{0pt}{}%
  \titlespacing*{\section}{0pt}{2.6ex plus 0.7ex minus 0.3ex}{0.9ex}%
  \titleformat{\subsection}[block]
    {\normalfont\rmfamily\bfseries\fontsize{10.8}{12.8}\selectfont\KGHeadingNoBreak}
    {}{0pt}{}%
  \titlespacing*{\subsection}{0pt}{2.0ex plus 0.5ex minus 0.2ex}{0.55ex}%
  \titleformat{\subsubsection}[block]
    {\normalfont\rmfamily\bfseries\normalsize\KGHeadingNoBreak}
    {}{0pt}{}%
  \titlespacing*{\subsubsection}{0pt}{1.5ex plus 0.4ex minus 0.2ex}{0.4ex}%
  \titleformat{\paragraph}[runin]
    {\normalfont\rmfamily\bfseries\normalsize}
    {}{0pt}{}[.]%
  \titlespacing*{\paragraph}{0pt}{1.2ex plus 0.3ex minus 0.2ex}{0.55em}%

  \begin{center}
    \vspace*{1.2em}
    \begin{minipage}{0.92\textwidth}
      \centering
      {\rmfamily\bfseries\fontsize{16.2}{19.0}\selectfont #1\par}
      \vspace{0.75em}
      {\rmfamily\bfseries\fontsize{13.2}{15.5}\selectfont Supplementary Information\par}
      \vspace{1.7em}
      {\normalsize #2\par}
      \vspace{0.9em}
      {\small #3\par}
      \vspace{1.45em}
      {\small #4\par}
      \vspace{0.7em}
      {\small\itshape #5\par}
    \end{minipage}
  \end{center}
  \vfill
  \clearpage
}

\newcommand{\knotlinewidth}{1.5pt}

\tikzset{
  knot/.style={
    line width=\knotlinewidth,
    baseline=-.5ex
  },
  move/.style={
    line width=\knotlinewidth
  },
  rest/.style={
    line width=\knotlinewidth,
    dashed
  },
  vertex/.style={
    circle,
    fill=black,
    inner sep=1.6pt
  },
  overcross/.style={
    double,
    line width=1.5,
    white,
    double=#1,
    double distance=\knotlinewidth
  },
  overcross/.default=black
}

\newcommand{\CrossPic}[1][]{%
  \tikz[knot,#1]{%
    \draw[move,blue] (-.48,.48) -- (.48,-.48);
    \draw[overcross=red] (-.48,-.48) -- (.48,.48);
  }%
}

\newcommand{\SidePic}[1][]{%
  \tikz[knot,#1]{%
    \draw[red,looseness=1.4] (-.5,-.5) to[out=0,in=0] (-.5,.5);
    \draw[blue,looseness=1.4] (.5,.5) to[out=180,in=180] (.5,-.5);
  }%
}

\newcommand{\TopPic}[1][]{%
  \tikz[knot,rotate=90,#1]{%
    \draw[red,looseness=1.4] (-.5,-.5) to[out=0,in=0] (-.5,.5);
    \draw[blue,looseness=1.4] (.5,.5) to[out=180,in=180] (.5,-.5);
  }%
}

\newcommand{\VertexPic}[1][]{%
  \tikz[knot,#1]{%
    \node[vertex] (v) at (0,0){};
    \draw[move,red]   (v) -- ++(45:0.55);
    \draw[move,green] (v) -- ++(135:0.55);
    \draw[move,black] (v) -- ++(225:0.55);
    \draw[move,blue]  (v) -- ++(315:0.55);
  }%
}

\newcommand{\Ypic}[1]{\vcenter{\hbox{\scalebox{0.55}{#1}}}}

\newcommand{\YCross}{\vcenter{\hbox{\scalebox{0.55}{\CrossPic}}}}
\newcommand{\YSide}{\Ypic{\SidePic}}
\newcommand{\YTop}{\Ypic{\TopPic}}
\newcommand{\YVertex}{\Ypic{\VertexPic}}

}
\renewcommand{\dbltopfraction}{0.92}
\renewcommand{\textfraction}{0.08}
\renewcommand{\dblfloatpagefraction}{0.75}

\newcommand{\ManuscriptTitle}{KnottedGraph: Scalable knotted-graph topology for scientific and mathematical discovery}
\newcommand{\ManuscriptAuthors}{
Hakan Akg\"un\textsuperscript{1,\hyperlink{corr-hakan}{\textcolor{myred}{*}}},
Xianquan Yan\textsuperscript{1,2},
Kehan Liu\textsuperscript{1},
Zhaoyun Chen\textsuperscript{1},
Ching Hua Lee\textsuperscript{1,\hyperlink{corr-lee}{\textcolor{myred}{\textdagger}}}
}
\newcommand{\ManuscriptAffiliations}{
\textsuperscript{1}Department of Physics, National University of Singapore, Singapore 117551, Singapore\\
\textsuperscript{2}Department of Computer Science, National University of Singapore, Singapore 117417, Singapore\\
}

\newcommand{\ManuscriptCorrespondence}{
Hakan Akg\"un, hakan.akgun@u.nus.edu;
Ching Hua Lee, phylch@nus.edu.sg
}
\newcommand{\SupplementaryVersionDate}{25 September 2026}

\title{\ManuscriptTitle}
\author{\ManuscriptAuthors}
\date{}

\hypersetup{
  pdftitle={\ManuscriptTitle},
  pdfauthor={Hakan Akg\"un; Xianquan Yan; Kehan Liu; Zhaoyun Chen; Ching Hua Lee}
}

\begin{document}

\twocolumn[
\begin{@twocolumnfalse}
\MakeManuscriptTitle{\ManuscriptTitle}{\ManuscriptAuthors}{\ManuscriptAffiliations}
\begin{abstract}
Scientific data span heterogeneous structures, including coordinates, networks, surfaces, volumes and fields, yet their topology can be quantified within a common framework through graph connectivity, cycle structure, genus and spatial embedding. Graph- and homology-based summaries do not determine spatial embedding, while standard knot and link polynomials require extensions to accommodate branching graphs. Here, we introduce \texttt{KnottedGraph}, a computational framework that converts such scientific representations to knotted graphs that retain graph connectivity and spatial embedding together. It constructs projected diagrams and PD codes, enabling various topological analyses, including Yamada-polynomial evaluation for topological classification. For scalable exact evaluation, it combines partial resolutions that leave the same unresolved connections and optimizes their processing order; the resulting algorithm is verified against published topological invariants of knotted graphs with up to 500 crossings. This scalability enables us to introduce an LLM-assisted mathematical-discovery methodology, in which computational topological data generated across knotted-graph families are used to identify candidate closed-form formulas. With this approach, we identify analytical Yamada-polynomials for generic graph motif families exhibiting Abelian and non-Abelian word sequences. Together, these scalable capabilities make knotted-graph topology computationally accessible across scientific domains, enabling large-scale classification and introducing a route from topological data to LLM-assisted AI4Math discovery.
\end{abstract}
\vspace{1em}
\end{@twocolumnfalse}
]
\MakeCorrespondenceFootnotes{hakan.akgun@u.nus.edu}{phylch@nus.edu.sg}

\section{Introduction}

\begin{figure*}[t!]
\centering
\includegraphics[
width=\linewidth,
keepaspectratio
]{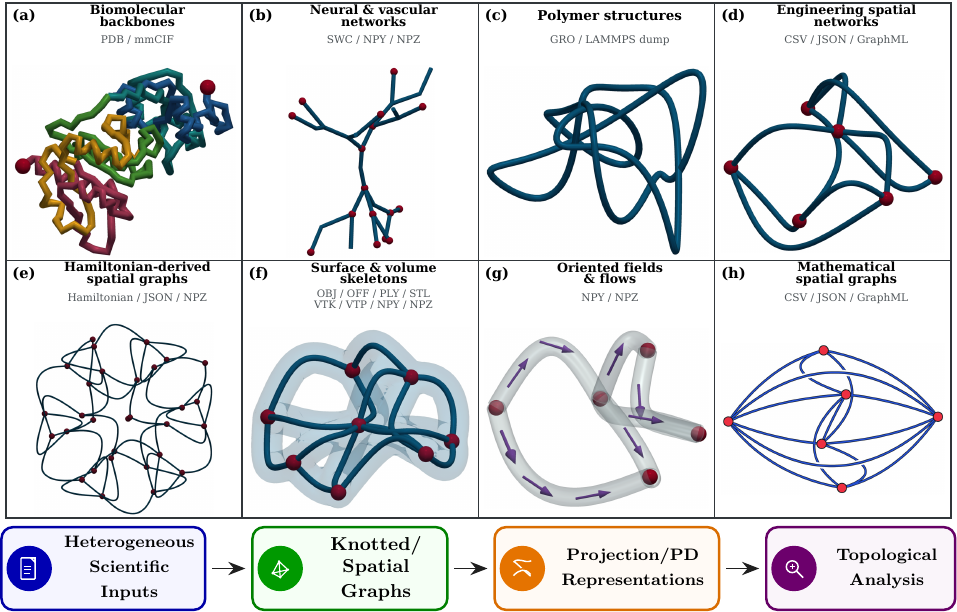}
\caption{
\textbf{Heterogeneous scientific geometry is brought to one knotted-graph representation.}
\textbf{(a)} Biomolecular backbones, represented by the C$\alpha$ trace of chain A of 1IPA \cite{kg_nureki2002rrma}, from the RCSB Protein Data Bank
\cite{kg_burley2019rcsb,kg_yeates2007protein_topology};
\textbf{(b)} neuronal and vascular morphologies
\cite{kg_cuntz2010neuronal_graphs,kg_aylward2005spatial_vascular};
\textbf{(c)} polymer configurations \cite{kg_deguchi2017topological_polymers,kg_topoly2021};
\textbf{(d)} engineered node--edge networks
\cite{kg_shai1999engineering_graphs,kg_peddada2023};
\textbf{(e)} Hamiltonian-derived level-set or occupied-region geometry
\cite{kg_fang2016nodal_line,kg_bi2017nodalknot,kg_lee2020nodalknots,akgun_yan_2026topologicalclassificationknottedgraphs};
\textbf{(f)} a volumetric regular-neighborhood input and a candidate graph spine
\cite{kg_ishii2008handlebody,kg_ishii2012handlebody};
\textbf{(g)} vector field curves \cite{kg_helman1989vector_field_topology};
and \textbf{(h)} mathematical spatial graphs
\cite{kg_kauffman1989spatial,kg_flapan2017spatial}.
The bottom workflow summarizes the common analysis path from heterogeneous
scientific inputs to knotted/spatial graphs, projection/PD representations and topological analysis.
Format labels in panels (b), (d) and (h) describe source data, not necessarily direct adapter inputs: SWC, GraphML and arbitrary graph-JSON records are converted externally to paired node--edge CSV before use (\SuppTabref{supptab:input_interfaces}).
}
\label{fig:input_formats_overview}
\end{figure*}

Many systems in biology, materials science and physics are governed by
graph connectivity and spatial embedding. Examples include
biomolecular backbones \cite{kg_burley2019rcsb,kg_yeates2007protein_topology}
(\Figpanelref{fig:input_formats_overview}{a}),
vascular \cite{kg_aylward2005spatial_vascular} and neuronal networks
\cite{kg_cuntz2010neuronal_graphs}
(\Figpanelref{fig:input_formats_overview}{b}), polymers
\cite{kg_deguchi2017topological_polymers,kg_topoly2021}
(\Figpanelref{fig:input_formats_overview}{c}), engineered spatial networks
\cite{kg_shai1999engineering_graphs,kg_peddada2023}
(\Figpanelref{fig:input_formats_overview}{d}), Hamiltonian-derived geometries
\cite{kg_fang2016nodal_line,kg_bi2017nodalknot,kg_lee2020nodalknots,
kg_yan2025hsg12m,yan_graphtransformer,
akgun_yan_2026topologicalclassificationknottedgraphs}
(\Figpanelref{fig:input_formats_overview}{e}) and vector flow curves
\cite{akgun_yan_2026topologicalclassificationknottedgraphs, kg_helman1989vector_field_topology}
(\Figpanelref{fig:input_formats_overview}{g}). Their physical origins differ,
but the computational question is the same when underlying topology matters: the representation must retain graph connectivity and spatial embedding, i.e., edge over/undercrossing information.

Graph connectivity and homological quantities capture abstract topological structure
\cite{carlsson2009topology,edelsbrunner2008persistent,kg_su2026tda}, but do not in general determine the ambient-isotopy class of a spatial graph
\cite{kg_kauffman1989spatial,kg_flapan2017spatial}. Classical knot and link polynomials are sensitive to embedding, but their standard definitions apply to knots and links
\cite{alexander1928topological,jones1985polynomial,kg_homfly1985,rolfsen2003knots},
and extensions are required to accommodate branching graph connectivity \cite{kg_mellor2018spatial}. On the other hand, knotted graphs\footnote{Following the terminology of our companion work \cite{akgun_yan_2026topologicalclassificationknottedgraphs}, we use \emph{knotted graph} and \emph{spatial graph} synonymously throughout this work, with \emph{spatial graph} retained where conventional mathematical terminology is useful.} retain both types of information, graph connectivity and spatial embedding, in the same object and can be distinguished using their associated invariants \cite{kg_kauffman1989spatial,kg_flapan2017spatial,kg_mellor2018spatial}, one of the most prominent being the Yamada polynomial \cite{kg_yamada1989,kg_mellor2018spatial}.

Previous work has established the usefulness of knotted-graph descriptions and
invariants in polymer systems, engineering design and condensed-matter physics
\cite{kg_deguchi2017topological_polymers,kg_peddada2023,
akgun_yan_2026topologicalclassificationknottedgraphs}. Extending these
analyses across heterogeneous scientific data requires a common knotted-graph
representation and scalable invariant evaluation. Explicit curves and networks
can enter this representation directly, whereas surfaces and volumes require
construction of the corresponding spatial graph while preserving graph
connectivity and three-dimensional edge geometry, including candidate graph
spines when a handlebody--spine description is appropriate
\cite{kg_ishii2008handlebody,kg_ishii2012handlebody}. The resulting knotted
graph can then be projected to a planar diagram retaining graph vertices,
projection crossings and edge over/undercrossing information
\cite{kg_kauffman1989spatial,kg_flapan2017spatial,kg_peddada2023}, converted
to PD encoding \cite{kg_mastin2015pdcode,kg_topoly2021}, and evaluated through
the Yamada polynomial \cite{kg_yamada1989,kg_mellor2018spatial}. Here, we
introduce a computational framework that integrates these stages and makes the
invariant calculation scalable enough for repeated analysis across derived
datasets, parameter sweeps and mathematically generated graph families, while
supporting additional analyses including vector-flow characterization \cite{akgun_yan_2026topologicalclassificationknottedgraphs},
repulsive-curve preconditioning \cite{kg_yu2021repulsive} and parameter-space analysis \cite{akgun_yan_2026topologicalclassificationknottedgraphs}.

For embedding-sensitive topological analysis at scale, one important algorithmic bottleneck is Yamada evaluation
\cite{kg_yamada1989,kg_mellor2018spatial}. Our algorithm combines resolutions whenever they leave the same unresolved connections and
orders local operations to keep the number of simultaneously unresolved arc ends small. On the structured published families in
Refs.~\cite{kg_dobrynin1996yamada,kg_li2018yamada}, this allows us to reproduce the
known formulas of spatial graphs with up to $500$ crossings. This efficiency enables repeated invariant evaluation across scientific parameter spaces and algorithmically generated graph families, where exact topological invariants can also serve as data for mathematical discovery.

Recent advances in AI-assisted mathematics have rapidly expanded
evaluation-guided search, mathematical construction and machine-verifiable
reasoning across increasingly challenging research settings
\cite{kg_funsearch2024,yan2025metric,kg_alphaevolve2025,kg_georgiev2025mathematical,
kg_alphaproof2026,burtsev2026ai}. In structured spatial-graph families,
Yamada-polynomial formulas are traditionally derived analytically from the
family construction, for example using recurrence or transfer-matrix methods
\cite{kg_dobrynin1996yamada,kg_li2018yamada,kg_lundstrom2022transfer}.
Such derivations are generally family-specific and are not automatically
inferred from computed invariant data. Here, we introduce an LLM-assisted
mathematical-discovery methodology in which Yamada values are generated
computationally, used to identify candidate family relations, converted to
explicit symbolic forms and tested on additional family members. Our framework
thereby extends this emerging direction to exact topological-invariant data and
the reconstruction of interpretable algebraic structure across knotted-graph
families.

More broadly, the framework opens several directions beyond the results developed here: embedding-sensitive analysis across scientific parameter spaces; extension from handlebody spines to more general volumetric topologies; implementation of additional knot, link and spatial-graph invariants, including the Alexander, Jones and HOMFLY--PT polynomials \cite{alexander1928topological,jones1985polynomial,kg_homfly1985}; and LLM-assisted mathematical discovery across broader structured families, including braid words, tangles, satellite operations and other recursive constructions \cite{kg_birman1974braids,kg_conway1970tangles,kg_schubert1953satellite,kg_lundstrom2022transfer}. Together, these directions connect scalable embedding-sensitive topological analysis with a general route from exact invariant data to automated mathematical formula discovery.

\section{Results}

\begin{figure*}[t!]
    \centering
    \includegraphics[width=\linewidth]
    {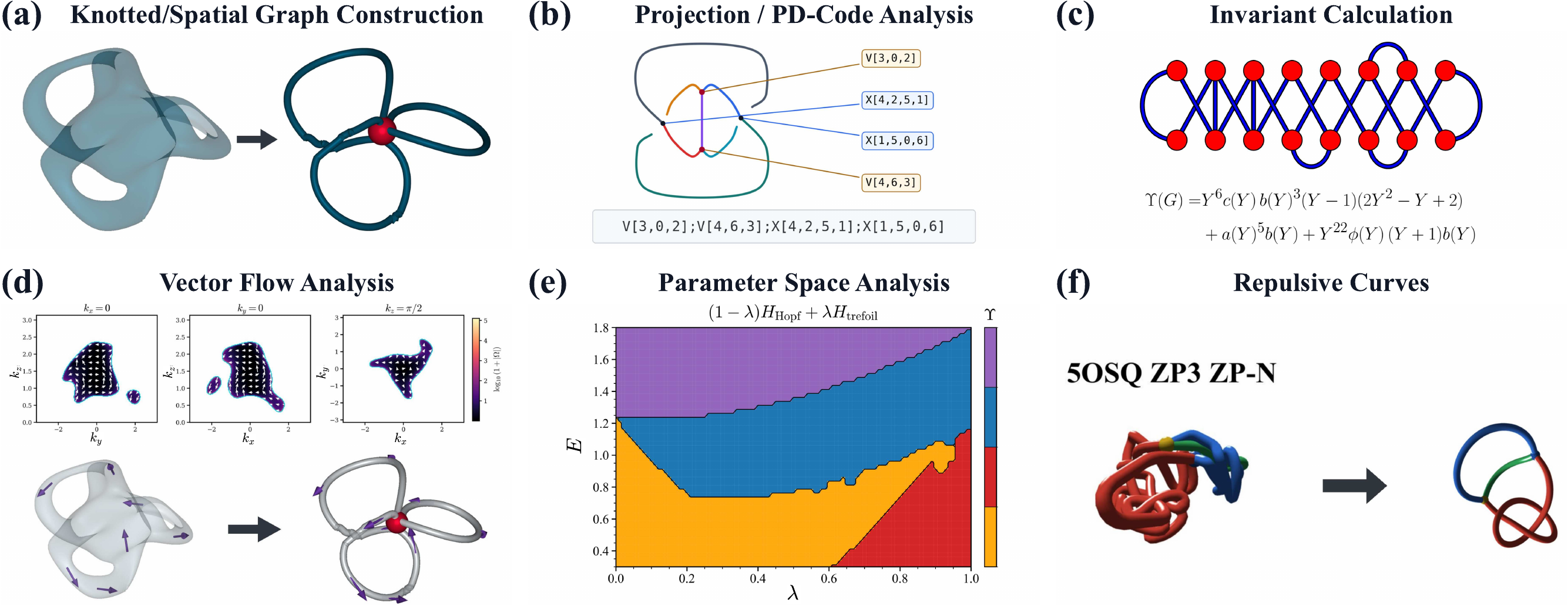}
    \caption{
    \textbf{One knotted-graph representation supports knotted-graph construction,
    invariant evaluation and repeated scientific analysis.}
    \textbf{(a)} Volumetric geometry is processed into a candidate graph spine;
    \textbf{(b)} the embedding is projected to a regular diagram and PD
    code; \textbf{(c)} the Yamada polynomial is evaluated;
    \textbf{(d)} vector flow analysis and graph-level orientation remain attached to the
    knotted graph \cite{akgun_yan_2026topologicalclassificationknottedgraphs}; \textbf{(e)} repeated calculations organize parameter-dependent graph outputs \cite{akgun_yan_2026topologicalclassificationknottedgraphs}; and \textbf{(f)} optional geometric preconditioning
    can regularize difficult embeddings before projection
    \cite{kg_yu2021repulsive}.
    }
    \label{fig:functionality_overview}
\end{figure*}

\subsection{A common knotted-graph representation for scientific geometry}
\label{sec:unified_representation}

We use the knotted graph as the common representation through which heterogeneous scientific geometries become directly comparable
(\Figref{fig:input_formats_overview}; \SuppFigref{suppfig:input_yamada_examples}). For inputs in which this graph is not given explicitly, a central contribution is knotted-graph construction: when a handlebody--spine representation is appropriate (\SuppSubsecref{supp:handlebody_basis}), surface or volumetric geometry is reduced to a one-dimensional skeleton
\cite{kg_lee1994thinning} and converted to a knotted graph through
multiscale junction identification and topology selection
\cite{kg_reinders2000skeletongraph}, as shown in
\Figpanelref{fig:functionality_overview}{a} and \SuppFigref{suppfig:skeletonization_steps}. The resulting graph retains connectivity and three-dimensional edge geometry for
projection and PD construction (\SuppNoteref{supp:projection}), invariant
evaluation (\SuppNoteref{supp:yamada_engine}), geometric and vector-flow
analysis, parameter-space analysis
(\SuppNoteref{supp:knot_fields}) and geometric preconditioning
(\SuppNoteref{supp:repulsive_layout}), as summarized in
\Figpanelref{fig:functionality_overview}{b--f}.

For surface or volumetric inputs representing a handlebody or handlebody-link,
the Yamada-invariant route is
\begin{equation}
\Omega
\longrightarrow
G_{\mathrm{spine}}
\longrightarrow
\operatorname{PD}(G_{\mathrm{spine}})
\longrightarrow
\overline{\Upsilon}(G_{\mathrm{spine}};Y),
\label{eq:main_pipeline}
\end{equation}
where $\Omega$ is the volumetric region,
$G_{\mathrm{spine}}$ the resulting candidate graph spine,
$\operatorname{PD}$ the corresponding PD code and
$\overline{\Upsilon}$ the normalized Yamada polynomial. Because these stages
are modular, applications can enter \Eqref{eq:main_pipeline} at the
representation appropriate to their input and use only the required downstream
operations.

The first transformation in \Eqref{eq:main_pipeline} is nontrivial because a
finite-resolution skeleton does not uniquely identify the physical junctions
of the underlying graph. A thick junction can thin to several nearby branch
voxels, while distinct physical junctions can also lie close in space. We
therefore construct candidate graphs from the same skeleton over neighbouring junction scales,
\begin{equation}
\Omega
\longrightarrow
S(\Omega)
\longrightarrow
\left\{G^{(h)}\right\}_{h=0}^{h_{\max}}
\longrightarrow
G_{\mathrm{spine}},
\label{eq:main_multiscale_reconstruction}
\end{equation}
where $S(\Omega)$ is the one-voxel-wide skeleton and $h$ controls the local
junction region. When a topology recurs across the tested scales, this cross-scale recurrence determines
the selected graph; if no valid topology recurs, the direct zero-radius trace
is retained. For large digital skeletons, sparse voxel connectivity is constructed once and reused across candidate scales, with three-dimensional edge coordinates generated only for the selected scale while preserving the same cross-scale consistency criterion (\SuppSubsecref{supp:extraction}).

The next transformation projects a spatial embedding to a regular
diagram \cite{kg_kauffman1989spatial,kg_flapan2017spatial}. For a sampled viewing direction $P_i$, let
\begin{equation}
D_i=P_i(G_{\mathrm{spine}}),
\qquad
n_{\mathrm{cr},i}=\#X(D_i),
\label{eq:main_projection_candidates}
\end{equation}
where $n_{\mathrm{cr},i}$ is the number of projection crossings. Constructing the complete
PD code for every sampled view is unnecessary. Instead,
\texttt{KnottedGraph} first compares the crossing counts $n_{\mathrm{cr},i}$ and constructs
complete diagrams only for the best candidates
(\SuppSubsecref{supp:projection_sampling}; \SuppFigpanelref{suppfig:pdcode_to_yamada}{a--c}). When a crossing is
found, its position along both parent graph edges is also recorded in the same
geometric search (\SuppSubsecref{supp:projection_crossings}). These positions are the cut locations required by
PD construction, so the same crossing does not need to be found again
(\SuppSubsecref{supp:pd_serialization}; \SuppFigref{suppfig:pd_code_generation}).

The final transformation is combinatorially more demanding. At each crossing,
the raw Yamada polynomial obeys the three-term local relation
\cite{kg_yamada1989},
\begin{equation}
\Upsilon\!\bigl(\YCross;Y\bigr)
=
Y\,\Upsilon\!\bigl(\YSide;Y\bigr)
+
Y^{-1}\,\Upsilon\!\bigl(\YTop;Y\bigr)
+
\Upsilon\!\bigl(\YVertex;Y\bigr),
\label{eq:main_yamada_local_relation}
\end{equation}
where \(\YSide\) and \(\YTop\) are the two smoothings and
\(\YVertex\) replaces the crossing by a graph vertex. For a complete resolution state $s$, let $G_s$ be
the resulting crossing-free graph and let
$n_{\mathrm{side}}(s)$ and $n_{\mathrm{top}}(s)$ count the two smoothing
choices. This is the standard Yamada state-sum construction
\cite{kg_yamada1989,kg_li2018yamada}; the diagram polynomial can be
written as
\begin{equation}
\Upsilon(D;Y)
=
\sum_{s\in\mathcal S(D)}
Y^{\,n_{\mathrm{side}}(s)-n_{\mathrm{top}}(s)}
\Upsilon(G_s;Y),
\label{eq:main_yamada_state_sum}
\end{equation}
For a crossing-free state, Yamada's graph polynomial is a specialization of
the Negami graph polynomial
\cite{kg_yamada1989,kg_negami1987,kg_li2018yamada}; equivalently, in the
included-edge form used by the library,
\begin{equation}
\Upsilon(G_s;Y)
=
(-1)^{|V_s|}
\sum_{F\subseteq E_s}
(-1)^{|F|}
\left(Y+2+Y^{-1}\right)^{\beta(F)} .
\label{eq:main_yamada_crossing_free}
\end{equation}
Here $V_s$ and $E_s$ are the vertices and edges of $G_s$, and
$\beta(F)=|F|-|V_s|+\omega(F)$ is the cycle rank of the spanning subgraph
$(V_s,F)$ and $\omega(F)$ is its number of connected components. Thus repeated
application of \Eqref{eq:main_yamada_local_relation} reduces the mathematical
problem to crossing-free graph contributions. Under the diagram moves
needed to pass from regular to ambient isotopy, the raw Yamada polynomial can
acquire a monomial factor \cite{kg_yamada1989,kg_mellor2018spatial}. To fix
that ambiguity consistently in the library, for a nonzero raw polynomial we
write
$\mu(D)=\min\deg_Y\Upsilon(D;Y)$ and use the convention
\begin{equation}
\overline{\Upsilon}(D;Y)
=
(-1)^{-\mu(D)}
Y^{-\mu(D)}
\Upsilon(D;Y),
\label{eq:main_yamada_normalization}
\end{equation}
with an evaluated zero kept as zero. For subcubic spatial graphs, this removes
the remaining monomial ambiguity so that
$\overline{\Upsilon}$ is independent of the accepted regular projection and
provides an ambient-isotopy invariant \cite{kg_yamada1989,kg_mellor2018spatial}
(\SuppNoteref{supp:yamada_engine}).

Since each crossing has the three local resolutions in
\Eqref{eq:main_yamada_local_relation}, a direct expansion of a diagram with
$n_{\mathrm{cr}}$ crossings contains
\begin{equation}
N_{\mathrm{res}}(n_{\mathrm{cr}})=3^{n_{\mathrm{cr}}}
\label{eq:main_yamada_resolution_space}
\end{equation}
complete resolution assignments
(\SuppSubsecref{supp:local_yamada_relation}; \SuppFigref{suppfig:yamada_resolution_states}).
Although \Eqref{eq:main_yamada_state_sum} sums over all of these resolutions,
many partial resolution histories become equivalent before completion when they
leave the same unresolved connectivity. Our algorithm therefore stores
one state for each such connectivity and combines the corresponding
Laurent-polynomial contributions.

\begin{figure*}[t!]
    \centering
    \includegraphics[width=\linewidth]
    {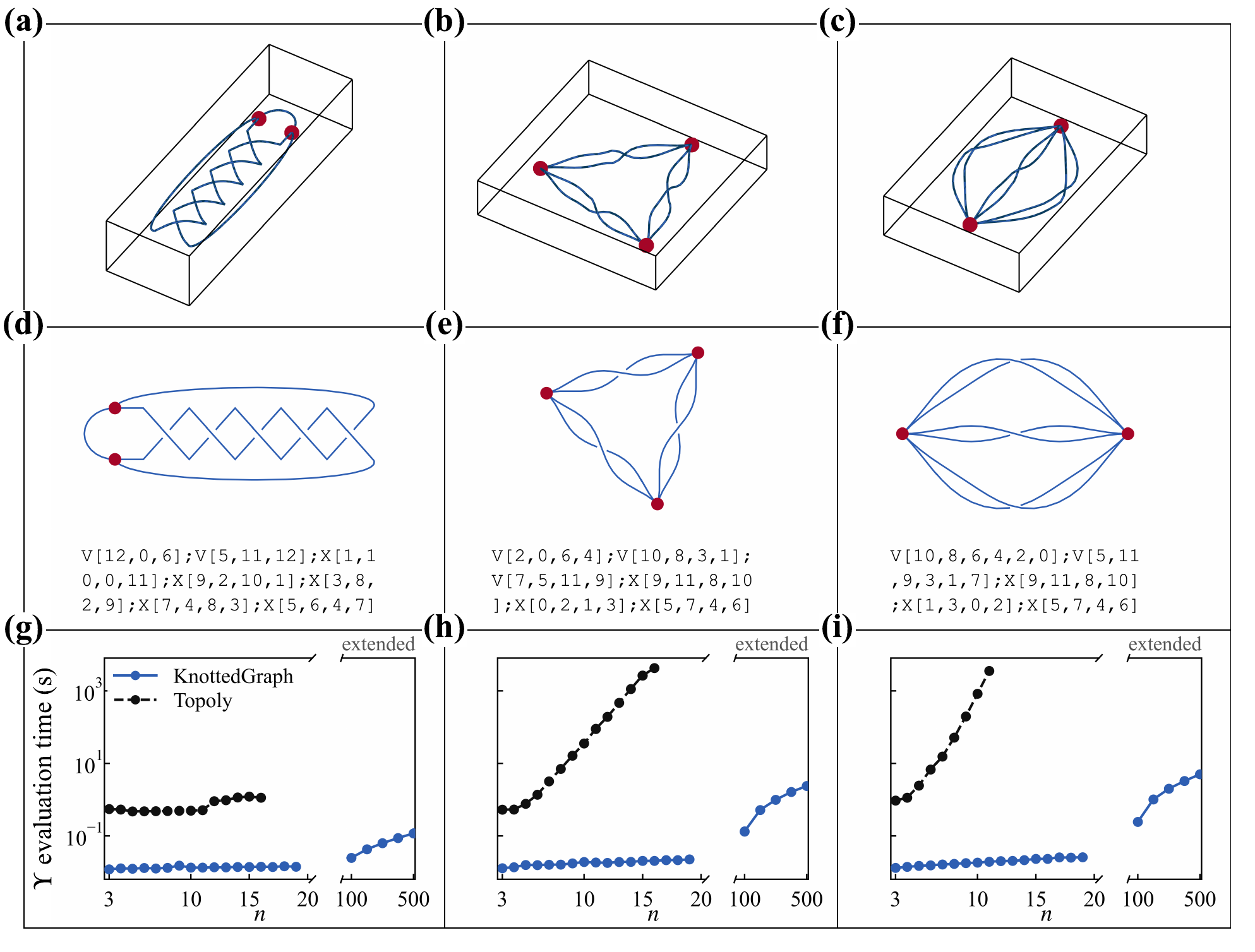}
    
\caption{
\textbf{Exact Yamada evaluation across published closed-form
spatial-graph families.}
Columns show the $\Theta(n)$ family
\cite{kg_dobrynin1996yamada}, the $C_n(\infty_{+})$ family and the
$\Theta_s(\infty_{+})$ family \cite{kg_li2018yamada}.
\textbf{(a,d)} Representative $\Theta(5)$ embedding and corresponding
projected diagram and PD code;
\textbf{(b,e)} representative $C_3(\infty_{+})$ embedding and corresponding
projected diagram and PD code;
\textbf{(c,f)} representative $\Theta_3(\infty_{+})$ embedding and
corresponding projected diagram and PD code.
\textbf{(g--i)} Invariant-evaluation time as a function of the corresponding
family parameter.
At $500$ crossings, the corresponding \texttt{KnottedGraph} evaluation
times are approximately $0.12\,\mathrm{s}$, $2.4\,\mathrm{s}$ and
$5.0\,\mathrm{s}$, with each result agreeing with the corresponding published
reference polynomial. External-software timings \cite{kg_topoly2021} are shown only where both
implementations reproduce the same published reference polynomial; the Topoly
series terminates once no further passing result is obtained under the benchmark
protocol. 
}
    \label{fig:knottedgraph_topoly_scaling}
\end{figure*}
The cost of this compressed calculation depends on $n_{\mathrm{cr}}$ and on how many
arc ends remain unresolved at the same time. \texttt{KnottedGraph} chooses a processing
order that keeps this number small and, for sufficiently difficult diagrams,
tests alternative starting choices. Vertices with many incident edges are
represented internally by equivalent smaller connection steps, further
reducing the number of unresolved arc ends.
Because the number of simultaneously unresolved arc ends also depends on the
projection, the projection with the fewest crossings need not give the fastest
Yamada evaluation. The state-combination and ordering procedures are developed in
\SuppSubsecref{supp:yamada_connectivity_states}, and the projection-refinement
procedure in \SuppSubsecref{supp:projection_sampling}. Thus, across the three transformations in \Eqref{eq:main_pipeline},
\texttt{KnottedGraph} retains only the information required by the next stage,
avoids constructing equivalent intermediate objects and uses the structure of
the downstream calculation when choosing among alternative representations.

\subsection{Independent tests of knotted-graph construction, projection and topological invariant evaluation}
\label{sec:validation_main}

Next, we validate the three steps individually: knotted graph construction in
\SuppSubsecref{supp:handlebody_validation}, projection and algebraic consistency in
\SuppSubsecref{supp:sanity_validation}, with the corresponding methods detailed in
\SuppSubsecref{supp:extraction}, \SuppNoteref{supp:projection} and
\SuppNoteref{supp:yamada_engine}, respectively.

Knotted-graph construction is tested through an inverse benchmark. Known knotted graphs
are thickened and voxelized, passed through the pipeline, and then compared
with their generating seeds
(\SuppFigref{suppfig:handlebody_stress}). The benchmark independently checks
the topology of the voxelized input, the graph connectivity and cycle structure of
the resulting graph, and the normalized Yamada polynomials of the seed and
output. Across all $4{,}400$ curated constructions, every case passes the stated
input and graph-level recovery checks, and all $4{,}400$ seed--output
normalized Yamada polynomials agree, showing consistent recovery of the
intended cycle structure and normalized Yamada invariant within the tested
handlebody regime and numerical conditions.

Then projection and invariant calculations are tested. Multiple
accepted generic views are generated for fixed subcubic spatial graphs.
Although different views can produce different crossing numbers, arc labels
and PD codes, all accepted regular projections of the same tested graph give the same
normalized Yamada polynomial, as required by its ambient-isotopy invariance in
the subcubic setting
\cite{kg_yamada1989,kg_mellor2018spatial}. Planar reference graphs also agree
with independently evaluated crossing-free graph polynomials. The Yamada
calculation is then tested against graph-polynomial identities, crossing and
mirroring relations, and published closed-form Yamada families
\cite{kg_yamada1989,kg_dobrynin1996yamada,kg_li2018yamada}. It reproduces the
tested tree, cycle, bouquet and theta relations
\cite{kg_yamada1989,kg_li2018yamada}, together with planar $K_4$ and other
published cubic-graph values
\cite{kg_dobrynin1996yamada}. Together, these tests verify projection consistency for the tested subcubic
embeddings and the Yamada calculation against independent mathematical
results. Stage-wise timing distributions for these steps and their subprocesses are reported in
\SuppFigref{suppfig:handlebody_timing_distributions}.

\subsection{Scalable Yamada polynomial evaluation across knotted graphs}
\label{sec:performance}
Having established the representation and its validation, we next test whether
spatial-graph invariants can be evaluated at a scale suitable for repeated
scientific and mathematical analysis. The defining local relation and full state sum
are given in \Eqsref{eq:main_yamada_local_relation}{eq:main_yamada_state_sum},
and the formal resolution-space growth in
\Eqref{eq:main_yamada_resolution_space}; the connectivity compression above
avoids explicit construction of that complete tree.

The structured-family test uses the three published families shown in
\Figref{fig:knottedgraph_topoly_scaling}
\cite{kg_dobrynin1996yamada,kg_li2018yamada}. The first column
(\Figpanelref{fig:knottedgraph_topoly_scaling}{a,d,g}) shows the
$\Theta(n)$ family; the second
(\Figpanelref{fig:knottedgraph_topoly_scaling}{b,e,h}) shows the
edge-replaced cycle family $C_n(\infty_{+})$, where each edge of the cycle
$C_n$ is replaced by the spatial part $\infty_{+}$; and the third
(\Figpanelref{fig:knottedgraph_topoly_scaling}{c,f,i}) shows the
corresponding edge-replaced theta family $\Theta_s(\infty_{+})$
\cite{kg_li2018yamada}. Across these families,
\texttt{KnottedGraph} reproduces the published formulas of knotted graphs up to $500$ crossings. At $500$ crossings, the measured evaluation
times are $0.12\,\mathrm{s}$, $2.4\,\mathrm{s}$ and $5.0\,\mathrm{s}$,
respectively, with each result verified against the corresponding published
formulas \cite{kg_dobrynin1996yamada,kg_li2018yamada} displayed in \SuppSubsecref{supp:sanity_validation}.

The invariant-evaluation scaling is shown in
\Figpanelref{fig:knottedgraph_topoly_scaling}{g--i}. Across the tested structured families, the algorithmic optimizations produce subexponential empirical runtime growth over the benchmarked range, enabling exact calculations through \(500\) crossings. In comparison, the external software implementation \cite{kg_topoly2021} exhibits approximately exponential growth over the accessible range, making direct comparison impractical at higher crossing numbers. The execution times at seven crossings are reported in \Tabref{tab:main_yamada_benchmark}.

\begin{table}[!t] \centering \small \setlength{\tabcolsep}{3.5pt} \begin{tabular}{lccc} \hline \textbf{Family} & \textbf{\texttt{KnottedGraph}} & \textbf{Topoly} & \textbf{Runtime ratio} \\ & \textbf{(s)} & \textbf{(s)} & \\ \hline $\Theta(n)$ & 0.0125 & 0.4758 & $38\times$ \\ $C_n(\infty_{+})$ & 0.0159 & 3.1880 & $200\times$ \\ $\Theta_s(\infty_{+})$ & 0.0162 & 15.4393 & $960\times$ \\ \hline \end{tabular} \caption{ \textbf{External-software timing comparison at seven crossings.} For each family, the two timings correspond to the same PD input and both implementations reproduce the corresponding published reference polynomial \cite{kg_dobrynin1996yamada,kg_li2018yamada} where the runtime ratio is $t_{\mathrm{Topoly}}/t_{\mathrm{KnottedGraph}}$. } \label{tab:main_yamada_benchmark} \end{table}

\begin{figure*}[t!]
    \centering
    \includegraphics[width=\linewidth]
    {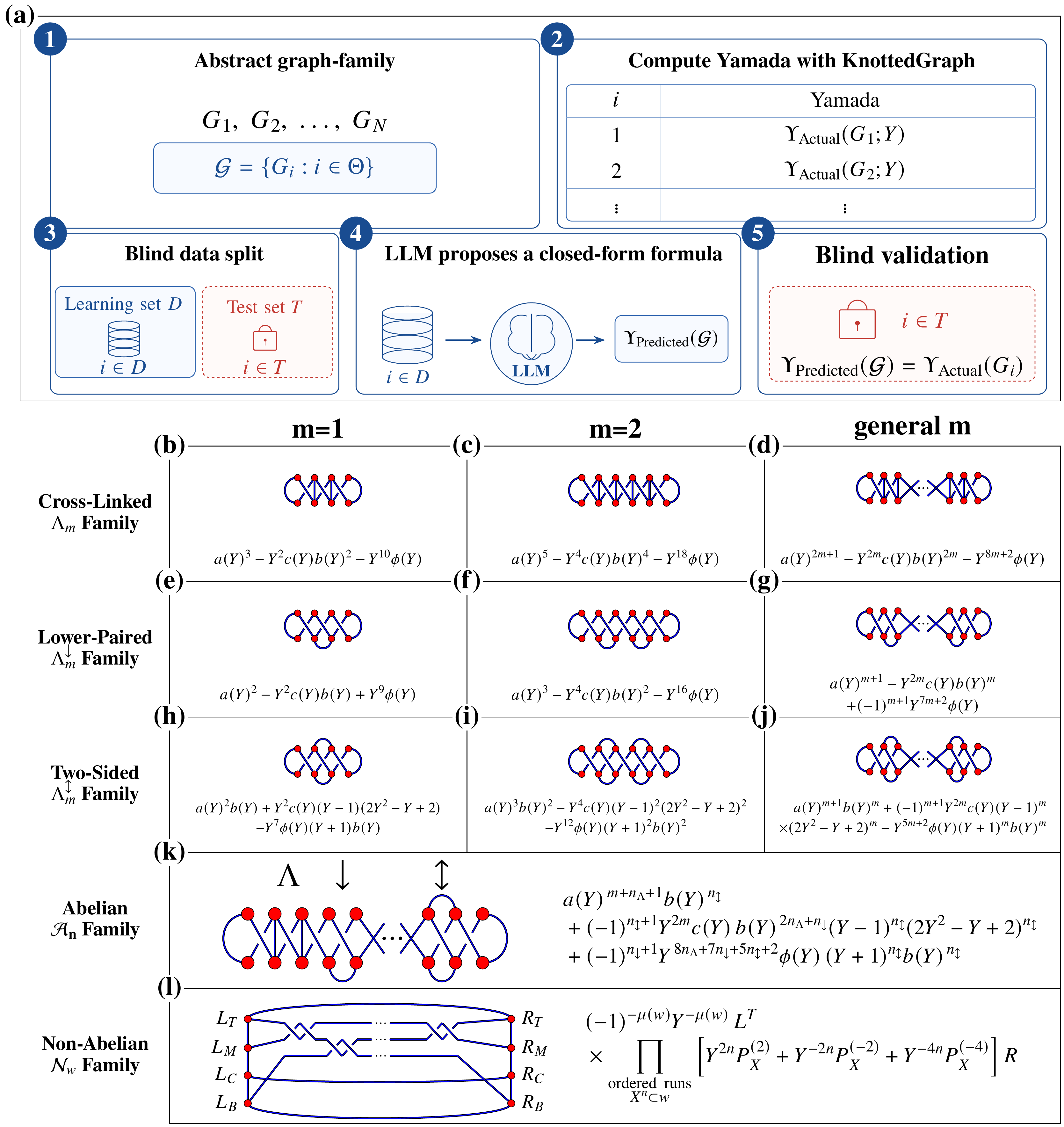}
\caption[Inferring and testing Yamada-polynomial family laws.]{
\textbf{Exact Yamada-polynomial data are used to infer and test commuting and
order-sensitive noncommuting family laws.}
\textbf{(a)} For a parameterized graph family, exact Yamada polynomials are
first generated with \texttt{KnottedGraph} and blindly divided into a discovery
set $D$ and a held-out test set $T$. Only $D$ is supplied during LLM-assisted
candidate generation; a retained relation is converted to a fixed symbolic
formula or transfer representation before the members of $T$ are evaluated.
\textbf{(b--d)} Cross-linked family $\Lambda_m$;
\textbf{(e--g)} lower-paired family $\Lambda_m^{\downarrow}$; and
\textbf{(h--j)} two-sided family $\Lambda_m^{\updownarrow}$, shown for
$m=1$, $m=2$ and the corresponding candidate general-$m$ law. The three
families exhibit a common three-channel transfer structure.
\textbf{(k)} Combining the three motifs within one family gives commuting
transfer operators, so the candidate invariant for an arbitrary motif word
depends only on the motif multiplicities $\mathbf n$ and is unchanged by
reordering.
\textbf{(l)} In contrast, on a fixed $8$-vertex, $12$-edge cubic multigraph
carrying ordered pure-braid blocks $A=\sigma_1^2$ and $B=\sigma_2^2$,
words with identical block multiplicities can yield different normalized
Yamada polynomials. Finite exact Yamada data reconstruct a $15$-state
noncommuting transfer representation whose generators obey a common cubic
relation, retaining braid-word order while reducing repeated runs to three
spectral channels.  
}
    \label{fig:three_theta_yamada_families}
\end{figure*}

\subsection{LLM-assisted mathematical discovery from exact knotted-graph data}
\label{sec:mathematical_discovery}

Recent advances in AI-assisted mathematics have shown how LLM-guided search,
computational evaluation and machine-verifiable reasoning can expose
mathematical structure across increasingly challenging problem settings
\cite{kg_funsearch2024,kg_alphaevolve2025,kg_georgiev2025mathematical,
kg_alphaproof2026}. Relations across parameterized mathematical families are
conventionally identified and derived through human-guided analysis of the
family construction and its computed examples
\cite{kg_dobrynin1996yamada,kg_li2018yamada,kg_lundstrom2022transfer}.
Previous Yamada-polynomial formulas for structured spatial-graph families have
likewise been obtained from analytically specified recurrences,
edge-replacement relations or transfer constructions
\cite{kg_dobrynin1996yamada,kg_li2018yamada,kg_lundstrom2022transfer}.
The scalability established above allows \texttt{KnottedGraph} to reverse this
direction of investigation: exact Yamada-polynomial data can be generated
before a family law is known and then used with LLM assistance to identify
candidate recurrences, closed forms and transfer structures. We apply this
strategy below to repeated-motif families with both commuting and
noncommuting composition laws.

The discovery and validation procedure is summarized in
\Figpanelref{fig:three_theta_yamada_families}{a}. A parameterized graph family
is first evaluated with \texttt{KnottedGraph} to generate exact Yamada
polynomials, which are then blindly divided into a discovery set $D$ and a
held-out test set $T$. Only $D$ is supplied during LLM-assisted candidate
generation. A retained proposal is converted to an explicit symbolic
expression or operator representation before the members in $T$ are evaluated
and compared with its predictions. Candidate generation and testing are
therefore separated by construction; agreement on $T$ provides a finite
out-of-sample test. The full protocol and completed
evaluation plans are detailed in
\SuppSubsecref{supp:llm_discovery_protocol}.

We first apply this workflow to the three homogeneous motif families in
\Figpanelref{fig:three_theta_yamada_families}{b--j}. For each family, the
first two panels show the $m=1$ and $m=2$ members together with normalized
Yamada polynomials computed directly by \texttt{KnottedGraph}, while the third
shows the corresponding general-$m$ relation obtained through LLM-assisted
candidate generation and symbolic reconstruction. To write these expressions
compactly, we define
\begin{equation}
\begin{gathered}
\begin{alignedat}{2}
a(Y)&=Y^2+1,        &\qquad b(Y)&=Y^2-Y+1,\\
c(Y)&=Y^4+Y^2+1,   &\qquad d(Y)&=2Y^2-Y+2,
\end{alignedat}\\
\phi(Y)=Y^4+Y^3+Y^2+Y+1.
\end{gathered}
\label{eq:main_family_factors}
\end{equation}
In each case, the inferred general relation agrees with the independently
evaluated family members used for verification.

The three candidate formulas have different coefficients but share the same
algebraic structure: each is reproduced by three transfer channels, with a
diagonal transfer operator associated with the repeated motif. The $m=1$ and
$m=2$ polynomials are direct evaluations, while the general-$m$ expressions
are the corresponding candidate analytical formulas proposed by LLM,
revealing a common three-channel mathematical structure across three distinct
repeated-motif families. Their matrices and algebraic expansions are given in
\SuppSubsecref{supp:transfer_commuting}.

The mixed construction in
\Figpanelref{fig:three_theta_yamada_families}{k} then tests how these
reconstructed motif transfers compose when different motifs occur within the
same knotted-graph family. Let
\begin{equation}
\begin{aligned}
w&=w_1\cdots w_m,
\qquad
w_j\in\{\Lambda,\downarrow,\updownarrow\},\\
\mathbf n&=(n_\Lambda,n_\downarrow,n_\updownarrow),
\qquad
m=n_\Lambda+n_\downarrow+n_\updownarrow ,
\end{aligned}
\label{eq:main_abelian_word_counts}
\end{equation}
where $\mathbf n$ records the motif multiplicities. Because the reconstructed
local transfer operators commute, the proposed invariant is independent of
motif order and therefore factors through the abelianization of the motif word,
\begin{equation}
w
\longmapsto
\mathbf n
\longmapsto
\overline{\Upsilon}(\mathcal A_{\mathbf n};Y).
\label{eq:main_abelianization}
\end{equation}
This gives a single closed-form Yamada-polynomial for arbitrary mixed
words in the three motifs: once the motif multiplicities are fixed, reordering
the word leaves the predicted invariant unchanged. The resulting
count-dependent candidate expression is displayed in
\Figpanelref{fig:three_theta_yamada_families}{k}, with its full symbolic
derivation and independent verification given in
\SuppSubsecref{supp:transfer_abelian_mixed}.

The fixed-graph pure-braid family in
\Figpanelref{fig:three_theta_yamada_families}{l} provides the complementary
order-sensitive case and a stronger departure from count-based family
descriptions. Here the underlying $8$-vertex, $12$-edge cubic
multigraph remains fixed, while a distinguished three-strand region carries an
ordered word
$w\in\{A,B\}^*$ with
$A=\sigma_1^2$ and $B=\sigma_2^2$, using the standard Artin braid generators
\cite{kg_birman1974braids}. The abstract graph is therefore unchanged as the
word varies; only the spatial embedding of the braid region is modified.
Already at length three, words with identical motif counts can have different
normalized Yamada polynomials,
\begin{equation}
\overline{\Upsilon}(\mathcal N_{AAB};Y)
=
\overline{\Upsilon}(\mathcal N_{BAA};Y)
\neq
\overline{\Upsilon}(\mathcal N_{ABA};Y).
\label{eq:main_nonabelian_example}
\end{equation}
Thus motif counts determine the proposed invariant for the commuting mixed
family, but not for the pure-braid family: the ordering of the local braid
blocks remains visible to the invariant. A length-five example of this
order-dependent braid construction is shown in
\SuppFigref{suppfig:nonabelian_worked_example}.

To reconstruct this order dependence directly from finite exact data, we use a
symbolic Hankel-based realization with finite-rank realization methods
\cite{kg_carlyle1971realizations,kg_balle2012spectral}. A
rank-fifteen finite Hankel block gives $15$-dimensional transfer operators
$T_A$ and $T_B$. This dimension is consistent with the three-strand
Yamada-algebra structure \cite{kg_chbili2016}, showing that the
LLM-proposed transfer description, once reconstructed exactly from finite
Yamada-polynomial data, recovers a structure compatible with the known
skein-algebraic setting. The reconstructed operators satisfy
\begin{equation}
[T_A,T_B]\neq0,
\label{eq:main_nonabelian_commutator}
\end{equation}
making the noncommutativity explicit. Each operator also
obeys the common cubic relation
\begin{equation}
(T_X-Y^2I)(T_X-Y^{-2}I)(T_X-Y^{-4}I)=0,
\label{eq:main_nonabelian_cubic}
\end{equation}
where $X\in\{A,B\}$. Since this cubic annihilating polynomial has three distinct
roots, the standard spectral decomposition gives
\begin{equation}
T_X^n
=
Y^{2n}P_X^{(2)}
+
Y^{-2n}P_X^{(-2)}
+
Y^{-4n}P_X^{(-4)}.
\label{eq:main_nonabelian_power}
\end{equation}
where $P_X^{(\alpha)}$ are the corresponding spectral projectors.

Writing an arbitrary braid word as its ordered maximal runs,
$w=X_1^{n_1}\cdots X_r^{n_r}$ with $X_j\in\{A,B\}$, gives the candidate
analytical family relation displayed in
\Figpanelref{fig:three_theta_yamada_families}{l}. Each of the two transfer
operators, $T_A$ and $T_B$, contributes the same three scalar spectral powers,
while the corresponding projectors remain ordered according to the braid word,
thereby compressing arbitrarily long repeated words. The boundary vectors
$L$ and $R$ are reconstructed from the Hankel representation, and $\mu(w)$ is
the minimum Laurent degree of the corresponding raw Yamada polynomial used for
normalization. The Hankel-based realization, $15$-dimensional linear
representation, projector construction, ordered-word formula and short- and
long-word verification are given in
\SuppSubsecref{supp:transfer_nonabelian_hankel}. These relations are supported by direct evaluation on all $459$ mixed-family words, $255$ distinct short pure-braid words and $20$ held-out pure-braid
words of lengths $101$, $125$, $150$ and $200$.

Taken together, the commuting mixed-family law above extends Yamada family formulas to arbitrary motif words whose invariant factors through motif multiplicities. More distinctively, the fixed-graph pure-braid construction provides, to our knowledge, the first Yamada-polynomial family law for arbitrary ordered words of noncommuting local motifs: the underlying cubic multigraph remains fixed, while words with identical braid-block multiplicities can yield different invariants. Finite exact Yamada-polynomial data further reconstruct the corresponding $15$-state noncommuting transfer representation and its cubic generator relations. This establishes a more general data-to-algebra methodology in which exact topological invariants can be used to infer candidate composition laws for entire parameterized graph families without prescribing their family-specific recurrence, transfer structure or closed form in advance. Coupled with LLM-assisted candidate generation, this provides an AI4Math route from exact invariant data to interpretable mathematical structure that can in principle be applied across broader motif-, word- and recursively generated graph families for which analytical formulas are not yet known.

\section{Discussion}
\label{sec:discussion} 

Overall, \texttt{KnottedGraph} establishes a unified, efficient computational route from
three-dimensional scientific geometry to embedding-sensitive topological
analysis, connecting problems across biology, materials science, engineering,
physics and mathematics. By
integrating knotted-graph construction, projection and exact invariant
evaluation, the framework makes spatial topology accessible for systematic
analysis across heterogeneous scientific structures, parameter spaces and
generated graph families. The scalability of the invariant calculation also
enables a different use of computational topology: exact spatial-graph
invariants can serve as structured data for mathematical discovery. Here,
LLM-assisted candidate generation identifies closed forms and transfer
structures from computed Yamada-polynomial data, followed by explicit symbolic
reconstruction and additional exact tests. The resulting commuting and
noncommuting transfer descriptions capture count-dependent and order-sensitive
family laws, respectively. Together, these capabilities connect large-scale embedding-sensitive topological analysis with a route from exact topological data to automated mathematical formula discovery.

Several computational and mathematical directions follow from the present
framework. In condensed-matter and materials settings, Hamiltonian parameters,
energy thresholds, strain \cite{kg_sunko2019lifshitz}, pressure
\cite{kg_xiang2015lifshitz} or composition
\cite{kg_dziawa2012topological} can generate families of three-dimensional
structures whose embedding-sensitive topology evolves across parameter space
\cite{akgun_yan_2026topologicalclassificationknottedgraphs}. Such physical
settings also motivate extending the framework beyond handlebodies to more
general volumetric topologies. A handlebody admits a graph-spine description,
whereas compression bodies \cite{kg_scharlemann2002heegaard} can contain
distinguished boundary components with direct physical relevance
\cite{akgun_yan_2026topologicalclassificationknottedgraphs}, together with
attachment and nesting information that cannot in general be encoded by a
single graph. This motivates extending the same automated computational
framework to spatial-graph invariants with set-valued constructions, such as
the Yamada set
\cite{akgun_yan_2026topologicalclassificationknottedgraphs}.

Related parameter-dependent analyses can also be developed for conformational
ensembles of biomolecules \cite{kg_yeates2007protein_topology} and polymers
\cite{kg_deguchi2017topological_polymers}, evolving vascular
\cite{kg_aylward2005spatial_vascular} or neuronal morphologies
\cite{kg_cuntz2010neuronal_graphs}, and families of engineered
three-dimensional networks
\cite{kg_shai1999engineering_graphs,kg_peddada2023}, whenever changes in
spatial embedding carry information beyond abstract graph connectivity.
Repeated application of the same knotted-graph and invariant interfaces can
then organize these systems into topological maps, providing a basis for
identifying topological transitions, comparing structural families and
incorporating embedding-sensitive topology into scientific classification
and design.

Further implementations of knot, link and spatial-graph
invariants could additionally provide complementary topological
classifications and independent exact datasets for mathematical discovery,
including the Alexander \cite{alexander1928topological}, Jones
\cite{jones1985polynomial} and HOMFLY--PT \cite{kg_homfly1985} polynomials,
allowing algebraic structures reconstructed from different invariants to be
compared. Braid words \cite{kg_birman1974braids}, tangles
\cite{kg_conway1970tangles}, satellite operations
\cite{kg_schubert1953satellite} and other recursive constructions
\cite{kg_lundstrom2022transfer} likewise provide broader family classes in
which the same LLM-assisted discovery strategy could be used to identify
candidate recurrences, transfer representations and composition laws from
exact invariant data, and to probe possible mathematical connections between
these different constructions.

\BackMatterHeading{Code and data availability}
\label{sec:code_availability}
Source code, documentation, workflows and data supporting this work are available at \url{https://github.com/HakanAkgn/KnottedGraph}, with additional reproducibility and software-architecture details provided in \SuppNoteref{supp:software_architecture}.

\BackMatterHeading{Acknowledgements}
\label{sec:acknowledgements}
We thank Henrik Schumacher and Keenan Crane for helpful guidance on the repulsive-curve framework and the associated \url{https://github.com/HenrikSchumacher/Repulsor} repository. During the preparation of this manuscript, the authors used OpenAI's GPT-5.6 Sol for language refinement, code optimization and candidate generation in the \hyperref[sec:mathematical_discovery]{\textcolor{myred}{LLM-assisted mathematical discovery from exact knotted-graph data}}. All calculations, validation and scientific interpretations remain the responsibility of the authors.

\FloatBarrier

\renewcommand{\bibfont}{\fontsize{9}{9}\selectfont}
\setlength{\bibsep}{2pt}
\bibliographystyle{naturemag}
\bibliography{GeneralVersion/references}

@article{kg_yamada1989,
  author  = {Yamada, Shuji},
  title   = {An invariant of spatial graphs},
  journal = {Journal of Graph Theory},
  volume  = {13},
  number  = {5},
  pages   = {537--551},
  year    = {1989},
  doi     = {10.1002/jgt.3190130503}
}

@article{kg_negami1987,
  author  = {Negami, Seiya},
  title   = {Polynomial invariants of graphs},
  journal = {Transactions of the American Mathematical Society},
  volume  = {299},
  number  = {2},
  pages   = {601--622},
  year    = {1987},
  doi     = {10.2307/2000516}
}

@incollection{kg_dobrynin1996yamada,
  author    = {Vesnin, Andrei Yu. and Dobrynin, Andrey A.},
  title     = {{Yamada} polynomial for graphs embedded knottedly in a three-dimensional space},
  booktitle = {Graph Theory and Applications},
  editor    = {Skorobogatov, V. A.},
  series    = {Vychislitel'nye Sistemy},
  volume    = {155},
  pages     = {37--86},
  publisher = {Institute of Mathematics, Siberian Branch of the Russian Academy of Sciences},
  address   = {Novosibirsk},
  year      = {1996},
  note      = {In Russian}
}

@article{kg_li2018yamada,
  author  = {Li, Miaowang and Lei, Fengchun and Li, Fengling and Vesnin, Andrei},
  title   = {The Yamada polynomial of spatial graphs obtained by edge replacements},
  journal = {Journal of Knot Theory and Its Ramifications},
  volume  = {27},
  number  = {9},
  pages   = {1842004},
  year    = {2018},
  doi     = {10.1142/S021821651842004X},
  eprint  = {1801.09075},
  archivePrefix = {arXiv},
  primaryClass  = {math.GT}
}

@article{kg_peddada2023,
  author  = {Peddada, Satya R. T. and Dunfield, Nathan M. and Zeidner, Lawrence E. and Givans, Zane R. and James, Kai A. and Allison, James T.},
  title   = {Enumeration and Identification of Unique 3D Spatial Topologies of Interconnected Engineering Systems Using Spatial Graphs},
  journal = {Journal of Mechanical Design},
  volume  = {145},
  number  = {10},
  pages   = {101708},
  year    = {2023},
  doi     = {10.1115/1.4062978},
  eprint  = {2107.13724},
  archivePrefix = {arXiv}
}

@article{kg_ishii2012handlebody,
  author  = {Ishii, Atsushi and Iwakiri, Masahide},
  title   = {Quandle Cocycle Invariants for Spatial Graphs and Knotted Handlebodies},
  journal = {Canadian Journal of Mathematics},
  volume  = {64},
  number  = {1},
  pages   = {102--122},
  year    = {2012},
  doi     = {10.4153/CJM-2011-035-0}
}

@article{kg_topoly2021,
  author  = {Dabrowski-Tumanski, Pawel and Rubach, Pawel and Niemyska, Wanda and Gren, Bartosz Ambrozy and Sulkowska, Joanna Ida},
  title   = {Topoly: Python package to analyze topology of polymers},
  journal = {Briefings in Bioinformatics},
  volume  = {22},
  number  = {3},
  pages   = {bbaa196},
  year    = {2021},
  doi     = {10.1093/bib/bbaa196}
}

@article{kg_bode2019prescribed,
  author  = {Bode, Benjamin and Dennis, Mark R.},
  title   = {Constructing a polynomial whose nodal set is any prescribed knot or link},
  journal = {Journal of Knot Theory and Its Ramifications},
  volume  = {28},
  number  = {1},
  pages   = {1850082},
  year    = {2019},
  doi     = {10.1142/S0218216518500827},
  eprint  = {1612.06328},
  archivePrefix = {arXiv},
  primaryClass  = {math.GT}
}

@article{kg_lee1994thinning,
  author  = {Lee, Ta-Chih and Kashyap, Rangasami L. and Chu, Chong-Nam},
  title   = {Building Skeleton Models via 3-D Medial Surface/Axis Thinning Algorithms},
  journal = {CVGIP: Graphical Models and Image Processing},
  volume  = {56},
  number  = {6},
  pages   = {462--478},
  year    = {1994},
  doi     = {10.1006/cgip.1994.1042}
}

@article{kg_vanderwalt2014skimage,
  author  = {van der Walt, St{\'e}fan and Sch{\"o}nberger, Johannes L. and Nunez-Iglesias, Juan and Boulogne, Fran{\c c}ois and Warner, Joshua D. and Yager, Neil and Gouillart, Emmanuelle and Yu, Tony and {the scikit-image contributors}},
  title   = {scikit-image: image processing in Python},
  journal = {PeerJ},
  volume  = {2},
  pages   = {e453},
  year    = {2014},
  doi     = {10.7717/peerj.453}
}

@article{kg_yu2021repulsive,
  author  = {Yu, Christopher and Schumacher, Henrik and Crane, Keenan},
  title   = {Repulsive Curves},
  journal = {ACM Transactions on Graphics},
  volume  = {40},
  number  = {2},
  pages   = {10:1--10:21},
  year    = {2021},
  doi     = {10.1145/3439429},
  eprint  = {2006.07859},
  archivePrefix = {arXiv}
}

@article{kg_ramer1972,
  author={Ramer, Urs}, title={An iterative procedure for the polygonal approximation of plane curves},
  journal={Computer Graphics and Image Processing}, volume={1}, number={3}, pages={244--256},
  year={1972}, doi={10.1016/S0146-664X(72)80017-0}
}

@article{kg_douglas1973,
  author={Douglas, David H. and Peucker, Thomas K.},
  title={Algorithms for the Reduction of the Number of Points Required to Represent a Digitized Line or Its Caricature},
  journal={Cartographica}, volume={10}, number={2}, pages={112--122}, year={1973},
  doi={10.3138/FM57-6770-U75U-7727}
}

@article{kg_lewiner2003marching,
  author={Lewiner, Thomas and Lopes, H{\'e}lio and Vieira, Ant{\^o}nio Wilson and Tavares, Geovan},
  title={Efficient Implementation of Marching Cubes' Cases with Topological Guarantees},
  journal={Journal of Graphics Tools}, volume={8}, number={2}, pages={1--15}, year={2003},
  doi={10.1080/10867651.2003.10487582}
}

@article{kg_sullivan2019pyvista,
  author={Sullivan, C. Bane and Kaszynski, Alexander A.},
  title={PyVista: 3D plotting and mesh analysis through a streamlined interface for the Visualization Toolkit (VTK)},
  journal={Journal of Open Source Software}, volume={4}, number={37}, pages={1450}, year={2019},
  doi={10.21105/joss.01450}
}

@misc{kg_shapely,
  author       = {Gillies, Sean and van der Wel, Casper and Van den Bossche, Joris and Taves, Mike W. and Arnott, Joshua and Ward, Brendan C. and others},
  title        = {{Shapely}},
  howpublished = {Software},
  year         = {n.d.},
  doi          = {10.5281/zenodo.5597138},
  note         = {All-version archive; no specific release is identified}
}

@inproceedings{kg_leutenegger1997str,
  title={STR: A simple and efficient algorithm for R-tree packing},
  author={Leutenegger, Scott T and Lopez, Mario A and Edgington, Jeffrey},
  booktitle={Proceedings 13th international conference on data engineering},
  pages={497--506},
  year={1997},
  organization={IEEE}
}

@inproceedings{kg_hagberg2008networkx,
  author={Hagberg, Aric A. and Schult, Daniel A. and Swart, Pieter J.},
  title={Exploring network structure, dynamics, and function using NetworkX},
  booktitle={Proceedings of the 7th Python in Science Conference}, pages={11--15}, year={2008}
}

@article{kg_harris2020numpy,
  author={Harris, Charles R. and Millman, K. Jarrod and van der Walt, St{\'e}fan J. and others},
  title={Array programming with NumPy}, journal={Nature}, volume={585}, number={7825},
  pages={357--362}, year={2020}, doi={10.1038/s41586-020-2649-2}
}

@article{kg_virtanen2020scipy,
  author={Virtanen, Pauli and Gommers, Ralf and Oliphant, Travis E. and others},
  title={SciPy 1.0: fundamental algorithms for scientific computing in Python},
  journal={Nature Methods}, volume={17}, pages={261--272}, year={2020},
  doi={10.1038/s41592-019-0686-2}
}

@article{kg_meurer2017sympy,
  author={Meurer, Aaron and Smith, Christopher P. and Paprocki, Mateusz and others},
  title={SymPy: symbolic computing in Python}, journal={PeerJ Computer Science},
  volume={3}, pages={e103}, year={2017}, doi={10.7717/peerj-cs.103}
}

@incollection{kg_ohser2002euler,
  author={Ohser, Joachim and Nagel, Werner and Schladitz, Katja},
  title={The Euler Number of Discretized Sets---On the Choice of Adjacency in Homogeneous Lattices},
  booktitle={Morphology of Condensed Matter}, series={Lecture Notes in Physics}, volume={600},
  pages={275--298}, publisher={Springer}, year={2002}, doi={10.1007/3-540-45782-8_12}
}

@article{kg_bi2017nodalknot,
  author={Bi, Ren and Yan, Zhongbo and Lu, Ling and Wang, Zhong},
  title={Nodal-knot semimetals}, journal={Physical Review B}, volume={96}, pages={201305},
  year={2017}, doi={10.1103/PhysRevB.96.201305}
}

@article{kg_lee2020nodalknots,
  author={Lee, Ching Hua and Sutrisno, Amanda and Hofmann, Tobias and Helbig, Tobias and Liu, Yuhan and Ang, Yee Sin and Ang, Lay Kee and Zhang, Xiao and Greiter, Martin and Thomale, Ronny},
  title={Imaging nodal knots in momentum space through topolectrical circuits},
  journal={Nature Communications}, volume={11}, pages={4385}, year={2020},
  doi={10.1038/s41467-020-17716-1}
}

@inproceedings{kg_yan2025hsg12m,
  author        = {Yan, Xianquan and Akg{\"u}n, Hakan and Kawaguchi, Kenji and Loh, N. Duane and Lee, Ching Hua},
  title         = {{HSG-12M}: A Large-Scale Benchmark of Spatial Multigraphs from the Energy Spectra of Non-Hermitian Crystals},
  booktitle     = {The Fourteenth International Conference on Learning Representations},
  year          = {2026},
  eprint        = {2506.08618v5},
  archivePrefix = {arXiv},
  doi           = {10.48550/arXiv.2506.08618},
  url           = {https://arxiv.org/abs/2506.08618v5}
}

@inproceedings{yan_graphtransformer,
  author    = {Yan, Xianquan and Akg{\"u}n, Hakan and Kawaguchi, Kenji and Loh, N. Duane and Lee, Ching Hua},
  title     = {Classifying the Graph Topology of Non-Hermitian Energy Spectra with Graph Transformer},
  booktitle = {Second Workshop on XAI4Science: From Understanding Model Behavior to Discovering New Scientific Knowledge},
  year      = {2026},
  url       = {https://openreview.net/forum?id=sH1WzXPzEr}
}

@inproceedings{yan2025metric,
  author    = {Yan, Xianquan},
  title     = {Automated Metric Discovery: Navigating Quantum Geometry with Symbolic Regression},
  booktitle = {AI4X 2025 International Conference},
  year      = {2025},
  url       = {https://openreview.net/forum?id=E0fa4FXUsW}
}

@article{kg_burley2019rcsb,
  author  = {Burley, Stephen K. and Berman, Helen M. and Bhikadiya, Charmi and others},
  title   = {{RCSB Protein Data Bank}: biological macromolecular structures enabling research and education in fundamental biology, biomedicine, biotechnology and energy},
  journal = {Nucleic Acids Research},
  volume  = {47},
  number  = {D1},
  pages   = {D464--D474},
  year    = {2019},
  doi     = {10.1093/nar/gky1004}
}

@article{kg_kauffman1989spatial,
  author  = {Kauffman, Louis H.},
  title   = {Invariants of graphs in three-space},
  journal = {Transactions of the American Mathematical Society},
  volume  = {311},
  number  = {2},
  pages   = {697--710},
  year    = {1989},
  doi     = {10.1090/S0002-9947-1989-0946218-0}
}

@incollection{kg_flapan2017spatial,
  author    = {Flapan, E. and Mattman, T. W. and Mellor, B. and Naimi, R. and Nikkuni, R.},
  title     = {Recent Developments in Spatial Graph Theory},
  booktitle = {Knots, Links, Spatial Graphs, and Algebraic Invariants},
  series    = {Contemporary Mathematics},
  volume    = {689},
  pages     = {81--102},
  year      = {2017},
  publisher = {American Mathematical Society},
  doi       = {10.1090/conm/689/13845}
}

@misc{kg_mellor2018spatial,
  title = {Invariants of {{Spatial Graphs}}},
  author = {Mellor, Blake},
  year = {2018},
  month = dec,
  number = {arXiv:1812.08885},
  eprint = {1812.08885},
  primaryclass = {math},
  publisher = {arXiv},
  doi = {10.48550/arXiv.1812.08885},
  urldate = {2025-03-06},
  archiveprefix = {arXiv},
  langid = {english}
}

@article{kg_ishii2008handlebody,
  author  = {Ishii, Atsushi},
  title   = {Moves and invariants for knotted handlebodies},
  journal = {Algebraic \& Geometric Topology},
  volume  = {8},
  number  = {3},
  pages   = {1403--1418},
  year    = {2008},
  doi     = {10.2140/agt.2008.8.1403}
}

@article{kg_funsearch2024,
  author  = {Romera-Paredes, Bernardino and Barekatain, Mohammadamin and Novikov, Alexander and others},
  title   = {Mathematical discoveries from program search with large language models},
  journal = {Nature},
  volume  = {625},
  pages   = {468--475},
  year    = {2024},
  doi     = {10.1038/s41586-023-06924-6}
}

@article{kg_yeates2007protein_topology,
  author  = {Yeates, Todd O. and Norcross, Todd S. and King, Neil P.},
  title   = {Knotted and topologically complex proteins as models for studying folding and stability},
  journal = {Current Opinion in Chemical Biology},
  volume  = {11},
  number  = {6},
  pages   = {595--603},
  year    = {2007},
  doi     = {10.1016/j.cbpa.2007.10.002}
}

@article{kg_deguchi2017topological_polymers,
  author  = {Deguchi, Tetsuo and Uehara, Erica},
  title   = {Statistical and Dynamical Properties of Topological Polymers with Graphs and Ring Polymers with Knots},
  journal = {Polymers},
  volume  = {9},
  number  = {7},
  pages   = {252},
  year    = {2017},
  doi     = {10.3390/polym9070252}
}

@inproceedings{kg_aylward2005spatial_vascular,
  author    = {Aylward, Stephen R. and Jomier, Julien and Vivert, Christelle and LeDigarcher, Vincent and Bullitt, Elizabeth},
  title     = {Spatial Graphs for Intra-cranial Vascular Network Characterization, Generation, and Discrimination},
  booktitle = {Medical Image Computing and Computer-Assisted Intervention -- MICCAI 2005},
  volume    = {3749},
  pages     = {59--66},
  year      = {2005},
  publisher = {Springer},
  doi       = {10.1007/11566465_8}
}

@article{kg_cuntz2010neuronal_graphs,
  author  = {Cuntz, Hermann and Forstner, Friedrich and Borst, Alexander and H{\"a}usser, Michael},
  title   = {One Rule to Grow Them All: A General Theory of Neuronal Branching and Its Practical Application},
  journal = {PLoS Computational Biology},
  volume  = {6},
  number  = {8},
  pages   = {e1000877},
  year    = {2010},
  doi     = {10.1371/journal.pcbi.1000877}
}

@article{kg_shai1999engineering_graphs,
  author  = {Shai, O. and Preiss, K.},
  title   = {Graph theory representations of engineering systems and their embedded knowledge},
  journal = {Artificial Intelligence in Engineering},
  volume  = {13},
  number  = {3},
  pages   = {273--285},
  year    = {1999},
  doi     = {10.1016/S0954-1810(99)00002-3}
}

@article{kg_fang2016nodal_line,
  author  = {Fang, Chen and Weng, Hongming and Dai, Xi and Fang, Zhong},
  title   = {Topological nodal line semimetals},
  journal = {Chinese Physics B},
  volume  = {25},
  number  = {11},
  pages   = {117106},
  year    = {2016},
  doi     = {10.1088/1674-1056/25/11/117106}
}

@article{kg_helman1989vector_field_topology,
  author  = {Helman, James and Hesselink, Lambertus},
  title   = {Representation and Display of Vector Field Topology in Fluid Flow Data Sets},
  journal = {Computer},
  volume  = {22},
  number  = {8},
  pages   = {27--36},
  year    = {1989},
  doi     = {10.1109/2.35197}
}

@inproceedings{kg_reinders2000skeletongraph,
  author    = {Reinders, Freek and Jacobson, Melvin E. D. and Post, Frits H.},
  title     = {Skeleton Graph Generation for Feature Shape Description},
  booktitle = {Data Visualization 2000},
  editor    = {de Leeuw, Wim C. and van Liere, Robert},
  pages     = {73--82},
  publisher = {Springer},
  address   = {Vienna},
  year      = {2000},
  doi       = {10.1007/978-3-7091-6783-0_8}
}

@article{kg_abraham2015gromacs,
  author  = {Abraham, Mark James and Murtola, Teemu and Schulz, Roland and P{\'a}ll, Szil{\'a}rd and Smith, Jeremy C. and Hess, Berk and Lindahl, Erik},
  title   = {{GROMACS}: High performance molecular simulations through multi-level parallelism from laptops to supercomputers},
  journal = {SoftwareX},
  volume  = {1--2},
  pages   = {19--25},
  year    = {2015},
  doi     = {10.1016/j.softx.2015.06.001}
}

@article{kg_thompson2022lammps,
  author  = {Thompson, Aidan P. and Aktulga, H. Metin and Berger, Richard and Bolintineanu, Dan S. and Brown, W. Michael and Crozier, Paul S. and in 't Veld, Pieter J. and Kohlmeyer, Axel and Moore, Stan G. and Nguyen, Trung Dac and Shan, Ray and Stevens, Mark J. and Tranchida, Julien and Trott, Christian and Plimpton, Steven J.},
  title   = {{LAMMPS}---a flexible simulation tool for particle-based materials modeling at the atomic, meso, and continuum scales},
  journal = {Computer Physics Communications},
  volume  = {271},
  pages   = {108171},
  year    = {2022},
  doi     = {10.1016/j.cpc.2021.108171}
}

@incollection{kg_bourne1997mmcif,
  author    = {Bourne, Philip E. and Berman, Helen M. and McMahon, Brian and Watenpaugh, Keith D. and Westbrook, John D. and Fitzgerald, Paula M. D.},
  title     = {Macromolecular crystallographic information file},
  booktitle = {Methods in Enzymology},
  volume    = {277},
  pages     = {571--590},
  publisher = {Academic Press},
  year      = {1997},
  doi       = {10.1016/S0076-6879(97)77032-0}
}

@inproceedings{kg_brandes2002graphml,
  author    = {Brandes, Ulrik and Eiglsperger, Markus and Herman, Ivan and Himsolt, Michael and Marshall, M. Scott},
  title     = {{GraphML} Progress Report: Structural Layer Proposal},
  booktitle = {Graph Drawing},
  series    = {Lecture Notes in Computer Science},
  volume    = {2265},
  pages     = {501--512},
  publisher = {Springer},
  year      = {2002},
  doi       = {10.1007/3-540-45848-4_59}
}

@article{kg_keinert2015fibonacci,
  author  = {Keinert, Benjamin and Innmann, Matthias and S{\"a}nger, Michael and Stamminger, Marc},
  title   = {Spherical Fibonacci Mapping},
  journal = {ACM Transactions on Graphics},
  volume  = {34},
  number  = {6},
  pages   = {193:1--193:7},
  year    = {2015},
  doi     = {10.1145/2816795.2818131}
}

@article{kg_cordella2004vf2,
  author  = {Cordella, Luigi P. and Foggia, Pasquale and Sansone, Carlo and Vento, Mario},
  title   = {A (Sub)Graph Isomorphism Algorithm for Matching Large Graphs},
  journal = {IEEE Transactions on Pattern Analysis and Machine Intelligence},
  volume  = {26},
  number  = {10},
  pages   = {1367--1372},
  year    = {2004},
  doi     = {10.1109/TPAMI.2004.75}
}

@article{kg_mastin2015pdcode,
  author  = {Mastin, Matt},
  title   = {Links and Planar Diagram Codes},
  journal = {Journal of Knot Theory and Its Ramifications},
  volume  = {24},
  number  = {3},
  pages   = {1550016},
  year    = {2015},
  doi     = {10.1142/S0218216515500169}
}

@incollection{kg_scharlemann2002heegaard,
  author    = {Scharlemann, Martin},
  title     = {Heegaard splittings of compact 3-manifolds},
  booktitle = {Handbook of Geometric Topology},
  editor    = {Daverman, R. J. and Sher, R. B.},
  pages     = {921--953},
  publisher = {North-Holland},
  address   = {Amsterdam},
  year      = {2002},
  doi       = {10.1016/B978-044482432-5/50019-6}
}

@article{kg_buck1995energy,
  author  = {Buck, Gregory and Orloff, J.},
  title   = {A simple energy function for knots},
  journal = {Topology and its Applications},
  volume  = {61},
  number  = {3},
  pages   = {205--214},
  year    = {1995},
  doi     = {10.1016/0166-8641(94)00024-W}
}

@misc{kg_schumacher_repulsor,
  author       = {Schumacher, Henrik},
  title        = {{Repulsor}: A package for handling tangent-point energy and other nonlocal energies},
  howpublished = {\url{https://github.com/HenrikSchumacher/Repulsor}},
  note         = {Software, source revision \nolinkurl{adc56b61f65f5958b59cbd7e1539f44ed0c5e993}}
}

@article{kg_carlyle1971realizations,
  author  = {Carlyle, J. W. and Paz, A.},
  title   = {Realizations by stochastic finite automata},
  journal = {Journal of Computer and System Sciences},
  volume  = {5},
  number  = {1},
  pages   = {26--40},
  year    = {1971},
  doi     = {10.1016/S0022-0000(71)80005-3}
}

@inproceedings{kg_balle2012spectral,
  author    = {Balle, Borja and Mohri, Mehryar},
  title     = {Spectral Learning of General Weighted Automata via Constrained Matrix Completion},
  booktitle = {Advances in Neural Information Processing Systems 25},
  pages     = {2168--2176},
  year      = {2012}
}

@article{kg_huh2024theta,
  author  = {Huh, Youngsik},
  title   = {Yamada polynomial and associated link of $\theta$-curves},
  journal = {Discrete Mathematics},
  volume  = {347},
  number  = {1},
  pages   = {113684},
  year    = {2024},
  doi     = {10.1016/j.disc.2023.113684}
}

@mastersthesis{kg_lundstrom2022transfer,
  author = {Lundstr{\"o}m, Teemu},
  title  = {Combinatorial Symmetries in Knot Theory: Yamada Polynomials from Transfer-Matrix Methods},
  school = {University of Helsinki},
  type   = {Master's thesis},
  year   = {2022},
  url = {https://hdl.handle.net/10138/339220}
}

@article{kg_li2019density,
  author  = {Li, Miaowang and Lei, Fengchun and Li, Fengling and Vesnin, Andrei},
  title   = {Density of Roots of the Yamada Polynomial of Spatial Graphs},
  journal = {Proceedings of the Steklov Institute of Mathematics},
  volume  = {305},
  number  = {1},
  pages   = {135--148},
  year    = {2019},
  doi     = {10.1134/S0081543819030076},
  eprint  = {1810.12749},
  archivePrefix = {arXiv},
  primaryClass  = {math.GT}
}

@article{kg_vesnin2025complete,
  author  = {Vesnin, Andrei Yu. and Oshmarina, Olga A.},
  title   = {Polynomials of Complete Spatial Graphs and Jones Polynomials of the Related Links},
  journal = {Sbornik: Mathematics},
  volume  = {216},
  number  = {5},
  pages   = {608--637},
  year    = {2025},
  doi     = {10.4213/sm10167e}
}

@misc{kg_alphaevolve2025,
  author        = {Novikov, Alexander and V{\~u}, Ng{\^a}n and Eisenberger, Marvin and Dupont, Emilien and Huang, Po-Sen and Wagner, Adam Zsolt and Shirobokov, Sergey and Kozlovskii, Borislav and Ruiz, Francisco J. R. and Mehrabian, Abbas and Kumar, M. Pawan and See, Abigail and Chaudhuri, Swarat and Holland, George and Davies, Alex and Nowozin, Sebastian and Kohli, Pushmeet and Balog, Matej},
  title         = {AlphaEvolve: A Coding Agent for Scientific and Algorithmic Discovery},
  year          = {2025},
  eprint        = {2506.13131},
  archivePrefix = {arXiv},
  primaryClass  = {cs.AI},
  doi           = {10.48550/arXiv.2506.13131}
}

@misc{kg_georgiev2025mathematical,
  author        = {Georgiev, Bogdan and G{\'o}mez-Serrano, Javier and Tao, Terence and Wagner, Adam Zsolt},
  title         = {Mathematical Exploration and Discovery at Scale},
  year          = {2025},
  eprint        = {2511.02864},
  archivePrefix = {arXiv},
  doi           = {10.48550/arXiv.2511.02864}
}

@article{kg_alphaproof2026,
  author  = {Hubert, Thomas and Mehta, Rishi and Sartran, Laurent and others},
  title   = {Olympiad-level formal mathematical reasoning with reinforcement learning},
  journal = {Nature},
  volume  = {651},
  pages   = {607--613},
  year    = {2026},
  doi     = {10.1038/s41586-025-09833-y}
}

@incollection{kg_chbili2016,
  author    = {Chbili, Nafaa},
  title     = {Ribbon Graphs and Temperley--Lieb Algebra},
  booktitle = {Knot Theory and Its Applications},
  series    = {Contemporary Mathematics},
  volume    = {670},
  pages     = {299--312},
  publisher = {American Mathematical Society},
  year      = {2016},
  doi       = {10.1090/conm/670/13452}
}

@article{kg_lifshitz1960,
  author  = {Lifshitz, I. M.},
  title   = {Anomalies of electron characteristics of a metal in the high pressure region},
  journal = {Soviet Physics JETP},
  volume  = {11},
  number  = {5},
  pages   = {1130--1135},
  year    = {1960}
}

@article{kg_varlamov2021lifshitz,
  author  = {Varlamov, A. A. and Galperin, Y. M. and Sharapov, S. G. and Yerin, Yuriy},
  title   = {Concise guide for electronic topological transitions},
  journal = {Low Temperature Physics},
  volume  = {47},
  number  = {8},
  pages   = {672--683},
  year    = {2021},
  doi     = {10.1063/10.0005556}
}

@article{kg_sunko2019lifshitz,
  author  = {Sunko, Veronika and Abarca Morales, Edgar and Markovi{\'c}, Igor and Barber, Mark E. and Milosavljevi{\'c}, Dijana and Mazzola, Federico and Sokolov, Dmitry A. and Kikugawa, Naoki and Cacho, Cephise and Dudin, Pavel and Rosner, Helge and Hicks, Clifford W. and King, Philip D. C. and Mackenzie, Andrew P.},
  title   = {Direct observation of a uniaxial stress-driven Lifshitz transition in Sr2RuO4},
  journal = {npj Quantum Materials},
  volume  = {4},
  pages   = {46},
  year    = {2019},
  doi     = {10.1038/s41535-019-0185-9}
}

@article{kg_xiang2015lifshitz,
  author  = {Xiang, Z. J. and Ye, G. J. and Shang, C. and Lei, B. and Wang, N. Z. and Yang, K. S. and Liu, D. Y. and Meng, F. B. and Luo, X. G. and Zou, L. J. and Sun, Z. and Zhang, Y. and Chen, X. H.},
  title   = {Pressure-Induced Electronic Transition in Black Phosphorus},
  journal = {Physical Review Letters},
  volume  = {115},
  pages   = {186403},
  year    = {2015},
  doi     = {10.1103/PhysRevLett.115.186403}
}

@article{kg_dziawa2012topological,
  author  = {Dziawa, P. and Kowalski, B. J. and Dybko, K. and Buczko, R. and Szczerbakow, A. and Szot, M. and {\L}usakowska, E. and Balasubramanian, T. and Wojek, B. M. and Berntsen, M. H. and Tjernberg, O. and Story, T.},
  title   = {Topological crystalline insulator states in Pb1-xSnxSe},
  journal = {Nature Materials},
  volume  = {11},
  number  = {12},
  pages   = {1023--1027},
  year    = {2012},
  doi     = {10.1038/nmat3449}
}

@article{kg_ichinokura2022lifshitz,
  author  = {Ichinokura, S. and Toyoda, M. and Hashizume, M. and Horii, K. and Kusaka, S. and Ideta, S. and Tanaka, K. and Shimizu, R. and Hitosugi, T. and Saito, S. and Hirahara, T.},
  title   = {Van Hove singularity and Lifshitz transition in thickness-controlled Li-intercalated graphene},
  journal = {Physical Review B},
  volume  = {105},
  pages   = {235307},
  year    = {2022},
  doi     = {10.1103/PhysRevB.105.235307}
}

@article{kg_galeski2022lifshitz,
  author  = {Galeski, S. and Legg, H. F. and Wawrzy{\'n}czak, R. and F{\"o}rster, T. and Zherlitsyn, S. and Gorbunov, D. and Uhlarz, M. and Lozano, P. M. and Li, Q. and Gu, G. D. and Felser, C. and Wosnitza, J. and Meng, T. and Gooth, J.},
  title   = {Signatures of a magnetic-field-induced Lifshitz transition in the ultra-quantum limit of the topological semimetal ZrTe5},
  journal = {Nature Communications},
  volume  = {13},
  pages   = {7418},
  year    = {2022},
  doi     = {10.1038/s41467-022-35106-7}
}

@article{kg_slizovskiy2015lifshitz,
  author  = {Slizovskiy, Sergey and Chubukov, Andrey V. and Betouras, Joseph J.},
  title   = {Magnetic Fluctuations and Specific Heat in NaxCoO2 Near a Lifshitz Transition},
  journal = {Physical Review Letters},
  volume  = {114},
  pages   = {066403},
  year    = {2015},
  doi     = {10.1103/PhysRevLett.114.066403}
}

@article{kg_shi2017lifshitz,
  author  = {Shi, X. and Han, Z.-Q. and Peng, X.-L. and Richard, P. and Qian, T. and Wu, X.-X. and Qiu, M.-W. and Wang, S. C. and Hu, J. P. and Sun, Y.-J. and Ding, H.},
  title   = {Enhanced superconductivity accompanying a Lifshitz transition in electron-doped FeSe monolayer},
  journal = {Nature Communications},
  volume  = {8},
  pages   = {14988},
  year    = {2017},
  doi     = {10.1038/ncomms14988}
}

@misc{akgun_yan_2026topologicalclassificationknottedgraphs,
      title={Topological classification through knotted graphs: Fermi surface dispersions and Lifshitz transitions}, 
      author={Hakan Akgün and Xianquan Yan and Ching Hua Lee},
      year={2026},
      eprint={2609.13390},
      archivePrefix={arXiv},
      primaryClass={cond-mat.mes-hall},
      url={https://arxiv.org/abs/2609.13390}, 
}

@article{li2019emergence,
  title={Emergence and full 3D-imaging of nodal boundary Seifert surfaces in 4D topological matter},
  author={Li, Linhu and Lee, Ching Hua and Gong, Jiangbin},
  journal={Communications physics},
  volume={2},
  number={1},
  pages={135},
  year={2019},
  publisher={Nature Publishing Group UK London},
  doi = {10.1038/s42005-019-0235-4}
}

@article{li2018realistic,
  title={Realistic floquet semimetal with exotic topological linkages between arbitrarily many nodal loops},
  author={Li, Linhu and Lee, Ching Hua and Gong, Jiangbin},
  journal={Physical review letters},
  volume={121},
  number={3},
  pages={036401},
  year={2018},
  publisher={APS},
  doi = {10.1103/physrevlett.121.036401}
}

@article{carlsson2009topology,
  title={Topology and data},
  author={Carlsson, Gunnar},
  journal={Bulletin of the American mathematical society},
  volume={46},
  number={2},
  pages={255--308},
  year={2009},
  doi = {10.1090/s0273-0979-09-01249-x}
}

@incollection{edelsbrunner2008persistent,
  title={Persistent homology---a survey},
  author={Edelsbrunner, Herbert and Harer, John},
  booktitle={Surveys on Discrete and Computational Geometry: Twenty Years Later},
  series={Contemporary Mathematics},
  volume={453},
  pages={257--282},
  year={2008},
  publisher={American Mathematical Society},
  doi={10.1090/conm/453/08802}
}

@book{kg_birman1974braids,
  author    = {Birman, Joan S.},
  title     = {Braids, Links, and Mapping Class Groups},
  series    = {Annals of Mathematics Studies},
  volume    = {82},
  publisher = {Princeton University Press},
  address   = {Princeton, NJ},
  year      = {1974},
  pages     = {ix+228},
  isbn      = {9780691081496}
}

@incollection{kg_conway1970tangles,
  author    = {Conway, John H.},
  title     = {An Enumeration of Knots and Links, and Some of Their Algebraic Properties},
  booktitle = {Computational Problems in Abstract Algebra},
  pages     = {329--358},
  publisher = {Pergamon Press},
  year      = {1970},
  doi       = {10.1016/B978-0-08-012975-4.50034-5}
}

@article{kg_schubert1953satellite,
  author  = {Schubert, Horst},
  title   = {Knoten und Vollringe},
  journal = {Acta Mathematica},
  volume  = {90},
  pages   = {131--286},
  year    = {1953},
  doi     = {10.1007/BF02392437},
  language = {German}
}

@article{kg_homfly1985,
  author  = {Freyd, Peter and Yetter, David and Hoste, Jim and Lickorish, W. B. R. and Millett, Kenneth C. and Ocneanu, Adrian},
  title   = {A New Polynomial Invariant of Knots and Links},
  journal = {Bulletin of the American Mathematical Society},
  volume  = {12},
  number  = {2},
  pages   = {239--246},
  year    = {1985},
  doi     = {10.1090/S0273-0979-1985-15361-3}
}

@book{rolfsen2003knots,
  title={Knots and links},
  author={Rolfsen, Dale},
  series={AMS Chelsea Publishing},
  volume={346},
  year={2003},
  publisher={American Mathematical Society},
  doi = {10.1090/chel/346}
}

@article{alexander1928topological,
  title={Topological invariants of knots and links},
  author={Alexander, James W},
  journal={Transactions of the American Mathematical Society},
  volume={30},
  number={2},
  pages={275--306},
  year={1928},
  publisher={American Mathematical Society},
  doi={10.1090/S0002-9947-1928-1501429-1}
}

@article{jones1985polynomial,
  author  = {Jones, Vaughan F. R.},
  title   = {A Polynomial Invariant for Knots via von Neumann Algebras},
  journal = {Bulletin of the American Mathematical Society},
  volume  = {12},
  number  = {1},
  pages   = {103--111},
  year    = {1985},
  doi     = {10.1090/S0273-0979-1985-15304-2}
}

@article{kg_nureki2002rrma,
  author  = {Nureki, O. and Shirouzu, M. and Hashimoto, K. and Ishitani, R. and Terada, T. and Tamakoshi, M. and Oshima, T. and Chijimatsu, M. and Takio, K. and Vassylyev, D. G. and Shibata, T. and Inoue, Y. and Kuramitsu, S. and Yokoyama, S.},
  title   = {An enzyme with a deep trefoil knot for the active-site architecture},
  journal = {Acta Crystallographica Section D: Biological Crystallography},
  volume  = {58},
  pages   = {1129--1137},
  year    = {2002},
  doi     = {10.1107/S0907444902006601}
}

@article{kg_su2026tda,
  author  = {Su, Zhe and Liu, Xiang and Bou Hamdan, Layal and Maroulas, Vasileios and Wu, Jie and Carlsson, Gunnar and Wei, Guo-Wei},
  title   = {Topological data analysis and topological deep learning beyond persistent homology: a review},
  journal = {Artificial Intelligence Review},
  volume  = {59},
  number  = {2},
  pages   = {58},
  year    = {2026},
  doi     = {10.1007/s10462-025-11462-w}
}

@article{kg_comoglio2011homfly,
author  = {Comoglio, Federico and Rinaldi, Maurizio},
title   = {A Topological Framework for the Computation of the {HOMFLY} Polynomial and Its Application to Proteins},
journal = {PLoS ONE},
volume  = {6},
number  = {4},
pages   = {e18693},
year    = {2011},
doi     = {10.1371/journal.pone.0018693}
}

@article{burtsev2026ai,
  title   = {How {AI} is reshaping discovery in maths and physics},
  author  = {Burtsev, Mikhail and He, Yang-Hui and Sobko, Evgeny and Bhattacharya, Ananyo and Graepel, Thore},
  journal = {Nature},
  volume  = {654},
  number  = {8118},
  pages   = {324--326},
  year    = {2026},
  doi     = {10.1038/d41586-026-01820-1},
  url     = {https://www.nature.com/articles/d41586-026-01820-1}
}

@inproceedings{kg_burton2018homfly,
  author    = {Burton, Benjamin A.},
  title     = {The {HOMFLY-PT} Polynomial is Fixed-Parameter Tractable},
  booktitle = {34th International Symposium on Computational Geometry (SoCG 2018)},
  series    = {Leibniz International Proceedings in Informatics (LIPIcs)},
  volume    = {99},
  pages     = {18:1--18:14},
  publisher = {Schloss Dagstuhl--Leibniz-Zentrum f{\"u}r Informatik},
  year      = {2018},
  doi       = {10.4230/LIPIcs.SoCG.2018.18},
  eprint    = {1712.05776},
  archivePrefix = {arXiv}
}

\setcounter{dbltopnumber}{2}
\renewcommand{\dbltopfraction}{0.70}
\renewcommand{\textfraction}{0.20}
\renewcommand{\dblfloatpagefraction}{0.50}

\BeginSupplementaryInformation
  {\ManuscriptTitle}
  {\ManuscriptAuthors}
  {\ManuscriptAffiliations}
  {Correspondence: \ManuscriptCorrespondence}
  {\SupplementaryVersionDate}

\setcounter{secnumdepth}{2}
\renewcommand{\thesubsection}{\thesection.\Alph{subsection}}
\ifcsname theHsection\endcsname
  \renewcommand*{\theHsection}{supp.\arabic{section}}
\fi
\ifcsname theHsubsection\endcsname
  \renewcommand*{\theHsubsection}{supp.\arabic{section}.\Alph{subsection}}
\fi
\titleformat{\subsection}[block]
  {\normalfont\rmfamily\bfseries\fontsize{10.8}{12.8}\selectfont\KGHeadingNoBreak}
  {Supplementary Note~\thesubsection\enspace\textbar\enspace}{0pt}{}
\titlespacing*{\subsection}{0pt}{2.0ex plus 0.5ex minus 0.2ex}{0.55ex}

\section{From heterogeneous inputs to knotted graphs}
\label{supp:representation}

\begin{figure}[!t]
    \centering
    \includegraphics[
        width=\linewidth,
        keepaspectratio
    ]{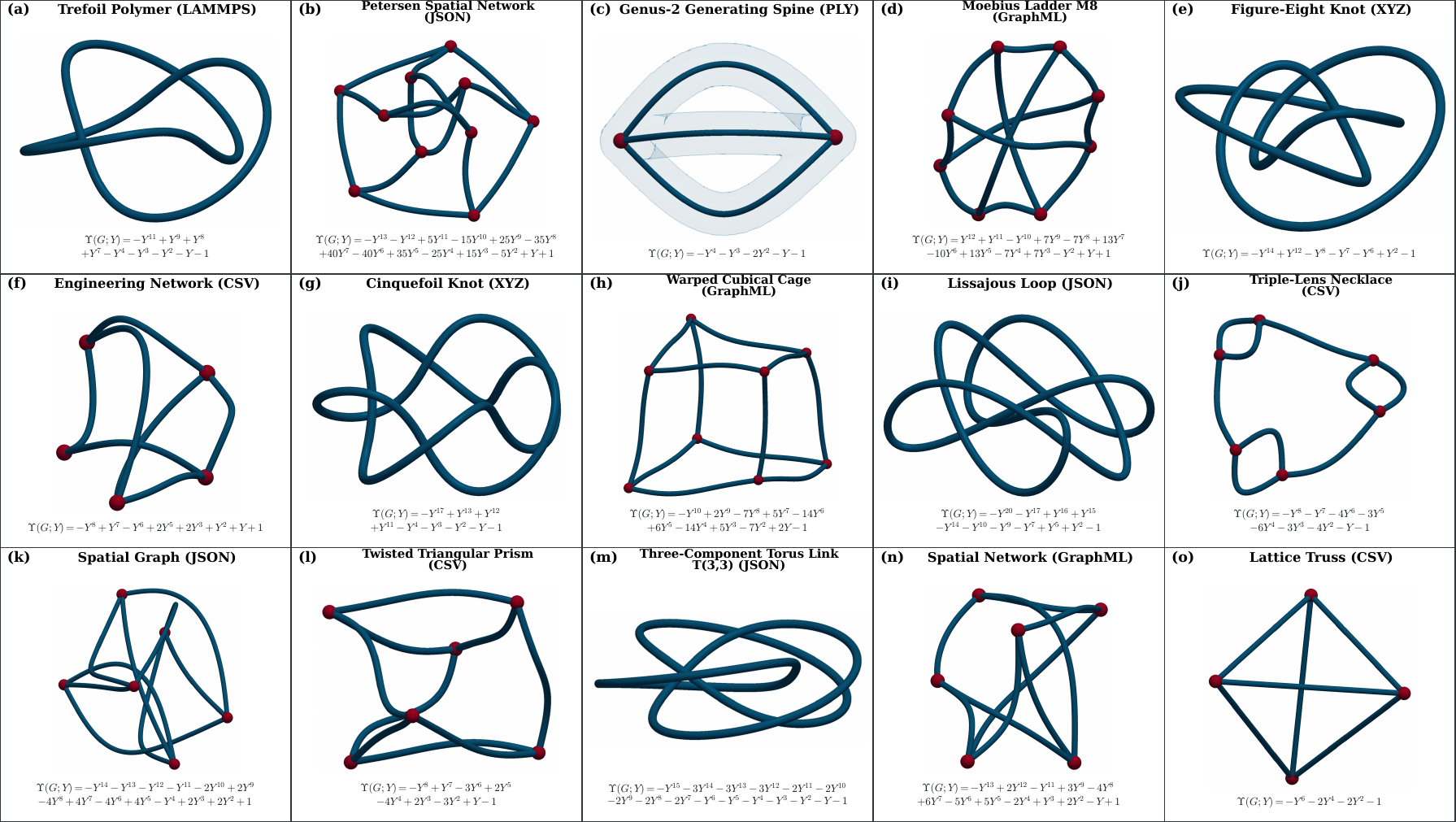}
    \caption{
    \textbf{Different input formats are converted to knotted graphs for Yamada evaluation.}
    Molecular structures, polymers, mathematical embeddings, engineered networks, knots, links and three-dimensional frameworks are normalized to the same knotted-graph representation and passed to the same Yamada-polynomial calculation.
    Representative inputs are \textbf{(a)} a trefoil polymer \cite{kg_deguchi2017topological_polymers}; \textbf{(b)} a Petersen spatial network; \textbf{(c)} a genus-two generating spine, illustrating the handlebody--spine viewpoint \cite{kg_ishii2008handlebody,kg_ishii2012handlebody}; \textbf{(d)} M\"obius ladder $M_8$; \textbf{(e)} a figure-eight knot; \textbf{(f)} an engineering network \cite{kg_shai1999engineering_graphs}; \textbf{(g)} a cinquefoil knot; \textbf{(h)} a warped cubical cage; \textbf{(i)} a Lissajous loop; \textbf{(j)} a triple-lens necklace; \textbf{(k)} a spatial $K_{3,3}$ graph; \textbf{(l)} a twisted triangular prism; \textbf{(m)} a three-component $T(3,3)$ torus link; \textbf{(n)} a spatial network originally stored in GraphML \cite{kg_brandes2002graphml} and converted externally to the knotted graph; and \textbf{(o)} a 3D truss network \cite{kg_shai1999engineering_graphs}. The polynomial displayed beneath each example is the output for that selected representation. Its invariant interpretation follows the graph and diagram conventions in \SuppNoteref{supp:yamada_engine}. Red spheres denote graph vertices and dark-teal curves denote the stored three-dimensional edges.
    }
    \label{suppfig:input_yamada_examples}
\end{figure}

\subsection{Heterogeneous inputs and the knotted-graph interface}
\label{supp:input_examples}

Each type of input enters the pipeline at the stage that matches the information it already contains. The supported routes, normalization rules and retained information
are summarized in \SuppTabref{supptab:input_interfaces}, with the corresponding
scientific input classes illustrated in \Figref{fig:input_formats_overview}.
Scientific structures of interest can be supplied as surface or volumetric
geometries even when an explicit one-dimensional graph is unavailable, for example
in tubular vascular geometries \cite{kg_aylward2005spatial_vascular} and
level-set regions such as finite-energy Fermi volumes
\cite{akgun_yan_2026topologicalclassificationknottedgraphs}. Related
spatial-network descriptions also arise in neuronal morphologies
\cite{kg_cuntz2010neuronal_graphs}, engineered systems
\cite{kg_shai1999engineering_graphs,kg_peddada2023}, and polymeric graph
structures \cite{kg_deguchi2017topological_polymers}.
After normalization, these routes converge on a common knotted-graph
representation retaining the available graph connectivity and three-dimensional
edge geometry, together with connected components and optional orientation
independently of any invariant. The resulting graph can then be projected and
passed to exact Yamada evaluation
(\Figpanelref{fig:functionality_overview}{b,c}); representative inputs and
their corresponding $\overline{\Upsilon}(G;Y)$ values are shown in
\SuppFigref{suppfig:input_yamada_examples}.

\begin{table*}[!t]
\centering
\setlength{\tabcolsep}{3.5pt}
\setlength{\arrayrulewidth}{0.5pt}
\setlength{\extrarowheight}{2.5pt}
\renewcommand{\arraystretch}{1.12}

\renewcommand{\tabularxcolumn}[1]{m{#1}}

\begin{tabularx}{\textwidth}{
|>{\centering\arraybackslash}m{0.17\textwidth}
|>{\centering\arraybackslash}m{0.24\textwidth}
|>{\centering\arraybackslash}X|}
\hline

\textbf{Input}
&
\textbf{Interface}
&
\textbf{How the input is checked and converted; what is kept}
\\
\hline

\shortstack[c]{\textbf{Molecular}\\\textbf{backbones}}
&
PDB/PDBx-mmCIF
\cite{kg_burley2019rcsb,kg_bourne1997mmcif}
&
The reader selects one model, one chain and one backbone atom type (for example C$\alpha$) and returns the selected atoms as an ordered backbone trace.
If the file contains several chains, the user must specify which chain to use.
For atoms with alternate locations, only the primary location is kept.
The selected model, chain and atom type are stored with the input.
mmCIF support is limited to RCSB-style \texttt{\_atom\_site} records in which each atom is written as one complete line.
\\
\hline

\shortstack[c]{\textbf{Explicit graphs}\\\textbf{and networks}}
&
Paired node--edge CSV; direct mathematical spatial graphs
&
Every node must have finite three-dimensional coordinates. Node and edge attributes are kept.
If an edge is supplied with its own polyline, the polyline endpoints are checked against the positions of the two nodes it connects.
This route covers vascular and neuronal networks \cite{kg_aylward2005spatial_vascular,kg_cuntz2010neuronal_graphs}, engineering networks \cite{kg_shai1999engineering_graphs,kg_peddada2023} and mathematical spatial graphs \cite{kg_kauffman1989spatial,kg_flapan2017spatial} (\Figpanelref{fig:input_formats_overview}{b,d,h}). GraphML, SWC and graph-JSON files in other layouts are converted to this node--edge form before being read.
\\
\hline

\textbf{Polymer snapshots}
&
GROMACS \cite{kg_abraham2015gromacs};
LAMMPS \cite{kg_thompson2022lammps}
&
Atoms from one snapshot are arranged into a chain. For GROMACS files, atoms are ordered by atom ID and coordinates are converted with an explicit unit scale (by default nm$\rightarrow$\AA).
For LAMMPS files, only the first frame is read, using the unscaled \texttt{x/y/z} columns. Atoms are ordered by a user-chosen column (by default atom ID), and if a \texttt{mol} column is present they are filtered by molecule.
Neither reader infers chemical bonds or unwraps periodic boundaries, so the selected atoms must already form the intended chain in the intended order.
\\
\hline

\textbf{Coordinate curves}
&
Finite $(N,3)$ arrays, including JSON or NPY
&
An open curve is stored as a single polyline. A closed curve is stored as a self-loop, an edge that starts and ends at the same vertex and carries the ordered coordinates.
The user must request closure explicitly; it is never inferred from how close the two ends are. Closure can be requested in two ways. Direct closure adds a final segment from the last point back to the first. Metadata-only closure marks the curve as closed without adding any segment.
\\
\hline

\shortstack[c]{\textbf{Surfaces,}\\\textbf{volumes and}\\\textbf{fields}}
&
Surface meshes; volumetric regions; sampled fields and Hamiltonians
&
Surface meshes are loaded as \texttt{PyVista PolyData} \cite{kg_sullivan2019pyvista}. Cleaning and triangulation are applied only when requested. Meshes that are empty, contain non-finite values or have open boundaries are reported, and open boundaries are not filled automatically.
The adapter returns the mesh together with a list of detected \texttt{issues}. Whether a given mesh can be reduced to the intended graph spine has to be judged for each application.
Hamiltonian-derived level sets or occupied regions (\Figpanelref{fig:input_formats_overview}{e}) \cite{akgun_yan_2026topologicalclassificationknottedgraphs}, volumetric regular neighborhoods (\Figpanelref{fig:input_formats_overview}{f}) \cite{kg_ishii2008handlebody,kg_ishii2012handlebody} and vector-field curves (\Figpanelref{fig:input_formats_overview}{g}) \cite{kg_helman1989vector_field_topology} are each converted to a knotted graph by their own geometry-to-graph route. The volumetric route is described in \SuppSubsecref{supp:extraction}.
\\
\hline

\shortstack[c]{\textbf{Analytic}\\\textbf{constructions}}
&
Named knots, links, torus types and Artin braid words
&
These inputs are generated directly from their mathematical definitions, so no file is read. When a braid word has to be realized as the zero set of a polynomial field, the field is built with the semiholomorphic construction of Bode and Dennis \cite{kg_bode2019prescribed}.
\\
\hline

\shortstack[c]{\textbf{Projected}\\\textbf{representations}}
&
Accepted diagrams or PD codes
&
No three-dimensional geometry is required. An existing diagram or PD code is passed directly to exact Yamada evaluation (\SuppNoteref{supp:yamada_engine}), skipping graph construction and projection.
\\
\hline

\end{tabularx}

\caption{
\textbf{Input routes and normalization.}
Each row lists one class of input, the formats or objects accepted (Interface), and how the input is checked and converted, together with the information kept (last column).
Each input enters at the stage that matches the information it already contains. Curves and explicit graphs enter directly as knotted graphs. Surfaces, volumes and fields first pass through graph construction. Diagrams and PD codes enter directly at invariant evaluation.
}
\label{supptab:input_interfaces}
\end{table*} 
\FloatBarrier

\subsection{Regular neighborhoods and spatial-graph spines}
\label{supp:handlebody_basis}

For the surface and volumetric inputs summarized in
\SuppSubsecref{supp:input_examples}, when a volumetric region is a regular
neighborhood of a spatial network, a one-dimensional spine
retains its handle structure, cycle organization and spatial entanglement while
discarding local thickness and boundary-shape details, as shown in
\Figpanelref{fig:input_formats_overview}{f}. Here we examine the corresponding
computational problem of recovering this graph spine from the
volumetric geometry, shown in
\Figpanelref{fig:functionality_overview}{a}. 

To formalize what information a recovered spine must preserve, it is useful first to distinguish graph connectivity from spatial embedding: \begin{quote}
\textbf{Abstract graph.}
A graph specified by its vertices, edges and incidence relations, independent
of any particular realization in space:

\medskip
\noindent\textbf{Spatial graph.}
An embedding of an abstract graph in three-dimensional space,
$G\hookrightarrow\mathbb R^3$, for which the embedding itself is part of the
mathematical object \cite{kg_kauffman1989spatial,kg_flapan2017spatial}.
\end{quote}

The same abstract graph can admit topologically distinct spatial realizations
when their spatial embeddings differ through knotting, linking or other
three-dimensional arrangements. More generally, spatial complexity
can increase independently of the local connectivity of the underlying abstract
graph. The structured families in
\Figpanelref{fig:knottedgraph_topoly_scaling}{a--f}, for example, represent
three distinct abstract graph families. Within each family, the
graph connectivity is preserved, while the spatial embedding becomes progressively more
crossing-rich. Their regular planar projections
therefore acquire increasing numbers of crossings even though the underlying
family-specific connectivity rule remains unchanged.
The equivalence is therefore an equivalence of embeddings rather
than of graph connectivity alone. Two spatial graphs are \emph{ambient-isotopic} when a continuous deformation of
the surrounding three-dimensional space carries one to the other without
cutting edges or passing nonincident edges through one another
\cite{kg_kauffman1989spatial,kg_flapan2017spatial}. Spatial-graph invariants such as the Yamada invariant, introduced in
\SuppNoteref{supp:yamada_engine}, probe this embedding-dependent topological
structure. \texttt{KnottedGraph} operates directly on the corresponding
spatial graphs and, when required, converts their spatial embedding into the
diagram needed for exact invariant evaluation.

For the surface and volumetric routes summarized in
\SuppSubsecref{supp:input_examples}, the spatial graph must be recovered from
the surrounding geometry. The connection between volumetric regular neighborhoods and spatial graphs can be
formalized through the following notions
\cite{kg_ishii2008handlebody,kg_ishii2012handlebody}:

\begin{quote}
\noindent\textbf{Regular neighbourhood.}
For a spatial graph $G$, a regular neighbourhood $N(G)$ is a sufficiently
small three-dimensional thickening of $G$, in which vertices expand into
compact junction regions and edges into connecting tubes.

\medskip
\noindent\textbf{Handlebody.}
For connected $G$, its regular neighbourhood $N(G)$ is a handlebody: a compact,
connected, orientable three-manifold obtained from a three-ball by attaching
$1$-handles.

\medskip
\noindent\textbf{Handlebody-link.}
For disconnected $G$, the regular neighbourhoods of its connected components
form a disjoint collection of embedded handlebodies, referred to as a
handlebody-link.

\medskip
\noindent\textbf{Spine.}
A spatial graph $G\subset\mathcal H$ is a spine of a handlebody $\mathcal H$
when $\mathcal H$ deformation retracts onto $G$.
\end{quote}

The deformation retract provides the mathematical idealization of reducing a
volumetric regular neighborhood to a spatial graph. In the regular-neighborhood
setting, the resulting spine represents the thickened object. Numerically,
we use the term candidate spine until the correspondence with the input
region has been established.

A handlebody generally admits many graph spines. Subdivision, local expansion or contraction and different geometric representatives can change the graph while preserving the represented handlebody-link. This equivalence is characterized by the following result:

\begin{quote}
\textbf{Theorem 2.3 \cite[Theorem~2.3]{kg_ishii2012handlebody}.}
Two spatial graphs without isolated vertices represent equivalent handlebody-links if and only if they
are related by a finite sequence of admissible spatial contraction moves and
ambient isotopies.
\end{quote}
This theorem establishes the equivalence but does not prescribe
how to recover a representative from sampled three-dimensional geometry.
Because one handlebody can admit many equivalent spines, the computational
target is not a unique geometric centerline but an admissible representative
of the same handlebody-link. The contraction moves in this
statement retain the spatial embedding. Graph contractions alone do
not establish the same equivalence, and the Yamada polynomial of a selected
spine need not be invariant under every change of spine.

\texttt{KnottedGraph}
implements numerical knotted-graph construction through voxelization, thinning,
multiscale skeleton-to-graph construction and scale-controlled cleanup
(\Figpanelref{fig:functionality_overview}{a};
\SuppFigref{suppfig:skeletonization_steps}). The recovered spine can then be
analyzed through its graph connectivity and cycle structure or projected (\Figpanelref{fig:functionality_overview}{b}) for topological invariant evaluation (\Figpanelref{fig:functionality_overview}{c}).
The benchmark in \SuppSubsecref{supp:handlebody_validation} tests this
construction against volumetric regular neighborhoods generated from known ground-truth
spatial graphs.

The same representation principle also points beyond the ordinary
handlebody setting. Volumetric geometries with distinguished inner
boundaries, enclosed voids or nested boundary components motivate a prospective
compression-body formulation. A compression body generalizes a handlebody by
allowing a distinguished positive boundary together with additional negative
boundary components \cite{kg_scharlemann2002heegaard}. Representing such
inputs therefore requires boundary-resolved graph data that record which graph
belongs to which boundary component and how those components are nested.

The boundary-resolved construction developed in the companion work
\cite{akgun_yan_2026topologicalclassificationknottedgraphs} motivates extending
these general-purpose computational interfaces. The
computational problem is also more difficult than the handlebody
construction treated here. In the handlebody case, the volumetric
region can be represented by a one-dimensional graph spine, allowing the
pipeline to proceed through volumetric thinning, junction identification and
knotted-graph cleanup. For geometries with multiple distinguished boundary
components, a single graph need not contain the full topological
information: the deformation retract can involve boundary surfaces
together with graph-like attachments. A computational representation must therefore preserve graph connectivity,
spatial embedding, boundary attachment, nesting and the relative topology of
the boundary components. Skeleton-to-graph reduction can erase
this information, so new boundary-aware graph construction, certification and
invariant-evaluation procedures are required.  

\subsection{Skeleton extraction and multiscale graph construction}
\label{supp:extraction}

\begin{figure}[!t]
    \centering
    \includegraphics[width=\linewidth]
    {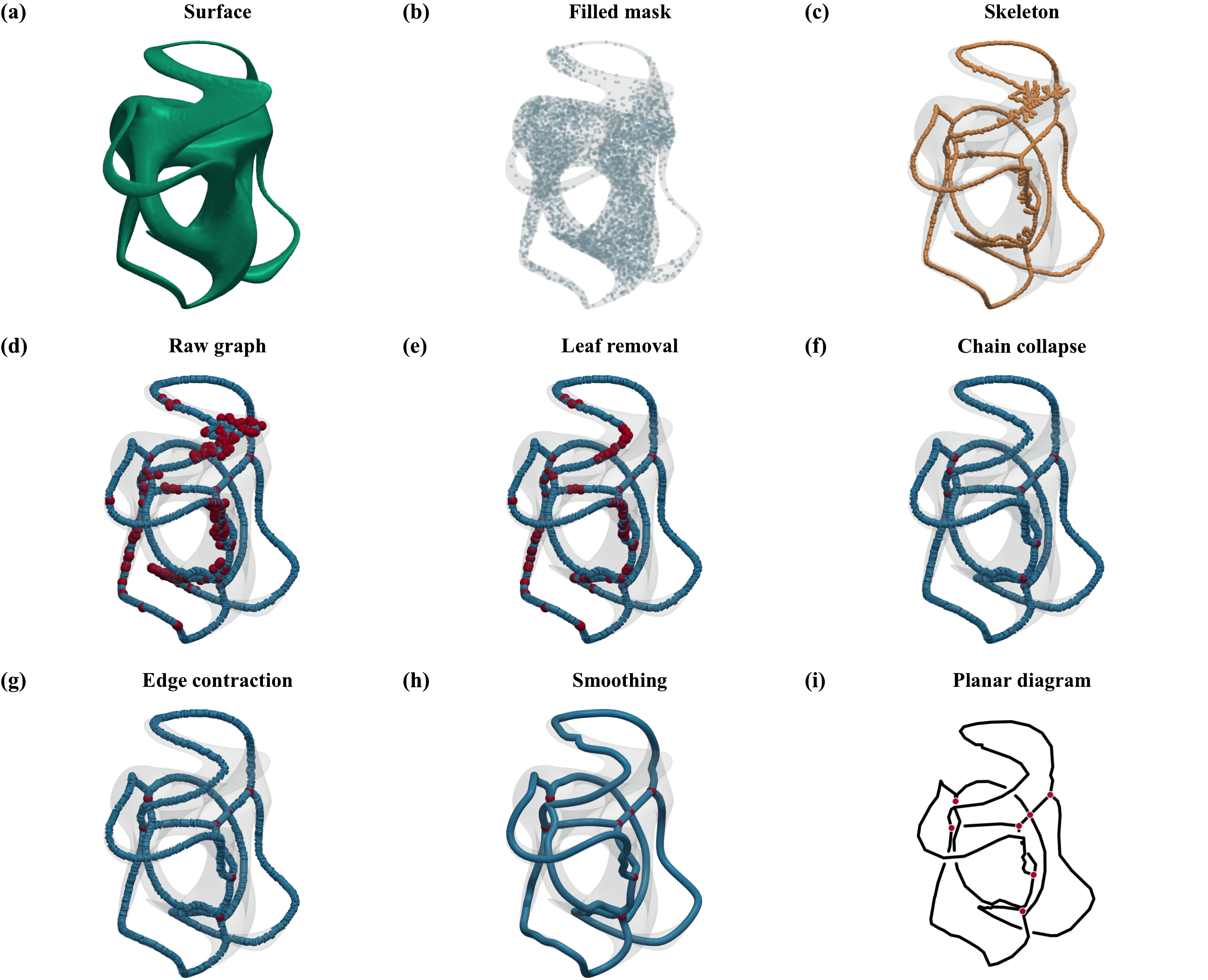}
    \caption{
    \textbf{A volumetric regular neighborhood is thinned, converted to a graph and cleaned before projection.}
    \textbf{(a)} Input boundary surface.
    \textbf{(b)} Filled volumetric mask used for three-dimensional thinning.
    \textbf{(c)} One-voxel-wide skeleton obtained with Lee thinning.
    \textbf{(d)} Raw knotted graph constructed from the voxel skeleton by
    identifying junction regions across neighbouring local scales, selecting
    graph connectivity recurring across the tested scales and tracing the intervening
    degree-two chains as three-dimensional edge polylines.
    \textbf{(e)} Removal of terminal leaf artifacts when these are not part of
    the intended graph.
    \textbf{(f)} Collapse of redundant degree-two chains into single graph
    edges, retaining their three-dimensional edge geometry.
    \textbf{(g)} User-controlled contraction of sufficiently short local edges,
    used to merge voxel-scale fragmentation of physical junctions.
    \textbf{(h)} Geometric simplification of the retained edge polylines to
    suppress voxel-scale jaggedness without changing graph connectivity.
    \textbf{(i)} Regular projection of the cleaned knotted graph for
    crossing detection and subsequent PD-code construction.
    Panels \textbf{(e--h)} isolate the effects of the cleanup primitives;
    in the thick-handlebody benchmark the call order is
    short-edge contraction, leaf removal, degree-two simplification and
    Ramer--Douglas--Peucker smoothing \cite{kg_ramer1972,kg_douglas1973}.
    Chain collapse changes only graph
    subdivision, whereas leaf removal and short-edge contraction are
    scale-dependent operations whose effect is validated against ground truth.
    }
    \label{suppfig:skeletonization_steps}
\end{figure}

The handlebody--spine correspondence \cite{kg_ishii2012handlebody} supplies the topological target for graph construction in \Figpanelref{fig:functionality_overview}{a}. Given a finite-resolution surface or volume, the task is to recover a knotted graph whose graph connectivity and three-dimensional edge geometry represent that object and remain suitable for subsequent topological analysis. The
reduction is shown panel by panel in
\SuppFigref{suppfig:skeletonization_steps}.

\paragraph{Surface to filled volume: panels (a,b)}
The pipeline begins from a surface or volumetric input
(\SuppFigpanelref{suppfig:skeletonization_steps}{a}). When the input is supplied
as a surface, the enclosed region is converted to the occupied
three-dimensional mask shown in
\SuppFigpanelref{suppfig:skeletonization_steps}{b}; field- or volume-derived
inputs can enter directly at this stage. Surface handling uses
\texttt{PyVista} \cite{kg_sullivan2019pyvista}, and level-set surface
extraction uses Lewiner marching cubes through \texttt{scikit-image}
\cite{kg_lewiner2003marching,kg_vanderwalt2014skimage}. The volume mask is the
object on which the subsequent digital thinning operation acts. Its
correspondence with the analytic source is a separate question.

\paragraph{Filled volume to one-dimensional skeleton: panel (c)}
Before thinning, empty margins are cropped so that computation is restricted
to the occupied bounding box. Then it is thinned with the three-dimensional
Lee algorithm through \texttt{scikit-image}
\cite{kg_lee1994thinning,kg_vanderwalt2014skimage}, producing the one-voxel-wide
backbone in \SuppFigpanelref{suppfig:skeletonization_steps}{c}. The cropped
result is restored to the original voxel coordinate frame before graph
construction. The implementation therefore separates the morphological
reduction itself from the later task of deciding which parts of the digital
skeleton constitute graph vertices and graph edges.

\paragraph{Voxel skeleton to raw knotted graph: panel (d)}
The one-voxel-wide skeleton in
\SuppFigpanelref{suppfig:skeletonization_steps}{c} is still a digital object:
its occupied voxels specify a centerline, but graph vertices and graph edges
have not yet been identified. The next stage converts this voxel skeleton into
the explicit knotted graph shown in
\SuppFigpanelref{suppfig:skeletonization_steps}{d}.

Following voxel-skeleton graph constructions
\cite{kg_reinders2000skeletongraph}, skeleton voxels that touch through a face,
edge or corner are first connected using the standard $26$-neighbour adjacency
of a three-dimensional voxel grid ($6$ face-, $12$ edge- and $8$
corner-sharing neighbours). \texttt{KnottedGraph} then suppresses a direct
diagonal lattice connection whenever the same local connection is already
represented through an occupied intermediate voxel by two strictly shorter
lattice steps. This additional cleanup prevents the discrete neighbourhood rule
from inserting redundant shortcuts that can create artificial local cycles or
junction structure.

For larger skeletons, the same construction is carried out by a compiled
C++ implementation. It builds the shortcut-reduced $26$-neighbour voxel graph once and
traces the candidate junction sizes directly on this sparse representation.
For each candidate size, it first records only the information needed to choose
between candidates---the graph connections, edge lengths, edge point counts and
whether the stored geometry is valid. After the cross-scale recurrence and multigraph
isomorphism tests select one scale, three-dimensional edge polylines
are constructed only for that selected graph. 

The cleaned voxel network is converted to a graph using the standard
degree-based skeleton interpretation \cite{kg_reinders2000skeletongraph}.
Degree-two voxels form the interiors of one-dimensional edges, whereas
connected regions containing voxels of degree different from two identify
endpoints or junctions. Each such region is represented by a graph vertex, and
the intervening degree-two chains are traced explicitly as graph edges. The
ordered three-dimensional coordinates along each chain are retained, so this
step preserves both graph connectivity and the three-dimensional edge geometry.
Components consisting entirely of degree-two voxels are retained separately as
closed loops. Isolated vertices and small tree components are also included
when accounting for the whole input; their removal would change its component
count.

The main difficulty is that voxelization does not provide a unique local
representation of a physical junction. A single thick junction can thin to a
cluster of nearby branch voxels and can therefore be misinterpreted as several
graph vertices connected by very short edges. Simply merging all nearby
vertices is not sufficient, because distinct junctions may also lie
close in space. To avoid choosing one arbitrary junction-clustering
scale, \texttt{KnottedGraph} follows the multiscale construction summarized
in \Eqref{eq:main_multiscale_reconstruction}: it constructs candidate graphs from the same sparse
voxel skeleton across neighbouring local junction-zone scales and uses
recurrence of the resulting graph connectivity across scales to resolve this ambiguity.

Starting from the zero-radius trace, candidate graphs are generated
through progressively expanded junction regions, up to a user-specified number
of voxel hops. Each candidate must contain nondegenerate three-dimensional edge geometry
and must avoid anomalously short junction connections relative to the
local edge scale. If an expected maximum valence is known, it is incorporated
as an additional admissibility constraint; when no such prior is supplied, no
valence bound is inferred. Thus multi-scale topology selection does not require
advance knowledge of the graph degree.

Candidate graph topologies are compared across neighbouring scales using
full multigraph isomorphism through NetworkX
\cite{kg_hagberg2008networkx,kg_cordella2004vf2}. Two consecutive clean,
isomorphic graph candidates provide the strongest cross-scale consistency signal, in which
case the smaller-scale representative is retained. The finer candidate that
will actually be returned must pass the short-edge check. The next coarser
candidate only needs to be a clean isomorphic graph, because it is being used
only to confirm that the same topology persists at the next scale. Requiring
that coarser graph to pass the same short-edge test can incorrectly reject a
stable finer graph just when two nearby junction regions are beginning to merge
at the coarser scale. If no consecutive pair is found, the most recurrent valid
isomorphism class across the tested scales is selected, provided that it occurs
at least twice, again preferring its smallest-scale representative. A topology observed at only one
expanded scale is not treated as evidence of cross-scale stability; if no valid topology
recurs, the direct zero-radius trace is retained; an unsupported scale-dependent
candidate is not promoted.

These comparisons test recurrence of the abstract graph over the chosen
scales; they do not establish equivalence of its spatial embedding. The procedure
combines local geometric validity with cross-scale graph-topology consistency and allows physical information about
vertex valence to strengthen the selection when such information is available.
For example, in the thick-handlebody construction benchmark
(\SuppSubsecref{supp:handlebody_validation}), the ground-truth graphs are
subcubic, so the optional maximum-degree prior is set to three and junction
expansion is tested through four voxel hops. At this stage the output is a knotted graph with explicit junctions and
three-dimensional edge geometry. Residual artifacts can remain at
the geometric scale of the voxelization.

\paragraph{Leaf removal: panel (e)}
Terminal degree-one edges require different treatment depending on the
object being represented. For a general knotted graph, a leaf can be part of
the intended graph and must not be removed. In the handlebody--spine setting,
however, a pendant tree edge does not introduce an additional cycle or
handle in the regular neighbourhood. An admissible spatial contraction of such an edge
changes the chosen graph spine without changing the represented
handlebody-link, consistent with the contraction-move equivalence of
Theorem~2.3 \cite{kg_ishii2012handlebody}. Accordingly, when the computational
target is the handlebody topology and not a prescribed abstract graph,
terminal tree protrusions may be removed to obtain a simpler equivalent spine,
as illustrated in
\SuppFigpanelref{suppfig:skeletonization_steps}{e}. This operation is not
applied universally: for biological trees, open polymers or other applications
in which degree-one edges are themselves part of the scientific object,
they are retained.

\paragraph{Degree-two chain collapse: panel (f)}
After terminal artifacts have been dealt with, long sequences of degree-two
vertices can remain solely because a continuous graph edge was sampled by
many voxel or graph points. Panel
\SuppFigpanelref{suppfig:skeletonization_steps}{f} removes this redundant
subdivision. Consecutive degree-two graph segments are concatenated into a
single edge between the surrounding significant vertices, and the ordered
three-dimensional polyline along the entire original chain is retained. Thus
this operation changes the combinatorial subdivision of the graph but not the
represented graph edge. Closed degree-two components remain represented as
self-loops and are not discarded.

\paragraph{Short-edge contraction: panel (g)}
A junction recovered from the voxel skeleton can be split across several nearby graph vertices by
finite-resolution thinning and skeleton-to-graph conversion. Panel
\SuppFigpanelref{suppfig:skeletonization_steps}{g} simplifies such local
structure by contracting sufficiently short edges between distinct vertices.
In the handlebody--spine setting, this operation has a direct topological
justification: an admissible edge contraction changes the chosen spatial-graph
spine but not the handlebody-link represented by its regular neighbourhood,
consistent with the contraction-move equivalence of
Theorem~2.3 \cite{kg_ishii2012handlebody}. The retained vertex is repositioned
between the original endpoints and the incident edge polylines are relinked to
the contracted junction.

The geometric threshold selects which short connections are proposed for
contraction. It does not by itself establish that the endpoint motion or the
relinked polylines realize an admissible spatial move. Those geometric checks
are needed in addition to the abstract contraction. The threshold must also
be chosen conservatively because two nearby junctions can represent
scientifically distinct graph connectivity even when their contraction is
permissible at the level of handlebody--spine equivalence. Accordingly, short-edge contraction is
an explicit, user-controlled operation and is not applied as a generic simplification.

\paragraph{Geometric smoothing: panel (h)}
Once graph connectivity has been fixed, the edge geometry is regularized without
changing connectivity. Panel
\SuppFigpanelref{suppfig:skeletonization_steps}{h} shows this geometric
simplification. The voxel-scale staircase structure of each stored polyline is
reduced using the Ramer--Douglas--Peucker algorithm
\cite{kg_ramer1972,kg_douglas1973}, implemented through
\texttt{fastrdp}. The retained polyline endpoints are then normalized to the
corresponding graph-vertex coordinates and consecutive duplicate samples are
removed. This operation improves the numerical conditioning of
later projection and crossing detection and leaves graph connectivity unchanged.  

\paragraph{Knotted graph to projected diagram: panel (i)}
The output of the graph-construction stage is therefore an explicit knotted graph,
not only a binary skeleton. Its vertices carry three-dimensional positions and
its edges retain ordered spatial polylines. This object can already be used for
graph connectivity, cycle and geometric measurements. When a
Yamada invariant is required, the cleaned graph is passed to the
generic-projection stage shown in
\SuppFigpanelref{suppfig:skeletonization_steps}{i}. A suitable viewing direction
is selected, edge over/undercrossing information is assigned at the projected crossings and
the resulting diagram is converted to the PD code used by the
exact Yamada calculation. The details of this final projection step are given
in \SuppNoteref{supp:projection}.

The panel sequence separates three operations:
volumetric topology reduction \textbf{(a--c)}, digital skeleton-to-graph
construction \textbf{(c,d)}, and explicit graph/geometric cleanup
\textbf{(e--h)} before the representation changes from three dimensions to a
projected diagram in \textbf{(i)}. The recovery benchmark in
\SuppSubsecref{supp:handlebody_validation} tests whether this entire sequence
returns the intended connectivity and normalized Yamada invariant for
volumetric regular neighborhoods generated from known spatial ground-truth graphs.

\section{From knotted graphs to projected diagrams}
\label{supp:projection}

Once graph construction has produced a knotted graph with the intended
connectivity and spatial geometry, the problem changes from recovering the
object to projecting its spatial embedding to a diagram for exact Yamada
evaluation (\Figpanelref{fig:functionality_overview}{b}). Projection produces the diagram required by the topological invariant calculation.

The input is first normalized to
the package's knotted-graph \texttt{MultiGraph} contract: every graph vertex has a
finite three-dimensional position and every graph edge is represented by an
ordered three-dimensional polyline whose endpoints coincide with its incident
vertices. Projection then changes only the representation of this fixed
embedding; it does not alter graph connectivity. This layer combines deterministic spherical-Fibonacci view sampling
\cite{kg_keinert2015fibonacci}, spatially indexed segment-intersection queries
\cite{kg_leutenegger1997str,kg_shapely}, and planar-diagram encoding
\cite{kg_mastin2015pdcode} under standard spatial-graph conventions
\cite{kg_kauffman1989spatial,kg_flapan2017spatial}.

The methodological
advance in \texttt{KnottedGraph} is to integrate these components into a
projection procedure with explicit acceptance checks and a retained record tailored to
knotted graphs: genuine graph vertices are kept distinct from
projection crossings, nongeneric views are rejected and are never forced
into a diagram, edge over/undercrossing information is determined from the original
three-dimensional edge geometry, duplicate crossing incidences are resolved
consistently along the stored edge paths, and the accepted projection is
retained together with its crossing structure and PD encoding for the exact
polynomial evaluated downstream. When the program finds a projected
intersection, it also records which two original graph edges produced it and
the distance of the intersection along each edge. These are exactly the
positions at which the edges must later be cut to build the PD diagram, so the
program does not need a second geometric search to locate the same crossings
again.

\subsection{Sampling and selecting regular projections}
\label{supp:projection_sampling}

A user may prescribe one Euler rotation explicitly, in which case that view is
used directly. Otherwise \texttt{KnottedGraph} evaluates a deterministic family
of candidate viewing directions, producing the candidate diagrams $D_i$ and
crossing counts $n_{\mathrm{cr},i}$ defined in \Eqref{eq:main_projection_candidates}. The default multi-view route uses an
upper-hemisphere spherical-Fibonacci/golden-angle construction
\cite{kg_keinert2015fibonacci}, as illustrated in \SuppFigpanelref{suppfig:pdcode_to_yamada}{a} with the projections
$P_1$, $P_2$ and $P_3$ for the same knotted graph. Opposite viewing directions are
not sampled separately and the in-plane roll is fixed, because these choices do
not generate distinct diagram topology. For each candidate direction, the
knotted graph is rigidly rotated and projected to the $xy$ plane. The
subsequent crossing detection, edge over/undercrossing determination and PD
serialization are described in
\SuppSubsecsref{supp:projection_crossings}{supp:pd_serialization} and illustrated
in \SuppFigref{suppfig:pd_code_generation}.

\begin{table}[!t]
\centering
\small
\begin{tabular}{p{0.27\linewidth}p{0.66\linewidth}}
\hline
\textbf{Projection event} & \textbf{Treatment} \\
\hline
Overlapping or collinear projected segments
&
Reject the candidate as nongeneric; do not assign an artificial
crossing.
\\[0.35em]

Near-tangent four-incidence intersection
&
Reject when cyclically adjacent projected half-edge directions approach
tangency within the angular tolerance, because the local crossing order is not
numerically well resolved.
\\[0.35em]

Unresolved edge height
&
Reject a projected crossing when the competing three-dimensional edge segments have
indistinguishable heights within the geometric tolerance, so that an
edge over/undercrossing assignment cannot be made reliably.
\\[0.35em]

Valid regular projection
&
Retain the projection record and continue to ordered crossing
detection and PD construction.
\\
\hline
\end{tabular}
\caption{
\textbf{Criteria for accepting a regular spatial-graph projection.}
Candidate views are rejected whenever the projected geometry does not support
an unambiguous diagram; invalid views are excluded rather
than coerced into a PD code.
}
\label{supptab:projection_acceptance}
\end{table}

The target is a \emph{regular, numerically resolvable diagram} suitable for exact evaluation, not simply a visually convenient projection.
Each candidate view is therefore subjected to explicit acceptance guards rather
than being forced into a PD code. When several candidate views are
requested, the program first counts the crossings in each view without building
its complete vertices, arcs and PD code. The views are ordered from fewest to
most crossings, with generation order breaking ties, and the complete diagram
is then built only for the best candidates until a valid one is obtained. The
same minimum-crossing rule is therefore retained, avoiding full PD
construction for views that will not be selected.

For normalized Yamada evaluation, the program performs one additional check on
difficult diagrams. If the minimum-crossing view has at least $12$ crossings
and its processing order would require at least $12$ arc ends to remain
unresolved at the same time, the other sampled views are examined as well. An
alternative view is accepted only if it lowers this maximum by at least two arc
ends. This matters because the exact calculation can sometimes be easier for a
diagram with slightly more crossings if those crossings can be processed with
fewer unresolved arc ends at once. If no candidate passes the
acceptance criteria in \SuppTabref{supptab:projection_acceptance}, projection
terminates with an error and does not manufacture a diagram.
 \subsection{Crossing detection and edge over/undercrossing determination}
\label{supp:projection_crossings}

Once a candidate view has passed the genericity checks in
\SuppSubsecref{supp:projection_sampling}, its planar geometry is fixed, but the
projection is not yet a combinatorial diagram. The apparent intersections must
first be identified, associated with their parent graph edges and assigned
edge over/undercrossing information using the original three-dimensional geometry. This is
the role of the crossing-recovery stage.

Crossing detection acts on the stored edge polylines; the graph edges are
not replaced by straight chords between graph vertices. Every polyline is split into two-point
segments. Along with each segment, the program stores which original graph edge
it belongs to and the distance from the start of that edge to the start of the
segment. An STR tree generates candidate segment pairs before vectorized
\texttt{Shapely} intersection tests
\cite{kg_shapely,kg_leutenegger1997str}. When an intersection is found, its
distance along both original edges is therefore obtained immediately from the
same calculation. The program can later split those edges at the crossing
without building another spatial index or searching for the crossing a second
time.

Line overlaps are treated as nongeneric projection degeneracies. Point
intersections are checked against the corresponding original
three-dimensional segments: the $z$ coordinate of each segment is linearly
interpolated at the projected $(x,y)$ intersection. If the two interpolated
heights agree within tolerance at a projected intersection of nonincident
edges, the view is rejected because it does not define a resolvable
edge over/undercrossing. Shared graph endpoints are treated as incidences rather
than ordinary crossings. Otherwise, the relative heights determine which
edge segment passes over the other.

\begin{figure}[!t]
    \centering
    \includegraphics[width=\linewidth]{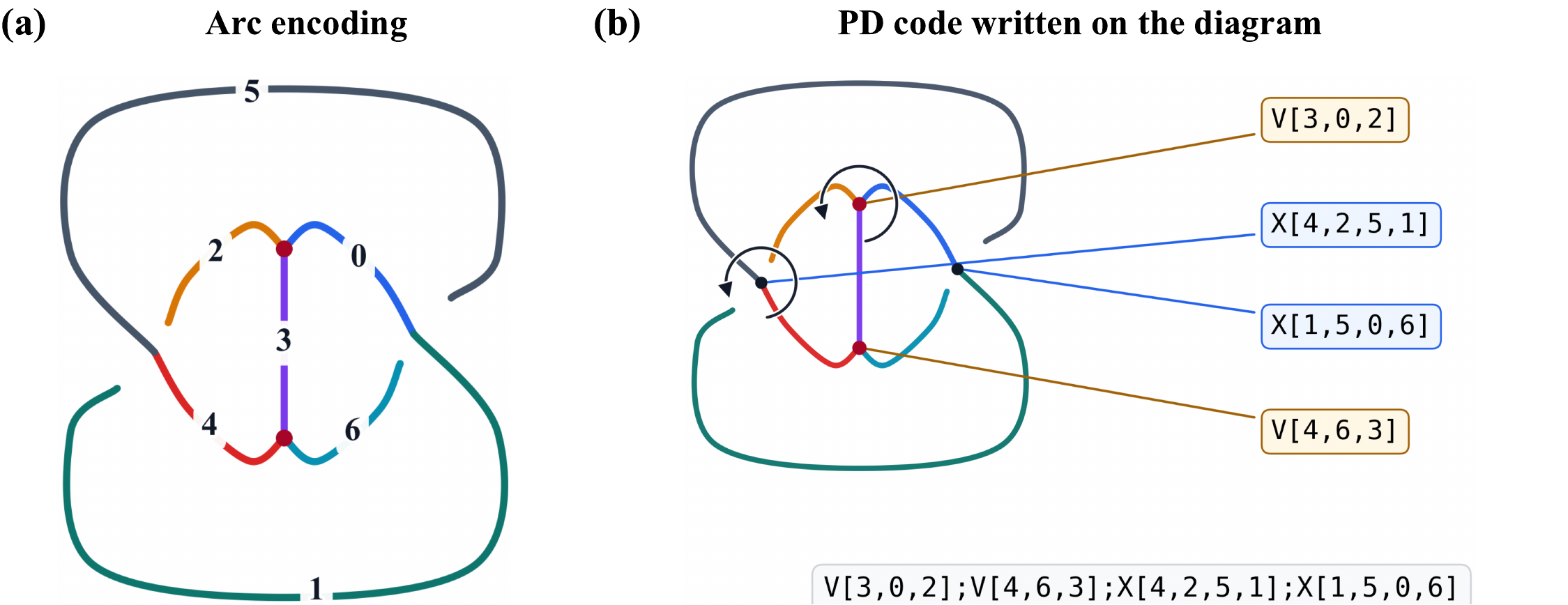}
    \caption{
    \textbf{PD code records graph vertices, crossings, cyclic order and edge over/undercrossing information.}
    \textbf{(a)} Every projected graph edge is cut at its ordered crossing
    positions, producing arcs whose endpoints are either graph vertices or
    projection crossings; each resulting arc receives a discrete label.
    \textbf{(b)} Graph vertices are serialized as $V[\cdots]$ records containing
    incident arc labels in counter-clockwise order; projection crossings
    are serialized as $X[\cdots]$ records. The four crossing incidences are
    ordered counter-clockwise and cyclically shifted, when necessary, so the
    overpassing arc pair occupies the convention used by the Yamada calculation.
    Numerical arc labels are arbitrary identifiers: incidence, cyclic order and
    edge over/undercrossing information is retained by the PD code.
    }
    \label{suppfig:pd_code_generation}
\end{figure}

The same geometric crossing can be encountered more than once when it lies at
an internal polyline sample shared by adjacent segments. Such duplicate
incidences are merged before diagram construction. Endpoint directions are
computed from distinct polyline samples, skipping degenerate successive
segments. The retained crossings are already ordered by their recorded
arclength along each original projected edge. The output of this stage is an
ordered crossing structure attached to the accepted projection: for every
graph edge, the software knows which crossings occur along it, in what
order, and which edge segment is over or under at each event. These ordered crossing
positions provide the cut locations used to construct the diagram arcs in
\SuppFigpanelref{suppfig:pd_code_generation}{a} and, across multiple accepted
views, lead to the distinct projection-specific diagram representations shown
in \SuppFigpanelref{suppfig:pdcode_to_yamada}{b}.

\subsection{Arc splitting, cyclic incidence and PD serialization}
\label{supp:pd_serialization}

\begin{figure}[!t]
    \centering
    \includegraphics[width=\linewidth]
    {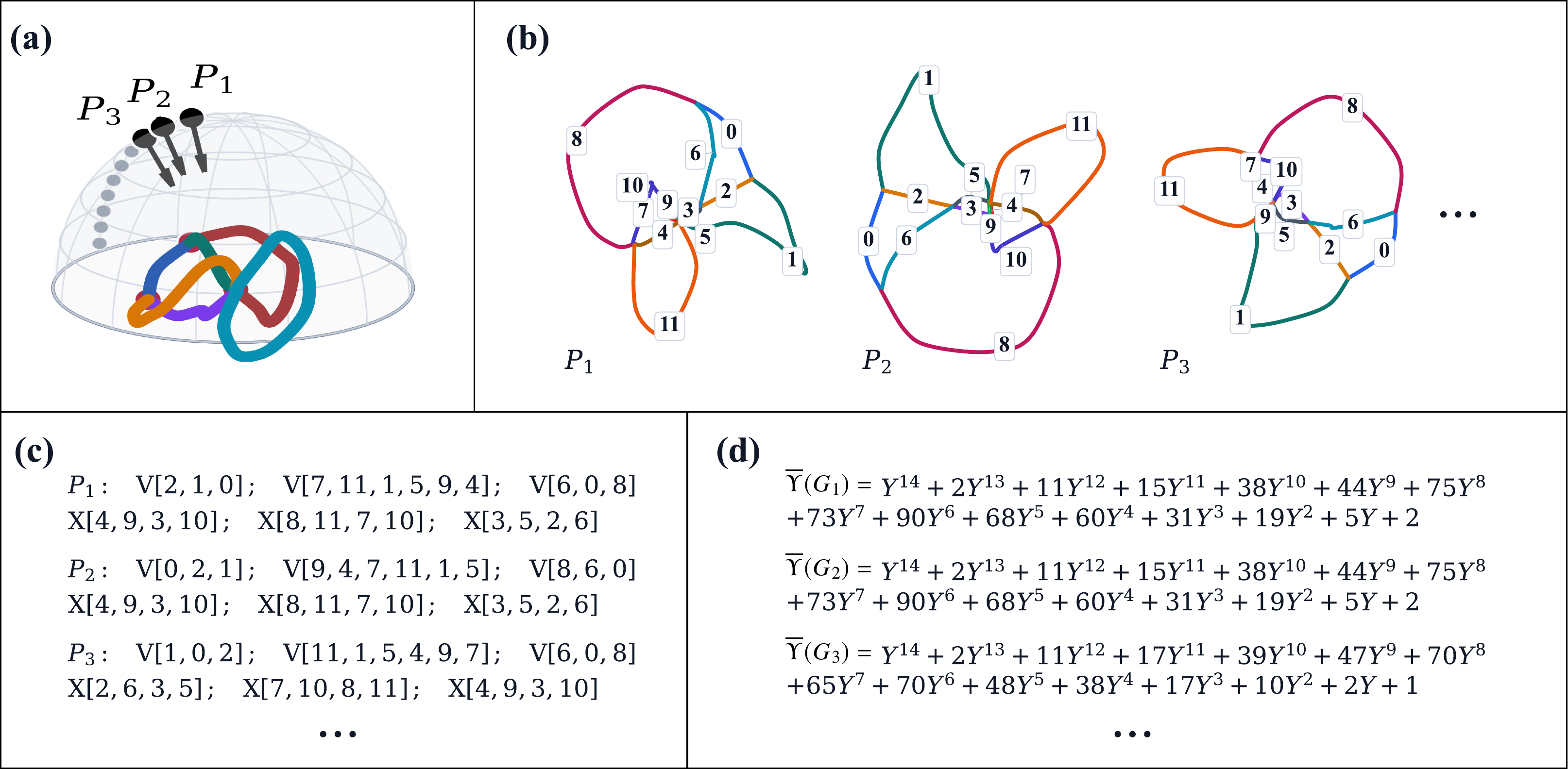}
\caption{
\textbf{Different regular projections give different PD codes; normalized Yamada values are projection independent for the tested subcubic graphs.}
\textbf{(a)} Three accepted viewing directions $P_1$, $P_2$ and $P_3$ are
selected for the same three-dimensional spatial graph using the procedure of
\SuppSubsecref{supp:projection_sampling}.
\textbf{(b)} Crossing detection and edge over/undercrossing determination
(\SuppSubsecref{supp:projection_crossings}) produce distinct diagrams.
\textbf{(c)} Each diagram is independently serialized according to
\SuppSubsecref{supp:pd_serialization}, producing its own arc labels and
$V[\cdots]$/$X[\cdots]$ records. The displayed graph contains a six-valent
vertex and is therefore not subcubic.
\textbf{(d)} The corresponding normalized Yamada outputs
$\overline{\Upsilon}(G_{P_i};Y)$ need not agree for this higher-valence
example when a common rigid-vertex structure has not been established.
Higher valence alone does not explain every discrepancy. Thus successful
projection and exact polynomial evaluation do not by
themselves imply projection-independent invariant characterization outside the
admissible subcubic regime used for the multi-view consistency tests in
\SuppSubsecref{supp:sanity_validation}.
}
 \label{suppfig:pdcode_to_yamada}
\end{figure}

The ordered crossing structure obtained in
\SuppSubsecref{supp:projection_crossings} is converted into the discrete
incidence data used by the exact Yamada calculation. The conversion
for one accepted projection is illustrated in
\SuppFigref{suppfig:pd_code_generation}; the same serialization applied
independently to several accepted views of one spatial embedding is shown in
\SuppFigpanelref{suppfig:pdcode_to_yamada}{c}. Panel
\SuppFigpanelref{suppfig:pd_code_generation}{a} shows the projected spatial
graph after its graph edges have been cut at the ordered crossing positions.
The resulting arc pieces are assigned the discrete labels
$0,1,\ldots,6$ shown directly on the diagram. For example, the upper graph
vertex is incident to arcs $3$, $0$ and $2$; the lower graph vertex is
incident to arcs $4$, $6$ and $3$. The longer outer edge segments are likewise
subdivided where they encounter the two projection crossings, so the labels
identify the individual arc segments between successive graph vertices or
crossing objects.

These labeled arcs are then converted into incidence records in
\SuppFigpanelref{suppfig:pd_code_generation}{b}. Genuine graph vertices are
serialized as
\[
V[a_1,a_2,\ldots,a_d],
\]
with the incident arc labels ordered counter-clockwise around the vertex. Thus
the upper and lower degree-three vertices in the example become, respectively,
\[
V[3,0,2]
\qquad\text{and}\qquad
V[4,6,3],
\]
exactly as indicated by the brown blocks in
\SuppFigpanelref{suppfig:pd_code_generation}{b}.

For a crossing, the four incident arcs are first ordered
counter-clockwise in the projection plane, and the three-dimensional height
comparison from \SuppSubsecref{supp:projection_crossings} determines which
opposite arc pair is overpassing. The $X[\cdots]$ record is then
written starting from an incidence belonging to the overpassing arc pair and
continuing counter-clockwise, so that the first and third entries correspond to
the overpassing arc pair and the second and fourth entries to the underpassing arc pair. In the
example, the two blue blocks in
\SuppFigpanelref{suppfig:pd_code_generation}{b} therefore give
\[
X[4,2,5,1]
\qquad\text{and}\qquad
X[1,5,0,6].
\]
For the first crossing, arcs $4$ and $5$ form the overpassing pair, so the
record starts from arc $4$; for the second, arcs $1$ and $0$ form the
overpassing pair, so the record starts from arc $1$. The particular starting
arc within the overpassing pair has no independent topological significance;
it fixes a consistent cyclic representative of the same crossing convention.

Then, these four local
records are assembled into the complete PD code displayed at the
bottom of panel \textbf{(b)},
\[
V[3,0,2];
V[4,6,3];
X[4,2,5,1];
X[1,5,0,6].
\]
This completes the representation change from a knotted graph to the
PD-code input required by the exact invariant calculation.
This representation is projection dependent: different generic
views of the same spatial embedding can produce different crossing layouts,
arc labels and PD strings, as shown in
\SuppFigpanelref{suppfig:pdcode_to_yamada}{c}. These projection-specific PD
representations are the inputs to the exact invariant calculation developed next.

\section{Exact Yamada evaluation from PD codes}
\label{supp:yamada_engine}

The preceding section converts each accepted projection into a complete
PD code. \texttt{KnottedGraph} then automates the
remaining exact Yamada-polynomial evaluation
(\Figpanelref{fig:functionality_overview}{c}). The Yamada polynomial
\cite{kg_yamada1989,kg_mellor2018spatial} was introduced as a
polynomial invariant for spatial graphs. At the diagram level, the
raw polynomial is a regular-isotopy quantity and can change under a
Reidemeister-I move by a monomial factor. After normalization, this dependence
is removed and the resulting polynomial is invariant under rigid-vertex
isotopy. When every graph vertex has degree at most three, i.e. when the graph
is \emph{subcubic}, the additional vertex-rotation move required for
pliable-vertex isotopy does not introduce a new ambiguity. For subcubic spatial
graphs, the normalized Yamada polynomial is therefore invariant under
pliable-vertex isotopy and can be used as an ambient-isotopy invariant of the
spatial embedding: different normalized polynomials certify topologically
distinct embeddings, although equality of the polynomial does not imply
equivalence \cite{kg_yamada1989,kg_mellor2018spatial}.

For a spatial graph $G$ represented by a chosen planar projection,
$\Upsilon(G;Y)$ denotes the raw regular-isotopy Yamada polynomial. The
normalization convention is defined in
\Eqref{eq:main_yamada_normalization}. For $\Upsilon(G;Y)=0$, the minimum
degree is undefined and we set $\overline{\Upsilon}(G;Y)=0$ directly. This
evaluated zero is distinct from an unavailable calculation.
When it is necessary to distinguish different projections of the same spatial
graph, we indicate the chosen projection by a subscript, writing
$G_{P_1},G_{P_2},\ldots$. For subcubic spatial graphs,
$\overline{\Upsilon}(G_{P_i};Y)$ is independent of the accepted generic
projection $P_i$, so the projection label can be suppressed and the invariant
written simply as $\overline{\Upsilon}(G;Y)$. For higher-valence graphs,
however, normalization alone need not remove dependence on the local cyclic
arrangement at a graph vertex, and the projection label must therefore be
retained.

The multi-view example in
\SuppFigref{suppfig:pdcode_to_yamada} shows this distinction directly. As described in the preceding section,
\SuppFigpanelref{suppfig:pdcode_to_yamada}{a--c} constructs three independent
PD codes for $G_{P_1}$, $G_{P_2}$ and $G_{P_3}$ from accepted regular
projections of the same three-dimensional spatial graph $G$. The graph contains
a six-valent vertex, visible for example through the six-entry vertex record in
\SuppFigpanelref{suppfig:pdcode_to_yamada}{c}, and is therefore not subcubic.
The quantities displayed in
\SuppFigpanelref{suppfig:pdcode_to_yamada}{d} are the normalized
polynomials
\[
\overline{\Upsilon}(G_{P_1};Y),\qquad
\overline{\Upsilon}(G_{P_2};Y),\qquad
\overline{\Upsilon}(G_{P_3};Y),
\]
yet they do not all coincide. The selected diagrams do not establish one
common rigid-vertex structure, so these outputs are retained with their
diagrams and are not compared as invariants of the underlying pliable graph.
Higher valence alone does not diagnose every disagreement: diagram
construction and the vertex convention still need to be checked. Multi-view
invariant-consistency tests are consequently performed on subcubic spatial
graphs, as described in \SuppSubsecref{supp:sanity_validation}.

The computational contribution developed here is complementary to this
topological distinction. \texttt{KnottedGraph} provides a general exact
algorithm for computing the Yamada polynomial of an arbitrary supplied PD
representation, including graphs containing higher-valence vertices; whether
the normalized result can subsequently be interpreted as a
projection-independent spatial-graph invariant depends on the graph class
described above. The performance of our library is
demonstrated in \Figref{fig:knottedgraph_topoly_scaling}, where the same algorithm reproduces independently published Yamada formulas
\cite{kg_dobrynin1996yamada,kg_li2018yamada} on structured diagrams containing
up to $500$ crossings.

\subsection{Local crossing relation and the formal resolution space}
\label{supp:local_yamada_relation}

The three local choices are shown directly in
\Eqref{eq:main_yamada_local_relation}: the side smoothing carries weight $Y$,
the top smoothing carries weight $Y^{-1}$ and the vertex resolution carries no
additional power of $Y$. The complete mathematical state sum and the
crossing-free graph contribution are given in
\Eqsref{eq:main_yamada_state_sum}{eq:main_yamada_crossing_free}. The algorithm
preserves these relations exactly; the computational improvement comes from
combining equivalent intermediate connectivity states instead of evaluating
every complete resolution independently.

\begin{figure}[!t]
    \centering
    \includegraphics[width=\linewidth]
    {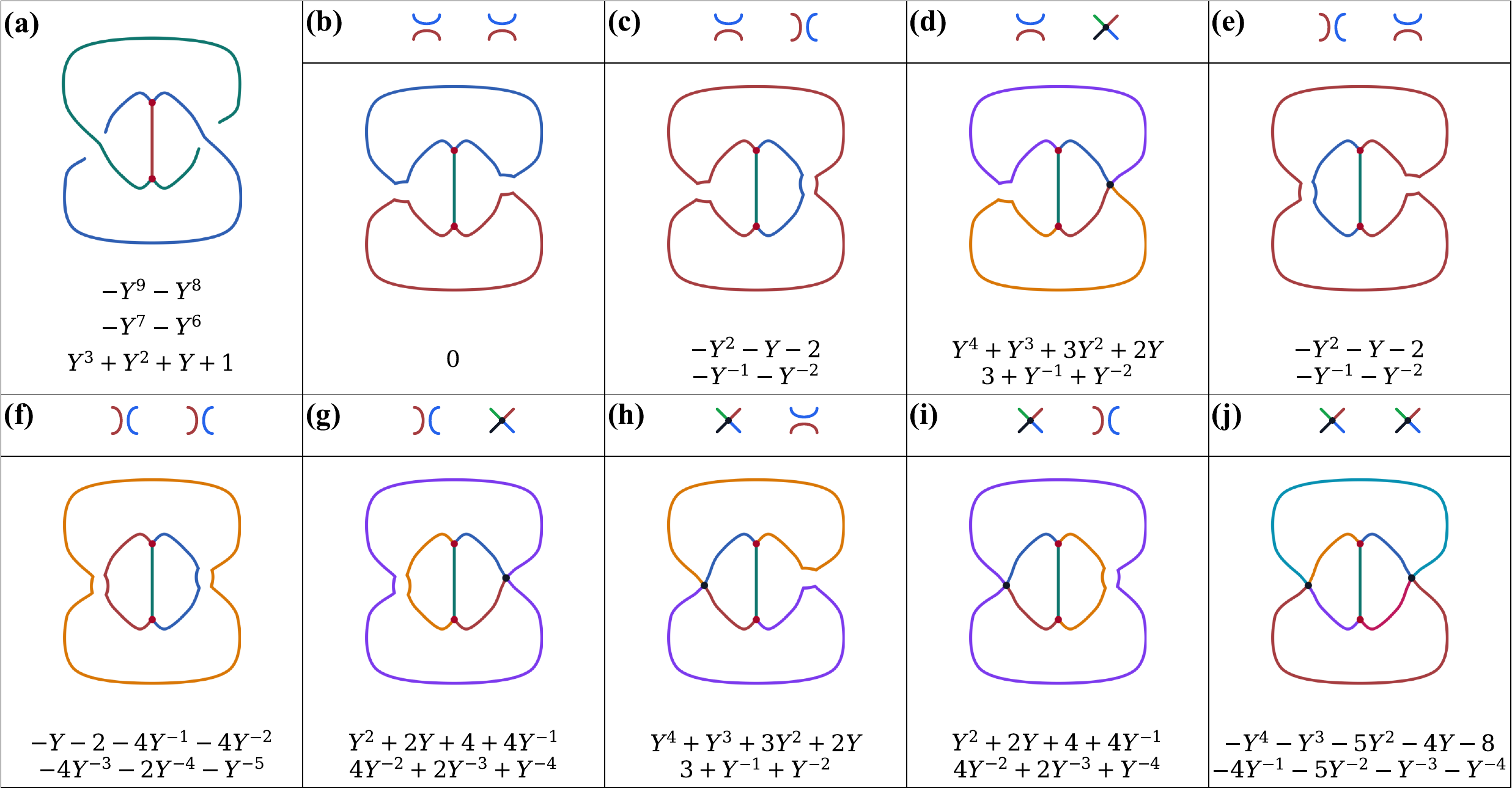}
    \caption{
    \textbf{The exact three-state Yamada relation is evaluated without
    calculating every resolution independently.}
    \textbf{(a)} A two-crossing diagram whose direct expansion contains
    $3^2=9$ complete resolutions.
    \textbf{(b--j)} The nine combinations obtained by assigning the
    side-smoothing, top-smoothing or vertex resolution independently to the two
    crossings, with local weights $Y$, $Y^{-1}$ and $1$, respectively,
    according to \Eqref{eq:main_yamada_local_relation}. Different resolutions already give
    repeated polynomial outcomes, for example
    \textbf{(c,e)}, \textbf{(d,h)} and \textbf{(g,i)}.
    During larger calculations, equivalent partial resolutions are combined
    as soon as their remaining connections are the same, and their exact
    Laurent-polynomial contributions are summed. This reduces repeated work
    without changing the Yamada algebra \cite{kg_yamada1989}.
    }
    \label{suppfig:yamada_resolution_states}
\end{figure}

The resulting formal resolution count is given in
\Eqref{eq:main_yamada_resolution_space}. Here we make this construction
explicit through the two-crossing example in
\SuppFigref{suppfig:yamada_resolution_states}. Because each crossing can
independently take the side-smoothing, top-smoothing or vertex state, the diagram in
\SuppFigpanelref{suppfig:yamada_resolution_states}{a} has $3^2=9$ complete
resolution assignments, displayed in
\SuppFigpanelref{suppfig:yamada_resolution_states}{b--j}.

Even this two-crossing example already contains repeated results. The
resolutions in \SuppFigpanelref{suppfig:yamada_resolution_states}{c} and
\SuppFigpanelref{suppfig:yamada_resolution_states}{e} give the same resolved-graph polynomial; the same occurs for
\SuppFigpanelref{suppfig:yamada_resolution_states}{d} and
\SuppFigpanelref{suppfig:yamada_resolution_states}{h}, and for
\SuppFigpanelref{suppfig:yamada_resolution_states}{g} and
\SuppFigpanelref{suppfig:yamada_resolution_states}{i}. The algorithm detects
this more generally during the calculation: whenever two partial resolutions
have become equivalent for everything that remains to be processed, their
weighted polynomial contributions are combined. Equality of a resolved-graph
polynomial must be distinguished from its local crossing weight.

\subsection{Compiling the PD code and removing redundancies}
\label{supp:yamada_compilation}

The state sum and crossing-free graph polynomial in
\Eqsref{eq:main_yamada_state_sum}{eq:main_yamada_crossing_free} are the
classical Yamada/Negami ingredients
\cite{kg_yamada1989,kg_negami1987,kg_li2018yamada}. The graph-vertex splitting,
state-combination and ordering procedures below are implementation choices of
\texttt{KnottedGraph}. The first computational improvement is to record the PD
connectivity once before the exact Yamada evaluation begins. From the $V[\cdots]$ and
$X[\cdots]$ records constructed in \SuppSubsecref{supp:pd_serialization}, the
algorithm stores each edge's endpoints and the two smoothing connections at
each crossing in \Eqref{eq:main_yamada_local_relation}.

Before the Yamada evaluation, the same stored connectivity is also used to
remove admissible Reidemeister-II redundancies \cite{kg_yamada1989}. In a
cancellable bigon, the
same edge segment passes over at both crossings. The edge segments can then be
separated without changing their topology, so explicitly resolving both
crossings would introduce unnecessary cases. The proposed cancellation below
illustrates why incidence alone is insufficient: as drawn, its crossing
assignment fails this edge condition.

\begin{center}
\begin{tikzpicture}[scale=0.9, line cap=round, line join=round]

\draw[line width=0.8pt]
    (0,0.45) -- (0.8,-0.45) -- (1.6,0.45) -- (2.4,0.45);
\draw[line width=0.8pt]
    (0,-0.45) -- (0.8,0.45) -- (1.6,-0.45) -- (2.4,-0.45);

\draw[white,line width=3.2pt] (0.30,0.1125) -- (0.50,-0.1125);
\draw[line width=0.8pt]       (0.30,0.1125) -- (0.50,-0.1125);

\draw[white,line width=3.2pt] (1.10,0.1125) -- (1.30,-0.1125);
\draw[line width=0.8pt]       (1.10,0.1125) -- (1.30,-0.1125);

\draw[->,line width=0.8pt] (2.8,0) -- (3.8,0)
    node[midway,above] {$\mathrm{RII}$};

\draw[line width=0.8pt] (4.2,0.45) -- (6.6,0.45);
\draw[line width=0.8pt] (4.2,-0.45) -- (6.6,-0.45);

\end{tikzpicture}

\small
\textbf{Rejected cancellation.} The displayed arrow is not a valid RII move:
different edge segments pass over at its two crossings. A cancellable bigon has the
same edge segment over at both crossings.
\end{center}

Reidemeister II is useful here because it preserves the raw
regular-isotopy Yamada polynomial and directly reduces the number of
crossings; the corresponding regular-isotopy behavior follows the Yamada
diagram relations \cite{kg_yamada1989}. Reidemeister I changes the raw
polynomial by a monomial factor, whereas Reidemeister III does not reduce the
crossing number. Each admissible
Reidemeister-II cancellation removes two crossings, reducing the formal
resolution space from $3^{n_{\mathrm{cr}}}$ to $3^{n_{\mathrm{cr}}-2}$.

The algorithm finds these configurations directly from the stored edges. It
searches for two crossings joined by exactly two edges and checks that these
edges occupy the required adjacent positions at both crossings and that the
same edge segment is over at both. The other two
edges at each crossing are then followed to determine how the edges reconnect
after the crossings are removed. If this reconnection is unambiguous and has
four distinct surviving endpoints, the two crossings and the two edges between
them are deleted and the external edges are reconnected pairwise. The remaining
edge and crossing indices and the corresponding smoothing connections are
updated, and the procedure is repeated until no further pair remains.

\subsection{Combining equivalent connectivity states in exact Yamada evaluation}
\label{supp:yamada_connectivity_states}

The calculation starts from the PD representation described in
\SuppSubsecref{supp:pd_serialization}. At this point the three-dimensional
embedding has already been converted to a combinatorial diagram: graph vertices
are stored as $V[\cdots]$ records, crossings as $X[\cdots]$ records, and the
arc labels specify how these local objects are connected
(\SuppFigref{suppfig:pd_code_generation}). The remaining problem is therefore
the exact evaluation of the Yamada polynomial from this fixed diagram.

The mathematical calculation to be reproduced is the state sum introduced in
\Eqref{eq:main_yamada_state_sum}. At every crossing, the local relation
\Eqref{eq:main_yamada_local_relation} gives three possibilities: the side
smoothing with weight $Y$, the top smoothing with weight $Y^{-1}$ and the
vertex resolution with unit local weight. A complete resolution state
$s\in\mathcal S(D)$ chooses one of these three possibilities at every crossing,
producing a crossing-free graph $G_s$. The corresponding contribution to the
raw diagram polynomial is
\begin{equation}
Y^{\,n_{\mathrm{side}}(s)-n_{\mathrm{top}}(s)}
\Upsilon(G_s;Y),
\end{equation}
and summing these contributions over all complete states gives
\Eqref{eq:main_yamada_state_sum}. Since every crossing independently admits
three choices, a diagram with $n_{\mathrm{cr}}$ crossings formally contains
$3^{n_{\mathrm{cr}}}$ complete resolution assignments, as stated in
\Eqref{eq:main_yamada_resolution_space}.

The two-crossing example in
\SuppFigref{suppfig:yamada_resolution_states} illustrates this direct
construction. Panel \textbf{(a)} contains two crossings and panels
\textbf{(b--j)} display the resulting $3^2=9$ complete assignments. The
repeated outcomes visible there already show that different resolution
histories can eventually contribute the same resolved-graph polynomial. The algorithm applies the same idea before the resolution tree is completed, processing the diagram one local object at a time and combining partial histories as soon as the unprocessed part can no longer distinguish them.

It is useful to view this procedure as a reordering of the sums already present
in \Eqsref{eq:main_yamada_state_sum}{eq:main_yamada_crossing_free}. A direct implementation would first generate each
complete crossing-resolution state $s$, construct its crossing-free graph
$G_s$, and then evaluate the edge-subset sum in
\Eqref{eq:main_yamada_crossing_free} separately for that graph. The present
algorithm instead accumulates the same local crossing weights, graph-vertex
signs, edge-subset signs and cycle factors progressively. Partial calculations
that have become identical with respect to everything still unprocessed are
summed immediately.

At an intermediate stage, imagine a conceptual boundary separating the part
of the diagram that has already been processed from the part that remains to
be processed. An arc that has entered the processed part but continues to a
crossing, vertex or connection that has not yet been processed intersects this
boundary in an \emph{open arc end}. These open ends are bookkeeping objects;
they are not physical dangling ends of the original graph. They identify the
only places through which later operations can interact with the part of the
diagram already incorporated into the calculation.

Here, ``processed'' means that the corresponding crossing or graph connection
has already been incorporated into the running state. Processing a crossing
means applying its three local choices from
\Eqref{eq:main_yamada_local_relation} to every state currently stored,
multiplying by the appropriate local weight and updating the connections among
the resulting open arc ends. Processing continues until every local object has
been incorporated. The calculation can therefore be summarized as
\begin{equation}
\begin{aligned}
\text{PD code}
&\longrightarrow
\text{take the next crossing or vertex connection}
\longrightarrow
\text{update open connections}
\\
&\longrightarrow
\text{combine states with the same open connections}
\longrightarrow
\text{remove finished arc ends}.
\end{aligned}
\label{suppeq:yamada_connectivity_workflow}
\end{equation}

The aggregation of partial states by unresolved boundary data has analogues in
dynamic-programming algorithms for knot polynomials, including the
tree-decomposition algorithm for HOMFLY--PT
\cite{kg_burton2018homfly}. Here, however, the stored boundary data and all
local updates are those required by the Yamada state sum in
\Eqref{eq:main_yamada_state_sum} and its crossing-free graph contribution in
\Eqref{eq:main_yamada_crossing_free}. If the input diagram has disconnected
components, the calculation is performed on each component separately and the
resulting Laurent polynomials are multiplied, hence we describe one
connected component below.

\paragraph{What is stored after part of the diagram has been processed}

Suppose that the current processing boundary contains the open arc ends
$p_1,\ldots,p_m$. The detailed geometry of the already processed region is no
longer needed. What matters for future steps is which of these open ends have
already become connected to one another through paths lying entirely inside the
processed part.

Two open ends therefore need not be joined by a single edge to be regarded as
connected. They are in the same stored component whenever a path through the
processed region already joins them. The algorithm records this information by
\begin{equation}
\kappa=(\kappa_1,\ldots,\kappa_m),
\qquad
\kappa_i=\kappa_j
\ \Longleftrightarrow
p_i\text{ and }p_j\text{ are joined through the processed part}.
\label{suppeq:yamada_connectivity_key}
\end{equation}
Thus $\kappa$ records a partition of the currently open arc ends into connected
components. The numerical labels themselves carry no information; only equality
or inequality of the labels matters. They are therefore renumbered
canonically from left to right, beginning with $0$.

For example,
\begin{equation}
\kappa=(0,1,0,1)
\label{suppeq:yamada_connectivity_example}
\end{equation}
means that $p_1$ and $p_3$ have already been joined by some path through the
processed part, while $p_2$ and $p_4$ have been joined through another
processed component. The state does not record the actual paths or which
earlier side-smoothing, top-smoothing or vertex choices created them. In
particular, the first and third entries being equal means only that the
processed portion already contains a path between them.

This is the information that remains relevant because every later local
operation acts on the processed region only through these open ends. Suppose,
for example, that two different partial Yamada-resolution histories have both
reached the same four open ends $p_1,p_2,p_3,p_4$ and both leave
\begin{equation}
\kappa=(0,1,0,1).
\end{equation}
The first history may have reached this connectivity through one sequence of
side, top and vertex resolutions, while the second may have reached it through
a completely different sequence. Their accumulated Laurent-polynomial
contributions can therefore be different; denote them by
$\mathcal W_1(Y)$ and $\mathcal W_2(Y)$.

Nevertheless, everything that remains to be processed sees exactly the same
boundary data in the two cases. In both histories, the same four arc ends are
present, $p_1$ is already connected to $p_3$, and $p_2$ is already connected
to $p_4$. Any later crossing resolution, graph-vertex connection or edge
inclusion therefore changes the two states in the same way. The future
calculation cannot distinguish which earlier sequence produced the
connectivity.

Consequently, the two histories do not need to be propagated separately.
Their accumulated polynomial contributions can be added immediately and
associated with the single boundary state,
\begin{equation}
(0,1,0,1)\longmapsto
\mathcal W_1(Y)+\mathcal W_2(Y).
\label{suppeq:yamada_merge_example}
\end{equation}
This is simply an early collection of terms that would otherwise remain
separate until the final sum in \Eqref{eq:main_yamada_state_sum}. If the
continuation of the diagram acts identically on both histories, propagating
$\mathcal W_1(Y)$ and $\mathcal W_2(Y)$ separately and adding them at the end
gives the same result as adding them first and propagating their sum.

More generally, at every stage the algorithm stores a table
\begin{equation}
\kappa\longmapsto \mathcal W_{\kappa}(Y),
\label{suppeq:yamada_state_table}
\end{equation}
where $\kappa$ specifies the connectivity of the currently open arc ends and
$\mathcal W_{\kappa}(Y)$ is the sum of the exact Laurent-polynomial
contributions of all partial histories that produce that same boundary
connectivity. Thus one stored state can represent many different prefixes of
the complete resolution tree in \Eqref{eq:main_yamada_state_sum}. The histories
are combined only after they have become indistinguishable to every operation
that remains unprocessed.

This also clarifies the distinction between the repeated outcomes in
\SuppFigref{suppfig:yamada_resolution_states} and the state compression used in
larger calculations. In that figure, different \emph{complete} assignments
can be seen to produce repeated final polynomial outcomes. The running
algorithm uses the stronger intermediate criterion above: two histories can be
combined before completion whenever they expose the same open arc ends with the
same connectivity and have already accumulated all algebraic factors arising
from their different processed histories.

\paragraph{Incorporating a crossing}

When a crossing is
processed, its four incident arc ends are introduced into every currently
stored boundary state. The three alternatives in
\Eqref{eq:main_yamada_local_relation} are then applied. For each incoming
state,
\begin{itemize}[leftmargin=*]
\item the side smoothing reconnects the four incidences according to the side
      pairing and multiplies its accumulated contribution by $Y$;
\item the top smoothing uses the complementary pairing and multiplies the
      contribution by $Y^{-1}$;
\item the vertex resolution joins the four incidences through a graph vertex
      and has unit weight in the local crossing relation.
\end{itemize}
The algorithm immediately computes the new boundary
connectivity $\kappa$ and merges any branches that have become identical
according to \Eqref{suppeq:yamada_state_table}. The vertex resolution must also be connected to the second part of the Yamada
calculation. Although its local crossing weight in
\Eqref{eq:main_yamada_local_relation} is $1$, it creates a graph vertex in the
resolved graph $G_s$. Every vertex of $G_s$ contributes through the factor
$(-1)^{|V_s|}$ in \Eqref{eq:main_yamada_crossing_free}. Thus a vertex created
by a crossing resolution contributes one factor $-1$ when the graph-polynomial
part of the calculation is incorporated. The local weight $1$ and the
graph-polynomial factor $-1$ arise from different formulas and must not be
conflated.

After the three local branches have been created and their connectivity has
been updated, states with identical boundary connectivity are immediately
combined. Thus the calculation still contains all three Yamada choices from
\Eqref{eq:main_yamada_local_relation}, but it does not retain two separate
copies of a future calculation once their processed parts have become
equivalent from the perspective of the unprocessed diagram.

\paragraph{Incorporating the crossing-free graph contribution}

After all crossings in a complete state $s$ have been resolved, the direct
state sum would require the evaluation of the crossing-free graph polynomial
\begin{equation}
\Upsilon(G_s;Y)
=
(-1)^{|V_s|}
\sum_{F\subseteq E_s}
(-1)^{|F|}
\left(Y+2+Y^{-1}\right)^{\beta(F)}
\end{equation}
from \Eqref{eq:main_yamada_crossing_free}. For each physical graph edge, the edge-subset sum in \Eqref{eq:main_yamada_crossing_free} introduces two possibilities: the edge is either excluded from $F$ or included in $F$. If it is excluded, the connectivity is unchanged and the contribution is multiplied by $+1$. If it is included, the contribution acquires the factor $-1$ from $(-1)^{|F|}$ and the connected components containing the two endpoints are joined. These include/exclude updates are applied directly within the running connectivity calculation, so the complete graph $G_s$ need not be constructed first.

The stored connectivity makes it possible to determine immediately whether
including the edge merely joins two components or creates a new cycle. If its
two endpoints belong to different components of the current state, including
the edge joins those components and contributes only the factor $-1$. If the
two endpoints are already connected by a path through the processed part,
including the edge closes an additional cycle. In that case the cycle rank
$\beta(F)$ in \Eqref{eq:main_yamada_crossing_free} increases by one and the
factor
\begin{equation}
Y+2+Y^{-1}
\end{equation}
is inserted at that moment. The complete contribution of such an included edge
is therefore
\begin{equation}
-(Y+2+Y^{-1}).
\end{equation}

In this way, the factor
$\left(Y+2+Y^{-1}\right)^{\beta(F)}$ is accumulated one completed cycle at a
time. The final value of $\beta(F)$ does not need to be stored explicitly:
every increase in cycle rank has already contributed its corresponding factor.
Likewise, the signs associated with included edges and graph vertices are
inserted when those objects are processed. The state weight
$\mathcal W_{\kappa}(Y)$ therefore contains all algebraic information from the
processed portion that is needed by
\Eqsref{eq:main_yamada_state_sum}{eq:main_yamada_crossing_free}.

\paragraph{Removing open ends that can no longer affect the calculation}

The set of open arc ends changes as processing advances. An open end is needed
only while it remains incident, through the original PD connectivity, to some
object that has not yet been processed. Once an open end lies entirely inside
the processed region and cannot participate in any later crossing, vertex or
edge operation, it can no longer affect the remainder of the state sum and is
removed from the boundary description.

For example, suppose five open ends have the connectivity
\begin{equation}
(0,1,0,2,1).
\end{equation}
This means that $p_1$ and $p_3$ are already connected, $p_2$ and $p_5$ are
connected through another component, and $p_4$ belongs to a third component.
If $p_1$ and $p_3$ no longer connect to any unprocessed object, both ends can
be deleted from the boundary. The remaining labels are $(1,2,1)$. Since the
numerical labels themselves are arbitrary, this state is canonically
renumbered as
\begin{equation}
(0,1,0).
\end{equation}

This removal can reveal additional equivalences. Two histories that previously
had different boundary states can become identical after their finished ends
are discarded. Their accumulated Laurent-polynomial contributions are then
combined according to \Eqref{suppeq:yamada_state_table}. Repeating this
operation ensures that the stored state describes only the portion of the
processed diagram that can still influence the unprocessed calculation.

\paragraph{Splitting a high-degree graph vertex into smaller connection steps}

A graph vertex of degree $d$ requires all $d$ of its incident arc ends to
belong to the same connected component in the resolved graph. Introducing all
$d$ incidences simultaneously can cause many open ends to be present at once,
even though the required final connectivity is simple. The algorithm therefore
enforces the same identification through a deterministic sequence of smaller
connection steps, each involving at most three arc ends.

These helper connections are bookkeeping operations only. They do not create
additional physical graph edges and do not alter the graph represented by the
PD code. Their purpose is to impose the original degree-$d$ vertex connection
without exposing all of its incident arc ends simultaneously.

The graph-polynomial factor must nevertheless remain exactly that of the
original vertex. The factor $(-1)^{|V_s|}$ in
\Eqref{eq:main_yamada_crossing_free} supplies one factor $-1$ for that graph
vertex, regardless of how many helper steps are used internally. One of the
smaller connection steps therefore carries this factor $-1$, while the
remaining helper steps have weight $+1$.

If a helper identification joins two arc ends that are already connected
through the processed state, the identification closes a cycle and contributes
the corresponding factor $+(Y+2+Y^{-1})$. It does not receive the additional
physical-edge sign $-1$ because a helper connection is not an edge belonging to
the subset $F$ in \Eqref{eq:main_yamada_crossing_free}. The sequence therefore
produces the same final connectivity and the same polynomial contribution as
processing the original degree-$d$ vertex directly.

\paragraph{Why the processing order matters}

The compression above removes dependence on the number of complete histories
that have already collapsed to the same boundary state, but the number of
possible distinct states still depends strongly on how many open arc ends are
present simultaneously. With only a few open ends there are relatively few
possible partitions into connected components. As the number of open ends
grows, many more distinct connectivities $\kappa$ can occur and fewer partial
histories can be combined.

This is why the computational cost is controlled not only by the total crossing
number in \Eqref{eq:main_yamada_resolution_space}, but also by the number of
connections exposed simultaneously during the running calculation. The main
text summarizes this dependence after
\Figref{fig:knottedgraph_topoly_scaling}, and the projection-selection
procedure in \SuppSubsecref{supp:projection_sampling} uses the same quantity
when deciding whether a slightly more crossing-rich projection may nevertheless
be easier to evaluate.

The algorithm therefore chooses the order in which crossings and graph-vertex
connection steps are processed. It preferentially continues from objects
already adjacent to the processed region, so that existing open connections
are closed before many unrelated new ones are introduced. For an unprocessed
step $j$, let $b_j$ denote the number of its connections to already processed
steps, $f_j$ the number of its connections to still-unprocessed steps and
$r_j$ the number of its incident arc ends. The first step is chosen with the
smallest total number of connections, using smaller $r_j$ to break ties.
Subsequent steps are ordered according to
\begin{equation}
\left(
\mathbf 1_{b_j=0},
f_j-b_j,
f_j,
-b_j,
r_j,
j
\right).
\label{suppeq:yamada_order_score}
\end{equation}
The entries are compared from left to right. The first term penalizes a step
that is disconnected from the currently processed region. The following terms
prefer steps that close existing connections while opening fewer new ones.
The remaining entries make the order deterministic when the earlier criteria
are tied.

For every candidate processing order, the algorithm records the maximum number
of open arc ends encountered, together with the largest number remaining after
a step and the sum of these post-step counts. For a diagram with at least
$12$ crossings whose initial order reaches at least $12$ simultaneous open
ends, up to $16$ alternative deterministic starting steps are tested. An
alternative is retained only when it lowers the maximum number of simultaneous
open ends by at least two. The projection refinement described in
\SuppSubsecref{supp:projection_sampling} uses the same criterion: for a
difficult calculation, a projection containing slightly more crossings can be
preferred when its PD connectivity can be processed with substantially fewer
open ends at once.

After every crossing, graph vertex, helper connection and physical graph edge
has been processed, no open arc ends remain. At that point all complete
crossing-resolution and edge-subset contributions represented by
\Eqsref{eq:main_yamada_state_sum}{eq:main_yamada_crossing_free} have been
accumulated into the final state. Its Laurent-polynomial contribution is
therefore the raw Yamada polynomial $\Upsilon(D;Y)$. In this sense, the running
table
$\kappa\mapsto\mathcal W_{\kappa}(Y)$ is only an intermediate organization of
the terms in the exact state sum; once the boundary becomes empty, their sum is
the same raw polynomial that would have been obtained by explicit enumeration.

The calculation first uses checked $64$-bit integer coefficients. If any
coefficient exceeds that range, the same exact calculation is repeated with
arbitrary-size integers, without approximation or truncation. Only after the
raw polynomial has been completed is the normalization in
\Eqref{eq:main_yamada_normalization} applied to obtain
$\overline{\Upsilon}(D;Y)$. Thus the complete computational sequence is
\begin{equation}
\begin{aligned}
\text{PD code}
&\longrightarrow
\text{local Yamada and graph-polynomial updates}
\longrightarrow
\{\kappa\mapsto\mathcal W_{\kappa}(Y)\}
\\
&\longrightarrow
\Upsilon(D;Y)
\longrightarrow
\overline{\Upsilon}(D;Y),
\end{aligned}
\end{equation}
with the state combination changing only how the exact terms are organized and
summed, not the Yamada relations being evaluated.

\subsection{Using Yamada values to test family formulas}
\label{supp:yamada_discovery_bridge}

The importance of this scaling extends beyond faster evaluation of individual
knotted graphs. The published formulas above exemplify the traditional
direction in which family-specific mathematics is first used to
derive a recurrence or closed form, after which large family members become
inexpensive to evaluate
\cite{kg_dobrynin1996yamada,kg_li2018yamada,kg_lundstrom2022transfer}. A
general exact algorithm allows the order of investigation to be reversed: a
family-specific recurrence need not be known before topological data are generated.
The scaling result in
\Figref{fig:knottedgraph_topoly_scaling} therefore serves two roles. It
validates the general exact algorithm against independently known mathematics
at a scale far beyond explicit enumeration of the full resolution tree, and it provides
the data regime used for the family-scale investigations developed in
\SuppNoteref{supp:transfer_derivation}.

\FloatBarrier

\section{Independent validation of knotted-graph construction, projection and Yamada evaluation}
\label{supp:validation}

 Knotted-graph construction is tested against known volumetric seeds; projection and PD construction are tested through multi-view consistency and planar references; and exact algebra is tested against published Yamada identities and closed-form formulas \cite{kg_yamada1989,kg_dobrynin1996yamada,kg_li2018yamada} (\Figpanelref{fig:functionality_overview}{a--c}). Matching each stage to an independent target prevents an error in one interface from being hidden by agreement elsewhere.

\subsection{Algebraic and projection checks}
\label{supp:sanity_validation}

The Yamada implementation is tested at four complementary levels, separating
graph-polynomial identities, published closed-form families
\cite{kg_dobrynin1996yamada,kg_li2018yamada}, crossing-dependent relations and
three-dimensional projection.

\begin{enumerate}[leftmargin=*,label=\textbf{(\arabic*)}]

\item \textbf{Graph-level algebra.}
The algorithm reproduces the tested tree family
$T_1,\ldots,T_6$, cycles $C_1,\ldots,C_7$, bouquets
$B_1,\ldots,B_6$ and theta graphs $\Theta_1,\ldots,\Theta_8$, together with
the isthmus and one-point-union identities, using the Yamada and
edge-replacement relations in
Refs.~\cite{kg_yamada1989,kg_li2018yamada}.
A planar $K_4$ value and additional cubic-graph values are checked against
Ref.~\cite{kg_dobrynin1996yamada}. The Negami
calculation is independently compared with Negami's defining edge-subset
polynomial \cite{kg_negami1987}, and its Yamada specialization is compared
with the direct graph calculation.

\item \textbf{Published closed-form families.}
The three structured families in
\Figref{fig:knottedgraph_topoly_scaling} provide independent algebraic
references because their Yamada polynomials were derived analytically. For the
$\Theta(n)$ family in
\Figpanelref{fig:knottedgraph_topoly_scaling}{a,d,g}, we use Theorem~2 of
Ref.~\cite{kg_dobrynin1996yamada},
\begin{equation}
\begin{aligned}
\Upsilon(\Theta(n);Y)
={}&
\left(
Y^2+Y+1+Y^{-1}+Y^{-2}
\right)Y^n
-
\left(
Y+Y^{-1}
\right)Y^{-2n}
-
\left(
Y^2+1+Y^{-2}
\right)(-1)^nY^{-n}.
\end{aligned}
\label{suppeq:dv_theta_oracle}
\end{equation}

For the $\infty_{+}$ edge-replaced families in
Ref.~\cite{kg_li2018yamada}, defining $\sigma(Y)=Y+1+Y^{-1}$,
Theorem~5.1, Corollary~5.2 and Example~5.4 of that work give the expression
used for \Figpanelref{fig:knottedgraph_topoly_scaling}{b,e,h},
\begin{equation}
\Upsilon\!\left(C_n(\infty_{+});Y\right)
=
\left[-Y^{-2}\sigma(Y)\right]^n
+
\sigma(Y)\left(Y^{-2}+1\right)^n.
\label{suppeq:li_cycle_oracle}
\end{equation}
For the corresponding edge-replaced theta family in
\Figpanelref{fig:knottedgraph_topoly_scaling}{c,f,i},
\begin{equation}
\Upsilon\!\left(\Theta_s(\infty_{+});Y\right)
=
\frac{
[-\sigma(Y)]^s
+
\sigma(Y)
\left[
(\sigma(Y)+1)Y^{-2}+1
\right]^s
}{
1+\sigma(Y)
}.
\label{suppeq:li_theta_oracle}
\end{equation}
Every displayed \texttt{KnottedGraph} benchmark point in
\Figref{fig:knottedgraph_topoly_scaling} agrees exactly with its corresponding
literature expression, including the largest tested structured diagrams with
$500$ crossings. This establishes the evaluated scale for these structured families. It does
not imply that every diagram with the same crossing number has comparable
cost.

Topoly \cite{kg_topoly2021} is used as an external software comparator
only where both implementations pass the same published reference.
The external software comparison is reported in
\SuppSubsecref{supp:benchmark_protocol}.

\item \textbf{Crossings and mirroring.}
Crossing-dependent tests use a planar theta graph, a crossed theta diagram and
its mirror. The expected one-crossing factor $(-Y)^{\pm1}$ is checked against
the behavior used by Peddada \emph{et al.} \cite{kg_peddada2023}.

\item \textbf{Projection invariance.}
Multiple regular projections are generated for each fixed
subcubic three-dimensional spatial graph $G$. Each accepted view is processed
independently through crossing detection, edge over/undercrossing determination and PD
construction. All accepted projections of the same subcubic spatial graph must
give the same normalized invariant,
$\overline{\Upsilon}(G;Y)$, even when their crossing numbers and raw diagram
polynomials differ. Planar-reference cases are additionally compared with the
independently evaluated crossing-free graph polynomial. The tested catalogue
includes theta, $K_4$, triangular-prism, $K_{3,3}$ and cube-type examples.

\end{enumerate}

These checks separate errors in the algebraic calculation from errors in the
three-dimensional-to-diagram conversion.

\subsection{Handlebody--knotted-graph construction tests}
\label{supp:handlebody_validation}

\begin{figure}[!t]
    \centering
    \includegraphics[width=\linewidth,keepaspectratio]
    {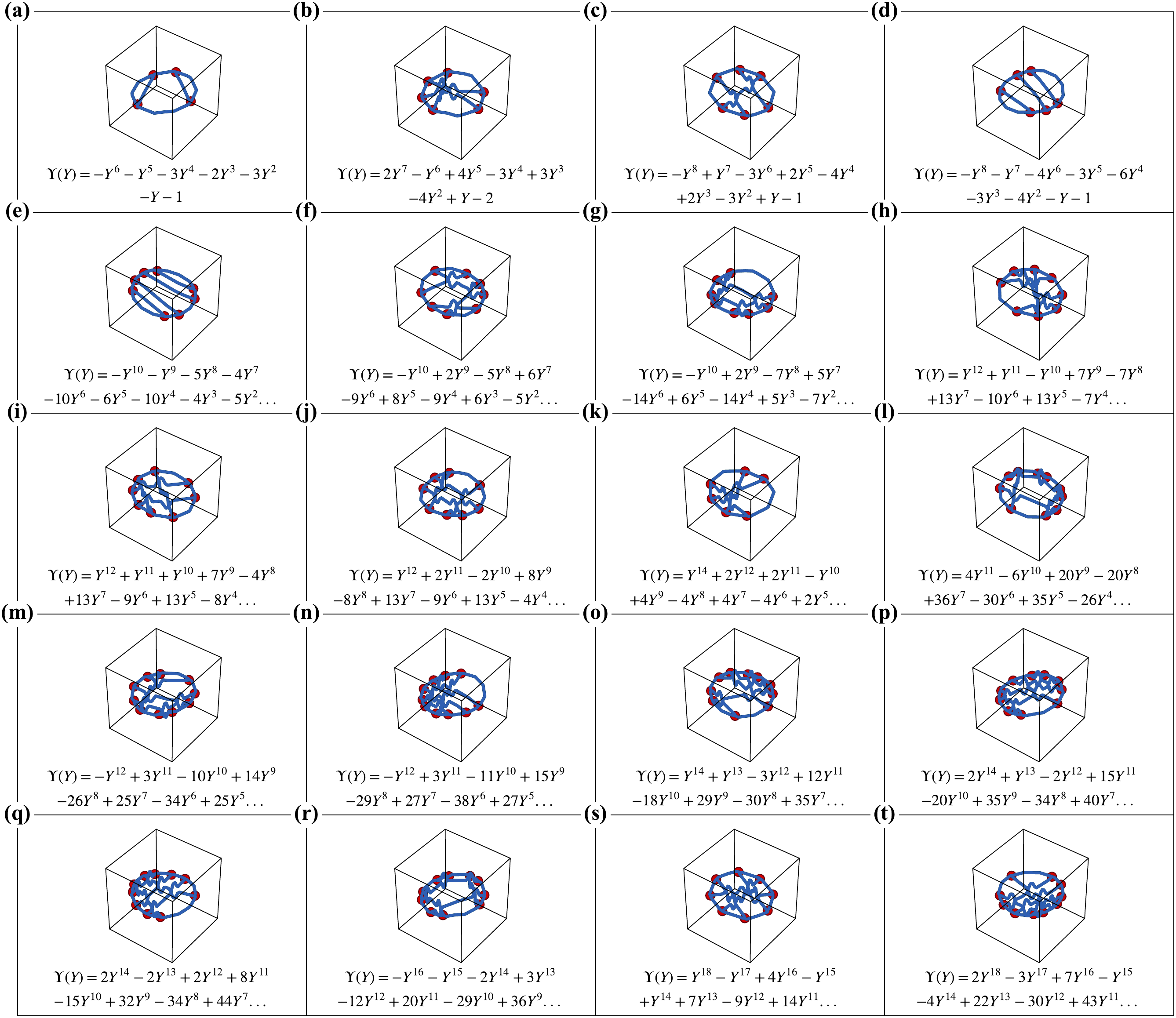}
    \caption{
    \textbf{Knotted-graph construction from volumetric inputs is tested against known generating graphs.}
    \textbf{(a--t)} Representative outputs from the recovery
    chain in \Eqref{suppeq:ground_truth_recovery}, with the normalized Yamada
    invariant $\overline{\Upsilon}$ evaluated on each resulting graph.
    The benchmark uses a $200^3$ voxel grid, tube radii of at least four voxels,
    and thickness and safety factors of $0.90$.
    Crossing-rich cases constitute approximately $60\%$ of the random ensemble
    and contain at least two verified projected crossings, with centerline
    separation exceeding $2.4$ tube radii at each prescribed crossing.
    The input acceptance uses the digital homology condition in
    \Eqref{suppeq:handlebody_betti}.
    }
    \label{suppfig:handlebody_stress}
\end{figure}

Here we test whether a knotted graph can be constructed from a volumetric
representation. We use an inverse benchmark in which a known knotted graph is
thickened and voxelized before the graph is recovered,
\begin{equation}
G_i
\longrightarrow
\Omega_i=N(G_i)
\longrightarrow
\Omega_i^{\mathrm{voxel}}
\longrightarrow
G_i^{\mathrm{rec}}.
\label{suppeq:ground_truth_recovery}
\end{equation}
Connected-component labeling and the lattice Euler characteristic are computed
with SciPy and \texttt{scikit-image}
\cite{kg_virtanen2020scipy,kg_vanderwalt2014skimage}, using the digital
Euler-number convention of Ohser, Nagel and Schladitz
\cite{kg_ohser2002euler}. \SuppFigref{suppfig:handlebody_stress} shows
representative outputs from the $4{,}400$-case program, ranging
from simple loop arrangements to crossing-rich embeddings. The polynomial
beneath each graph is a spatial-embedding comparison quantity, not a visual
measure of construction quality. All $4{,}400$ cases pass the stated input
and graph-level recovery checks, and all $4{,}400$ seed--output
normalized Yamada comparisons agree. The benchmark procedure is as follows:

\begin{tcolorbox}[
  breakable,enhanced,sharp corners,colback=white,colframe=black,
  boxrule=0.4pt,width=0.92\linewidth,center,
  left=4pt,right=4pt,top=4pt,bottom=4pt,
  before skip=8pt,after skip=8pt]
\textbf{Inverse handlebody-to-graph construction benchmark}

\begin{enumerate}[leftmargin=*,label=\textbf{\arabic*.},itemsep=3pt,topsep=4pt]

\item \textbf{Generate a known knotted graph.}
Construct a connected, bridge-free and subcubic knotted graph $G_i$. At every prescribed crossing, the
three-dimensional centerlines must satisfy $d_{\min}>2.4\,r_{\mathrm{tube}},$
and the package's standard $xy$ projection must resolve the prescribed
edge over/undercrossing information.
\item \textbf{Thicken and voxelize the graph.}
First estimate the minimum non-contact separation $d_{\mathrm{clear}}$ between
distinct parts of the knotted graph. Edge curves are resampled densely for
this estimate, with neighborhoods of shared vertices excluded when comparing
incident edges and the tube radius $r_i$ is chosen. The regular neighbourhood is defined as
\[
\Omega_i
=
N_{r_i}(G_i)
=
\left\{
\mathbf{x}\in\mathbb{R}^3:
\operatorname{dist}(\mathbf{x},G_i)\le r_i
\right\}.
\]
A cubic $200^3$ grid is placed around the complete embedding with an additional
margin of $2.25r_i$. For every piecewise-linear edge segment
$[\mathbf a,\mathbf b]$, the closest point to a voxel center $\mathbf x$ is
computed as
\[
\mathbf c
=
\mathbf a+
\operatorname{clip}_{[0,1]}
\left[
\frac{(\mathbf x-\mathbf a)\cdot(\mathbf b-\mathbf a)}
{\|\mathbf b-\mathbf a\|^2}
\right]
(\mathbf b-\mathbf a).
\]
The voxel is assigned to $\Omega_i^{\mathrm{voxel}}$ whenever
$\|\mathbf x-\mathbf c\|\le r_i
$ for at least one edge segment. Thus the binary input is constructed directly
from the geometric $r_i$-neighbourhood of the known knotted graph; no
pre-existing surface or mesh is used.

\item \textbf{Check the voxelized input.}
The voxelized object is accepted only when
\begin{equation}
\bigl(
\beta_0(\Omega_i^{\mathrm{voxel}}),
\beta_1(\Omega_i^{\mathrm{voxel}}),
\beta_2(\Omega_i^{\mathrm{voxel}})
\bigr)
=
\bigl(1,\beta_1(G_i),0\bigr),
\label{suppeq:handlebody_betti}
\end{equation}
and its tube radius is at least four voxels. This checks the measured component, cycle and cavity counts against the seed.
It is a necessary consistency test, not a proof that the voxel object is a
manifold or an ambient-isotopic regular neighbourhood.

\item \textbf{Construct the knotted graph.}
Apply the \texttt{KnottedGraph} pipeline to
$\Omega_i^{\mathrm{voxel}}$ to obtain $G_i^{\mathrm{rec}}$.

\item \textbf{Check recovered graph connectivity.}
The graph-level recovery check passes when
\[
G_i^{\mathrm{rec}}
\ \text{is connected and subcubic},
\qquad
\beta_1(G_i^{\mathrm{rec}})
=
\beta_1(G_i).
\]
Abstract isomorphism between $G_i$ and $G_i^{\mathrm{rec}}$ is recorded
separately as an additional diagnostic.

\item \textbf{Compare normalized Yamada invariants.}
Compute the normalized Yamada invariant independently for the original and
resulting knotted graphs,
\[
\overline{\Upsilon}_{\mathrm{seed}}
=
\overline{\Upsilon}(G_i;Y),
\qquad
\overline{\Upsilon}_{\mathrm{rec}}
=
\overline{\Upsilon}(G_i^{\mathrm{rec}};Y).
\]
Whenever both evaluations succeed, the polynomial comparison passes only if $\overline{\Upsilon}_{\mathrm{seed}}
=
\overline{\Upsilon}_{\mathrm{rec}}.$

\item \textbf{Assign the benchmark result.}
A case passes the recovery checks when the voxelized input
passes the digital-homology and thickness conditions and the resulting
graph passes the graph-level conditions above. Yamada preservation is recorded
as a separate spatial-embedding check whenever both polynomial evaluations
are available. A failed applicable condition is retained as a failed check, and
an unavailable polynomial remains a separate outcome. Equal polynomials are
not a complete embedding-equivalence certificate.

\end{enumerate}
\end{tcolorbox}

\begin{figure*}[!t]
    \centering
    \includegraphics[
        width=\linewidth,
        keepaspectratio
    ]{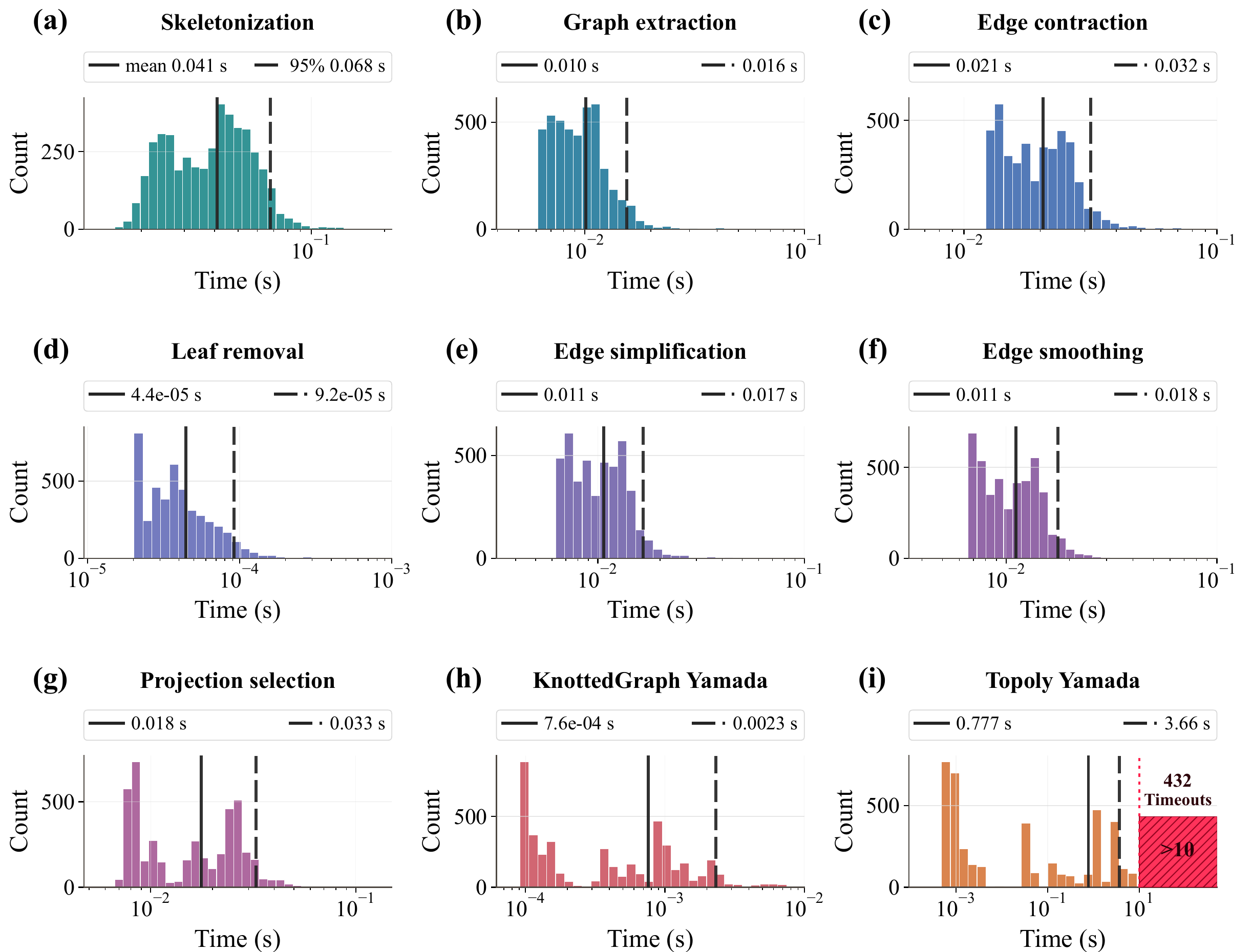}
    \caption{
    \textbf{Runtime distributions for the \texttt{KnottedGraph} stages of the handlebody-to-graph construction benchmark.}
    Histograms show per-case execution times for the $4{,}400$ curated cases in
    the inverse benchmark of \Eqref{suppeq:ground_truth_recovery}. The plotted
    stages correspond to the
    graph construction, cleanup, projection-selection and invariant-evaluation
    operations described in \SuppSubsecref{supp:extraction},
    \SuppNoteref{supp:projection} and
    \SuppNoteref{supp:yamada_engine}.
    \textbf{(a)} Skeletonization of the filled voxel volume to a one-voxel-wide
    backbone, corresponding to \SuppFigpanelref{suppfig:skeletonization_steps}{c};
    \textbf{(b)} multiscale extraction of the raw knotted graph from the
    voxel skeleton, corresponding to
    \SuppFigpanelref{suppfig:skeletonization_steps}{d};
    \textbf{(c)} short-edge contraction, corresponding to
    \SuppFigpanelref{suppfig:skeletonization_steps}{g};
    \textbf{(d)} leaf removal, corresponding to
    \SuppFigpanelref{suppfig:skeletonization_steps}{e};
    \textbf{(e)} degree-two edge simplification, corresponding to
    \SuppFigpanelref{suppfig:skeletonization_steps}{f};
    \textbf{(f)} geometric edge smoothing by Ramer--Douglas--Peucker
    simplification, corresponding to
    \SuppFigpanelref{suppfig:skeletonization_steps}{h};
    \textbf{(g)} selection of a regular projection and associated diagram
    for the recovered knotted graph;
    \textbf{(h)} normalized Yamada-invariant evaluation on the recovered graph
    using the exact \texttt{KnottedGraph} calculation described in
    \SuppNoteref{supp:yamada_engine}; and
    \textbf{(i)} the corresponding Yamada calculation using \texttt{Topoly} on
    the same selected PD codes. Completed Topoly evaluations are shown in the
    histogram; timeout and other non-completed cases are excluded from the
    completed-time statistics and reported separately. The horizontal axis gives
    time in seconds and the vertical axis gives the number of benchmark cases.
    Solid vertical lines mark the mean runtime for each stage, and dashed
    vertical lines mark the 95th percentile.
    }
    \label{suppfig:handlebody_timing_distributions}
\end{figure*}

The Topoly distribution is right-censored by the configured runtime limit:
completed evaluations contribute to the histogram and its summary statistics,
whereas non-completed cases are retained separately by status. The
\texttt{KnottedGraph} projection and Yamada timings are available for all
$4{,}400$ cases. The marginal software distributions are therefore not
interpreted as a paired speedup measurement. Pointwise software ratios are reported separately in
\SuppSubsecref{supp:benchmark_protocol} using the structured published
families in \Figref{fig:knottedgraph_topoly_scaling}.
\FloatBarrier

\subsection{Timing protocol}
\label{supp:benchmark_protocol}

Two timing protocols are used for the results reported in
\Figref{fig:knottedgraph_topoly_scaling},
\Tabref{tab:main_yamada_benchmark} and
\SuppFigref{suppfig:handlebody_timing_distributions}. The structured-family
benchmark isolates invariant evaluation on fixed PD inputs, whereas the
handlebody benchmark reports stage-wise runtimes across the full
$4{,}400$-case construction workload.

The distributions in
\SuppFigref{suppfig:handlebody_timing_distributions} characterize the
computational cost of the successive stages of the handlebody-to-graph
pipeline. Panels \textbf{(a--f)} report skeletonization, multiscale graph
extraction and the graph-cleanup operations; panels \textbf{(g,h)} report
projection selection and exact \texttt{KnottedGraph} Yamada evaluation; and
panel \textbf{(i)} reports the corresponding Topoly calculation on the same
selected PD codes. The displayed mean runtimes obey the approximate ordering
\begin{equation}
\begin{aligned}
T_{\mathrm{Topoly},\overline{\Upsilon}}
&\gg
T_{\mathrm{skeletonization}}
>
T_{\mathrm{projection}}
>
T_{\mathrm{contraction}}
>
T_{\mathrm{simplification}}
\\
&\approx
T_{\mathrm{smoothing}}
>
T_{\mathrm{extraction}}
>
T_{\mathrm{KG},\overline{\Upsilon}}
\gg
T_{\mathrm{leaf}} .
\end{aligned}
\end{equation}
The observed ordering follows the size and complexity of the representation handled at each stage. Skeletonization acts on the full voxel volume, while extraction and cleanup operate on the much smaller graph representation. Projection remains comparatively costly because it requires geometric crossing detection and validity checks across candidate views. The exact \texttt{KnottedGraph} Yamada calculation is smaller in this benchmark because equivalent partial resolution states are merged during evaluation, and leaf removal is the simplest graph-level operation. This ordering is specific to the tested benchmark.

\section{Parameter sweeps of volumetric geometries}
\label{supp:knot_fields}

The computational efficiency above supports repeated application of the same
construction and invariant interfaces across a parameter space. For a family of volumetric regions
$\Omega_{\boldsymbol{\theta}}$, we use the common notation
\begin{equation}
\boldsymbol{\theta}
\longmapsto
\Omega_{\boldsymbol{\theta}}
\longmapsto
G_{\boldsymbol{\theta}}
\longmapsto
\mathcal O(G_{\boldsymbol{\theta}}),
\label{suppeq:generic_phase_map}
\end{equation}
where $G_{\boldsymbol{\theta}}$ is the resulting knotted graph and
$\mathcal O$ records the quantity used for that analysis: an admissible
spatial-graph invariant, a fixed-diagram polynomial, an abstract graph summary or
an unavailable result. A region with one recorded label need not be a region
of equivalent analytic solids. The displays below illustrate the
parameter-scan interface; their grouping and filtering are specified before
any physical interpretation. The construction, projection and Yamada stages are those
established in
\SuppSubsecsref{supp:handlebody_basis}{supp:extraction},
\SuppNoteref{supp:projection} and \SuppNoteref{supp:yamada_engine}; here we
report the parameter-dependent outputs obtained by repeated application of
those stages.

\begin{figure}[!t]
    \centering
    \includegraphics[width=\linewidth]
    {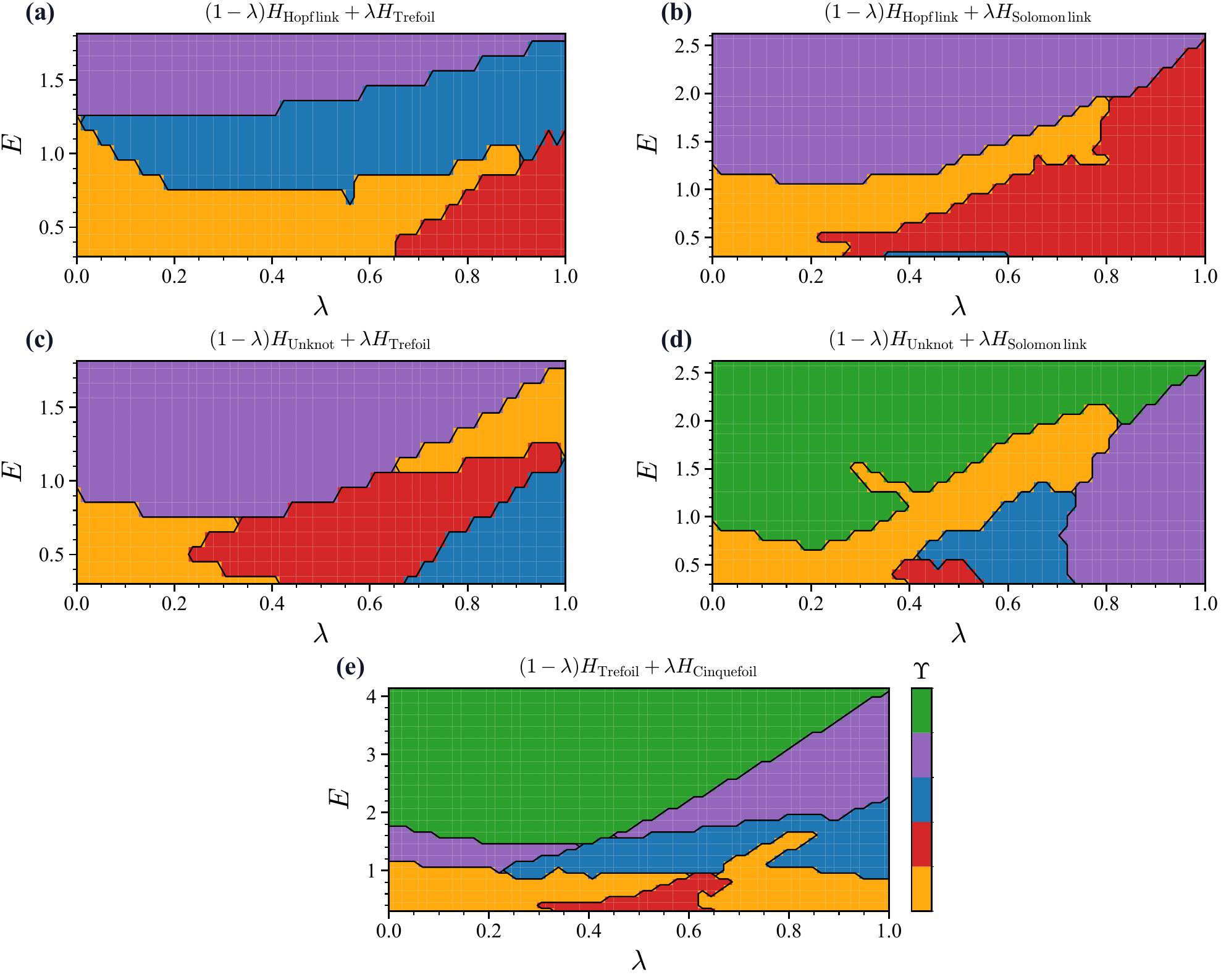}
    \caption{
\textbf{Hamiltonian-derived parameter scans across five interpolating families.}
The displayed colors are postprocessed output labels and can mix spatial
evaluations with abstract fallback values. Black curves separate displayed
labels; they are not established Lifshitz boundaries.
\textbf{(a)} Hopf link $\rightarrow$ trefoil;
\textbf{(b)} Hopf link $\rightarrow$ Solomon link;
\textbf{(c)} unknot $\rightarrow$ trefoil;
\textbf{(d)} unknot $\rightarrow$ Solomon link;
\textbf{(e)} trefoil $\rightarrow$ cinquefoil.
Each panel samples the interpolation and energy axes over the range used for
that family. The finite-domain and postprocessing conventions are specified in
\SuppSubsecref{supp:hamiltonian_phase_maps}.
}
    \label{suppfig:additional_phase_diagrams}
\end{figure}
\subsection{Volumetric regions in momentum space in condensed matter contexts}
\label{supp:hamiltonian_phase_maps}

A physical example is provided by finite-energy Fermi volumes, defined in momentum space, which is Fourier conjugate to the real-space lattice in condensed matter. Lattice couplings and periodic driving can generate intricate momentum-space nodal links \cite{li2018realistic}, while higher-dimensional circuit constructions provide access to nodal-boundary Seifert surfaces \cite{li2019emergence}.

Changes
in Fermi-surface topology define Lifshitz transitions
\cite{kg_lifshitz1960,kg_varlamov2021lifshitz}, and experimentally accessible
controls such as strain \cite{kg_sunko2019lifshitz}, pressure
\cite{kg_xiang2015lifshitz} can continuously deform a material Hamiltonian and move these transition
boundaries. Composition-driven topological band evolution
\cite{kg_dziawa2012topological} provides another example of a tunable
Hamiltonian, without identifying that band transition with a particular
Fermi-surface Lifshitz event.
The finite-energy Fermi-volume knotted-graph construction is introduced in
\cite{akgun_yan_2026topologicalclassificationknottedgraphs}. Here we use
the same geometry-to-graph interface to organize finite-window calculations
across the corresponding two-parameter scans.

The analytic benchmark models combine the complex-variable nodal-knot
construction of Bi \emph{et al.} with the Bloch lattice map used by Lee \emph{et al.}
\cite{kg_bi2017nodalknot,kg_lee2020nodalknots}:
\begin{equation}
\begin{aligned}
z(\mathbf k)
&=
\cos(2k_z)+c_0
+i\!\left(\cos k_x+\cos k_y+\cos k_z-m\right),\\
w(\mathbf k)
&=
\sin k_x+i\sin k_y,\\
f_{p,q}(\mathbf k)
&=
z(\mathbf k)^p-w(\mathbf k)^q ,
\end{aligned}
\label{suppeq:nodal_pq_field}
\end{equation}
with
\begin{equation}
H_{p,q}(\mathbf k)
=
\operatorname{Re}f_{p,q}(\mathbf k)\,\sigma_x
+
\operatorname{Im}f_{p,q}(\mathbf k)\,\sigma_z .
\label{suppeq:nodal_pq_hamiltonian}
\end{equation}
The recovered endpoint constructors use $m=2$ and $c_0=0.5$, except for the
Solomon endpoint, which uses the literal value $c_0=0.333$ (not exactly
$1/3$). A path to that endpoint therefore also changes the endpoint map
parameter. The unknot, Hopf link, trefoil, Solomon link and
cinquefoil correspond to $(p,q)=(1,1)$, $(2,2)$, $(2,3)$, $(2,4)$ and
$(2,5)$, respectively. The corresponding interpolations are Hopf link
$\rightarrow$ trefoil in
\SuppFigpanelref{suppfig:additional_phase_diagrams}{a}, Hopf link
$\rightarrow$ Solomon link in
\SuppFigpanelref{suppfig:additional_phase_diagrams}{b}, unknot
$\rightarrow$ trefoil in
\SuppFigpanelref{suppfig:additional_phase_diagrams}{c}, unknot
$\rightarrow$ Solomon link in
\SuppFigpanelref{suppfig:additional_phase_diagrams}{d}, and trefoil
$\rightarrow$ cinquefoil in
\SuppFigpanelref{suppfig:additional_phase_diagrams}{e}.

Since the positive-energy band is
\begin{equation}
\varepsilon_{p,q}(\mathbf k)=|f_{p,q}(\mathbf k)|,
\end{equation}
the filled Fermi region up to energy $E$, restricted to the sampled
momentum-space domain specified below, is
\begin{equation}
\Omega_{\lambda,E}
=
\left\{
\mathbf k:
\varepsilon_{\lambda}(\mathbf k)\leq E
\right\},
\label{suppeq:fermi_volume_phase_map}
\end{equation}
where $\varepsilon_{\lambda}$ is the positive-energy dispersion of
\begin{equation}
H_{\lambda}(\mathbf k)
=
(1-\lambda)H_0(\mathbf k)
+
\lambda H_1(\mathbf k),
\qquad
0\leq\lambda\leq1 .
\label{suppeq:phase_map_interpolation}
\end{equation}

The calculation uses a finite sampling of
$B_{\mathbf k}=[-\pi,\pi]\times[-\pi,\pi]\times[0,\pi]$. Opposite faces are not
periodically identified. The presence of both $\cos k_z$ and $\cos2k_z$
means that $[0,\pi]$ is not a periodic fundamental interval for the full
field. Boundary-touching inputs are consequently finite-window objects.

For the two real components $d_1,d_3$ of $H_\lambda$, the computational
constructor also uses the auxiliary non-Hermitian matrix
$\widetilde H_{\lambda,E}=H_\lambda+iE\sigma_y$. Its eigenvalues are
$\pm\sqrt{d_1^2+d_3^2-E^2}$. Thus its ideal purely imaginary-spectrum region
has the geometric inequality $d_1^2+d_3^2\leq E^2$ used in
\Eqref{suppeq:fermi_volume_phase_map}, although the auxiliary Hamiltonian
itself is not Hermitian.

The five scans sample a common interpolation grid in $\lambda$ together with
family-dependent energy ranges used in the corresponding panels of
\SuppFigref{suppfig:additional_phase_diagrams}. The scan can substitute an
abstract graph polynomial when a spatial evaluation is unavailable. The
displayed labels are postprocessed for clearer view.

A verified Lifshitz boundary would provide a natural point of comparison
with density-of-states, transport, thermodynamic and superconducting
responses \cite{kg_galeski2022lifshitz,kg_ichinokura2022lifshitz,
kg_slizovskiy2015lifshitz,kg_shi2017lifshitz}. The same construction extends directly to material-derived multiband
Hamiltonians. A microscopic Hamiltonian parameter can play the role of
$\lambda$, and $E$ defines a volumetric region from a selected
band-energy or interband-gap condition; constructing $G_{\lambda,E}$ across
this plane then organizes its parameter-dependent graph outputs. For material
scans, the program loops over $\lambda$ first because changing the energy or
gap threshold does not change the Hamiltonian at that fixed $\lambda$. The
Hamiltonian is therefore diagonalized once for each sampled $\lambda$, and the
resulting band energies are reused for every energy or gap threshold on the
second scan axis. Only the spectrum for the current $\lambda$ is kept in
memory, and the records are returned in the public output layout. This avoids repeating the same diagonalization many times without
storing spectra for all sampled $\lambda$ values. Their interpretation as
source-topology labels requires the corresponding graph construction and
equivalence checks.  
\FloatBarrier

\section{Geometric preconditioning of difficult embeddings}
\label{supp:repulsive_layout}
\raggedbottom

Even after a spatial graph has been constructed, its geometry can make a
useful planar diagram difficult to obtain. A folded edge may take long
detours, pass close to other edges or contain many small bends, all of which
can obscure the crossing structure in projection. Geometric preconditioning
addresses this problem by moving the graph's curves in three dimensions
before projection (\Figpanelref{fig:functionality_overview}{f}). It optimizes
the coordinates along the edges while retaining the graph's vertices and
incidence relations. The resulting curves provide a geometric representation
for visualization and topological analysis.

The nonlocal repulsion is based on the tangent-point energy introduced by
Buck and Orloff \cite{kg_buck1995energy} and the computational approach of
Yu, Schumacher and Crane \cite{kg_yu2021repulsive}. Using Schumacher's
\texttt{Repulsor} library \cite{kg_schumacher_repulsor},
\texttt{KnottedGraph} converts graph edges to sampled curves, constrains the
graph vertices, checks the geometric updates and maps the relaxed curves
back to the graph for projection and invariant evaluation.

\subsection{Protein-derived theta graphs}
\label{supp:repulsive_inputs}

\SuppFigref{suppfig:repulsive_curves_pipeline} illustrates the workflow with
three protein-derived $\theta$ graphs, labelled by their source structures
1AOC, 3ULK and 5OSQ in the RCSB Protein Data Bank \cite{kg_burley2019rcsb}.
Their initial geometries are shown in
\SuppFigpanelref{suppfig:repulsive_curves_pipeline}{a},
\SuppFigpanelref{suppfig:repulsive_curves_pipeline}{g} and
\SuppFigpanelref{suppfig:repulsive_curves_pipeline}{m}, respectively.
A $\theta$ graph has two vertices joined by three internally disjoint paths.
The red, blue and green curves identify those three paths throughout each
row. Each path can combine backbone portions with selected bridges, so the
colours track path identity across the processing stages. These constructions
select a $\theta$ graph from each protein's geometry and connections; the
complete molecular and crosslink networks contain additional structure.

The protein examples supplied with the library use chain A of the first
structural model. Backbone portions follow the ordered C$\alpha$ coordinates.
Selected bridges connect these portions directly or through a metal-ion
coordinate. For example, the 3ULK construction joins residues D217 and E393
by three routes: the intervening backbone, a bridge through Mg498, and the
remaining selected backbone portions with an added closure. The 5OSQ
construction similarly joins D437 and C469 using a backbone route, a route
through the selected C376--C469 disulfide bridge, and a longer route through
Ca503, Ca504 and a closure. The 1AOC construction combines selected backbone
portions and cysteine bridges into three paths, with a closure in one path.
An artificial closure joins selected backbone endpoints to complete one of
the paths. This modelling connection is specified separately from the
structure-derived bridges. Its coordinates are part of the input embedding,
and the subsequent analysis is conditional on that closure choice.

Each path is represented by a polyline. The two graph vertices are shared
endpoints of all three polylines, and intermediate samples discretize each
path. Their assignment to the original graph edge is retained throughout
relaxation. Subdivision that retains every original corner leaves the path
unchanged. Resampling that replaces corners changes the geometry, including
when it increases the sample count, and requires validation at that stage.
The checks on subsequent solver steps begin from the resulting sampled
embedding. Sample removal by checked shortcuts is described below.

In these $\theta$ graphs, the branch vertices have degree three and the
intermediate samples have degree two. The selected graphs are thus subcubic,
and their normalized Yamada polynomials are ambient-isotopy invariants
\cite{kg_yamada1989,kg_mellor2018spatial}.

\subsection{Unfolding, compaction and smoothing}
\label{supp:repulsive_stages}

\begin{figure}[!t]
    \centering
    \includegraphics[width=\linewidth]{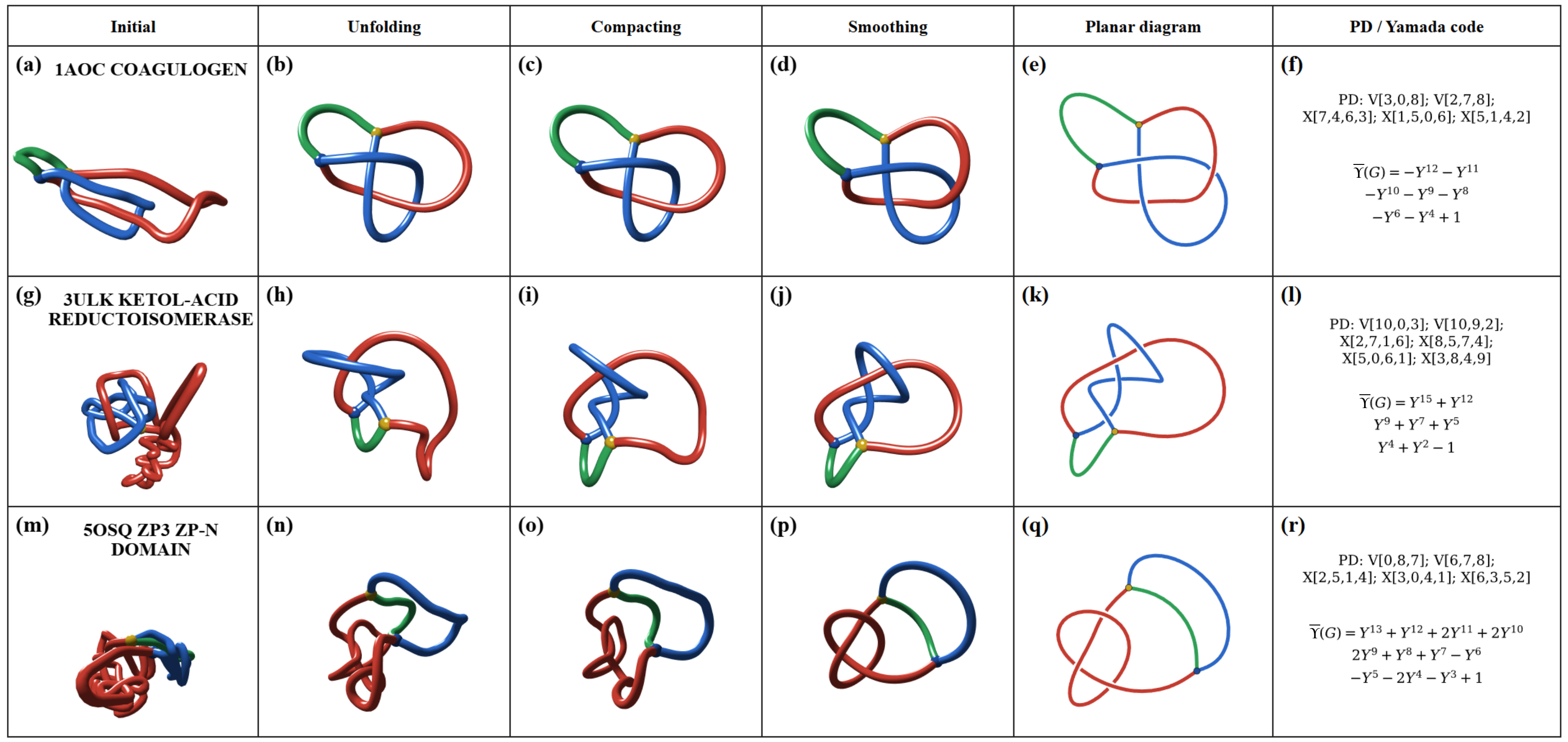}
    \caption{
    \textbf{From a crowded protein-derived embedding to a readable graph diagram.}
    The rows show selected $\theta$ graphs derived from 1AOC
    (\textbf{a--f}), 3ULK (\textbf{g--l}) and 5OSQ (\textbf{m--r}).
    Red, blue and green track the three paths between the two branch vertices;
    a path may contain both backbone portions and selected bridges.
    The initial geometries (\textbf{a}, \textbf{g}, \textbf{m}) are followed
    by unfolding (\textbf{b}, \textbf{h}, \textbf{n}), compaction
    (\textbf{c}, \textbf{i}, \textbf{o}) and smoothing
    (\textbf{d}, \textbf{j}, \textbf{p}). These operations open close
    approaches, shorten detours and regularize the curves, respectively,
    while retaining the three paths and their common endpoints. The planar diagrams
    (\textbf{e}, \textbf{k}, \textbf{q}) retain depth information at crossings,
    and the last column (\textbf{f}, \textbf{l}, \textbf{r}) gives the
    corresponding PD codes and Yamada polynomials.
    }
    \label{suppfig:repulsive_curves_pipeline}
\end{figure}

Each row of \SuppFigref{suppfig:repulsive_curves_pipeline} follows the selected
paths from their protein-derived geometry through three relaxation stages
and then to a planar diagram. The stages emphasize separation, length control
and local regularity, respectively.

\paragraph{Unfolding}
Nonlocal repulsion opens crowded regions by moving nearby strands apart,
giving the curves room to move before substantial shortening is attempted.
The unfolded 1AOC, 3ULK and 5OSQ geometries are shown in
\SuppFigpanelref{suppfig:repulsive_curves_pipeline}{b},
\SuppFigpanelref{suppfig:repulsive_curves_pipeline}{h} and
\SuppFigpanelref{suppfig:repulsive_curves_pipeline}{n}, respectively.
Here, \emph{unfolding} describes a geometric layout operation rather than a
simulation of molecular unfolding. The motion is subject to the
strand-crossing checks specified in \SuppNoteref{supp:repulsive_topology}.

\paragraph{Compaction}
Length control draws in long detours present in the initial geometry or
introduced by the opening stage. Repulsion remains active during shortening
to maintain strand separation. Compaction shortens the geometric paths while
retaining every graph edge and its endpoints. The corresponding compacted
geometries are shown in
\SuppFigpanelref{suppfig:repulsive_curves_pipeline}{c},
\SuppFigpanelref{suppfig:repulsive_curves_pipeline}{i} and
\SuppFigpanelref{suppfig:repulsive_curves_pipeline}{o}.

\paragraph{Smoothing}
A local bending penalty reduces residual polyline oscillations. A separate
clearance penalty can encourage enough separation for curves drawn as tubes
to remain visually distinct. The smoothed geometries are shown in
\SuppFigpanelref{suppfig:repulsive_curves_pipeline}{d},
\SuppFigpanelref{suppfig:repulsive_curves_pipeline}{j} and
\SuppFigpanelref{suppfig:repulsive_curves_pipeline}{p}.
The smoothed curves are then projected, their
over--under crossings are recorded, and the resulting PD code is evaluated.
The corresponding planar diagrams are shown in
\SuppFigpanelref{suppfig:repulsive_curves_pipeline}{e},
\SuppFigpanelref{suppfig:repulsive_curves_pipeline}{k} and
\SuppFigpanelref{suppfig:repulsive_curves_pipeline}{q}, with their PD codes
and Yamada polynomials in
\SuppFigpanelref{suppfig:repulsive_curves_pipeline}{f},
\SuppFigpanelref{suppfig:repulsive_curves_pipeline}{l} and
\SuppFigpanelref{suppfig:repulsive_curves_pipeline}{r}.
The objective acts on the three-dimensional curves; projection selection is
performed after relaxation.

\clearpage

\subsection{Tangent-point-energy relaxation of knotted graphs}
\label{supp:repulsive_objective}

Let $\mathcal X=\{\mathbf x_i\}\subset\mathbb R^3$ denote the sampled curve
vertices and $\mathcal S$ their line segments. A segment
$s=[\mathbf x_i,\mathbf x_j]$ has length
$\ell_s=\|\mathbf x_j-\mathbf x_i\|$, midpoint
$\mathbf c_s=(\mathbf x_i+\mathbf x_j)/2$, unit tangent
$\boldsymbol\tau_s=(\mathbf x_j-\mathbf x_i)/\ell_s$, and normal projector
$\Pi_s^\perp=I-\boldsymbol\tau_s\boldsymbol\tau_s^T$.
The projector extracts the part of a displacement perpendicular to the
segment, making the interaction sensitive to both segment separation and
tangent direction.

The symmetric midpoint interaction is
\begin{equation}
E_{\mathrm{mid}}^{p,q}(\mathcal X)
=
\sum_{\{s,t\}\subset\mathcal S,\ s\ne t}
\ell_s\ell_t
\frac{\|\Pi_s^\perp(\mathbf c_t-\mathbf c_s)\|^q+
\|\Pi_t^\perp(\mathbf c_s-\mathbf c_t)\|^q}
{\|\mathbf c_t-\mathbf c_s\|^p}.
\label{suppeq:tangent_point_energy}
\end{equation}
Each unordered pair of distinct segments is counted once; the two numerator
terms account for the two tangent directions. Self-pairs are omitted. In the
Repulsor discretization, neighbouring segments that share a vertex also
contribute to the midpoint interaction. Collision and tube-clearance checks
instead use pairs with no shared vertex, allowing the intentional contacts
at graph junctions.

\SuppEqref{suppeq:tangent_point_energy} describes the basic midpoint
kernel. The native calculation uses \texttt{Repulsor}'s
\texttt{TangentPointEnergy0}, with hierarchical far-field evaluation and
adaptive treatment of close, non-neighbouring segments
\cite{kg_schumacher_repulsor}. This evaluates the interactions through a
spatial hierarchy. The midpoint kernel differs from the endpoint-based
trapezoidal quadrature of Yu, Schumacher and Crane \cite{kg_yu2021repulsive}.
The default exponents are $p=8$, $q=4$.

Length control uses the sum of squared segment lengths,
\begin{equation}
E_{\mathrm{len}}=\sum_{s\in\mathcal S}\ell_s^2,
\end{equation}
which encourages shorter segments. Unlike the total arc length
$\sum_s\ell_s$, this quantity changes when a straight segment is subdivided.
Local smoothing uses the second-difference penalty
\begin{equation}
E_{\mathrm{bend}}
=
\sum_{i\in\mathcal I_2}
\|\mathbf x_{a(i)}-2\mathbf x_i+\mathbf x_{b(i)}\|^2,
\end{equation}
where $\mathcal I_2$ contains only degree-two samples and $a(i),b(i)$ are
their two neighbours. Restricting the sum to these samples gives each term a
unique pair of neighbours and excludes branch vertices and endpoints. Both
penalties depend on the sampling density and coordinate scale, which are
recorded together with their weights.

For tubes of display radius $r$ and a desired gap $g$, the optional clearance
term is
\begin{equation}
E_{\mathrm{tube}}
=\sum_{\{s,t\}\in\mathcal N}
\bigl[\max\{0,2r+g-d(s,t)\}\bigr]^2,
\end{equation}
where $\mathcal N$ consists of unordered segment pairs with no shared vertex
and $d(s,t)$ is their minimum Euclidean distance. This soft penalty encourages
a centreline separation of $2r+g$; crossing prevention is handled by the
motion checks. The objective combines the
backend tangent-point energy and these penalties with non-negative weights,
$E=w_{\mathrm{tp}}E_{\mathrm{tp}}+
w_{\mathrm{len}}E_{\mathrm{len}}+
w_{\mathrm{bend}}E_{\mathrm{bend}}+
w_{\mathrm{tube}}E_{\mathrm{tube}}$.
The weights set the emphasis of each stage. The same formulation also
supports a single relaxation with fixed weights. Because crossing number
is absent from the objective, a minimum-crossing projection is not guaranteed.

\subsection{Relaxation and topology checks}
\label{supp:repulsive_topology}

Edges can pass through one another during an update even when its initial
and final configurations are free of self-intersections. The solver checks
the intervening motion of each proposed update. Starting from a valid
embedded polyline network, it computes a descent direction using the
tangent-point metric and sets an initial step to a fraction of
\texttt{Repulsor}'s maximum-safe-step estimate. A trial is accepted when both
the motion check and an Armijo energy-decrease test pass; otherwise the step
is halved and tested again.

During a trial step, every sample moves linearly from its old to its proposed
position. For each pair of segments with no shared vertex, the additional
motion check first tests whether their swept bounding boxes are separated.
For the remaining pairs it finds candidate contact times from the cubic
coplanarity condition of the four moving endpoints and tests segment
distances at those times, as well as at the step endpoints and midpoint.
A contact within the specified tolerance rejects the step. Degenerate
coplanarity cases not excluded by the bounding-box test are rejected
conservatively. The resulting verification is numerical, with its contact
tolerance defined in the input coordinate units.

All original graph vertices are pinned by default;
interior curve samples can move. Optional pinned \emph{collar} samples are
the first few interior samples on every incident edge. Fixing a vertex
preserves its position, whereas also fixing a collar preserves a local
piece of each incident edge. Pairs sharing a sample are omitted from the
additional motion check because their common endpoint is intentional.
For higher-valence graphs, the local junction geometry therefore requires
separate constraints and validation. Invariant comparisons in that setting
also require a common rigid-vertex structure.

Per-step motion checking is enabled by default. When intermediate states are
saved, an optional independent verifier replays the successive swept motions;
a failed replay raises an error. Static clearance measurements at the
initial, relaxed and final-simplified states report the geometric separation
at each stage. They complement the motion checks by showing how strand
separation changes during the relaxation.

Simplification is checked separately from relaxation. Replacing the two
segments $[a,b]$ and $[b,c]$ by $[a,c]$ sweeps the triangle with vertices
$a,b,c$. The decimator accepts this shortcut only when its conservative
clearance tests against segments sharing none of $a,b,c$ pass and its pinning
constraints are respected. Configurations at shared vertices require the
local junction constraints described above. Related intersection-tested
polygonal reductions have also been used in protein knot-polynomial workflows
\cite{kg_comoglio2011homfly}.

\subsection{Mapping relaxed curves back to the knotted graph}
\label{supp:repulsive_return}

After relaxation and any accepted simplification, the retained sample-to-edge
mapping transfers the coordinates to a copy of the original graph.
Vertex labels, edge keys and non-geometric attributes are preserved; positions
and edge polylines are updated. The result enters the usual projection
search, crossing detection, PD-code construction and Yamada evaluation, with
each projected arc still associated with its original graph edge.

The retained mapping establishes the correspondence of vertices and edges
before and after relaxation. Swept-motion and shortcut checks assess the
geometric changes along that correspondence. For subcubic graphs, agreement
of the normalized Yamada values supplies a further consistency check;
polynomial equality alone is insufficient to establish isotopy because
distinct embeddings can share an invariant. Failed or resource-limited
evaluations are recorded as unresolved. The output is a relaxed geometric
representative in the same graph data structure, ready for the standard
projection and invariant pipeline.

\subsection{Solver settings and reproducibility}
\label{supp:repulsive_settings}

\SuppTabref{supptab:repulsive_defaults} lists the default configuration of the
native solver. Repulsion is active by default; compaction, bending and tube
clearance are enabled by setting their weights. The solver performs at most
100 iterations and terminates earlier when no acceptable descent step can
be found.

\begin{table}[H]
\centering
\small
\caption{\textbf{Default native relaxation settings.}
Distances and geometric tolerances refer to the supplied coordinate units.
The inner-solve tolerance controls the iterative calculation of the descent
direction.}
\label{supptab:repulsive_defaults}
\begin{tabularx}{\linewidth}{@{}>{\raggedright\arraybackslash}p{0.28\linewidth}>{\raggedright\arraybackslash}p{0.23\linewidth}>{\raggedright\arraybackslash}X@{}}
\toprule
Control & Default & Meaning \\
\midrule
Tangent-point exponents & $p=8$, $q=4$ & Distance and normal-displacement powers. \\
Objective weights & $w_{\mathrm{tp}}=1$; other penalties $=0$ & Compaction, bending and tube clearance are opt-in. \\
Outer iterations & At most 100 & Maximum number of attempted descent iterations. \\
Safe-step fraction & $0.95$ & Initial fraction of the backend's safe-step estimate. \\
Line search & 12 backtracking reductions; Armijo coefficient $10^{-4}$ & A rejected trial is halved. \\
Metric solve & At most 60 iterations; tolerance $10^{-4}$ & Controls the iterative descent-direction solve. \\
Swept-motion check & Enabled; tolerance $10^{-7}$ & Checks segment pairs with no shared sample. \\
Graph constraints & Original vertices pinned; collar depth 0 & Local incident-edge directions are not pinned by default. \\
Independent replay & Off & Requires both saved states and explicit verification. \\
\bottomrule
\end{tabularx}
\end{table}

Reproducing a relaxation requires the input graph, closure coordinates,
sample counts, coordinate scale, stage weights, pinning choices and
simplification settings. The run records include the graph--curve mapping,
solver parameters, accepted-step history, static clearance summaries, total
curve length and bounding-box scale. Intermediate states and the independent
replay result are saved when requested. The Repulsor source revision is
specified in the software reference \cite{kg_schumacher_repulsor}. Together,
the input, settings and geometric records allow the relaxed curves and their
projected diagrams to be traced back to the original graph.

\FloatBarrier
\flushbottom
\section{Inferring and testing Yamada family formulas from Yamada data}
\label{supp:transfer_derivation}

The preceding sections establish scalable exact Yamada evaluation across
knotted-graph families. This enables a second use of the calculation: exact
invariant values can serve as data from which family-level mathematical
structure is inferred. Instead of evaluating a family only after its
analytical law is known, we consider the inverse problem of reconstructing
candidate formulas from finitely many evaluated family members.

Existing Yamada-polynomial family formulas have been derived from
family-specific mathematical structure. An explicit formula for the
$\Theta(n)$ family is given in Ref.~\cite{kg_dobrynin1996yamada}, while
formulas for spatial graphs obtained by replacing the edges of cycle, theta
and bouquet graphs by spatial parts are derived in
Ref.~\cite{kg_li2018yamada}. These relations have also been used in the study
of Yamada-polynomial roots \cite{kg_li2019density}. Transfer-matrix
constructions for repeated layers of symmetric spatial graphs are developed in
Ref.~\cite{kg_lundstrom2022transfer}. Further analytical relations connect the
Yamada polynomial to knot and link invariants, including the Jones polynomial
of associated links of $\theta$-curves \cite{kg_huh2024theta} and relations
among Yamada, Jaeger and Jones polynomials for spatial $K_4$-graphs and their
subgraphs \cite{kg_vesnin2025complete}. These results begin from a specified
family construction, recurrence, transfer structure or invariant relation and
derive the corresponding polynomial law. They do not provide a general
procedure for starting from exact Yamada values when that family-specific
algebra is not known in advance.

Recent AI-assisted mathematics provides an evaluation-guided setting for this
inverse direction. FunSearch combines LLM-generated proposals with systematic
evaluation to obtain new mathematical constructions
\cite{kg_funsearch2024}, while AlphaEvolve uses an evaluation-guided
evolutionary loop for scientific and algorithmic discovery
\cite{kg_alphaevolve2025}. Related work reports cases in which finite
computational data are generalized to candidate formulas across mathematical
problem classes \cite{kg_georgiev2025mathematical}, while AlphaProof
illustrates the complementary role of machine-verifiable reasoning
\cite{kg_alphaproof2026}.

Here we apply this data-to-structure strategy to topological-invariant
data. \texttt{KnottedGraph} first generates exact Yamada polynomials for
selected members of a parameterized knotted-graph family. An LLM then uses
only a designated discovery subset to propose candidate recurrences, closed
forms or transfer structures. A retained proposal is converted to an explicit
symbolic formula or operator representation before additional family members
are evaluated and used as held-out tests. The methodological novelty is
therefore the direction of inference: candidate family algebra is reconstructed
from exact Yamada-polynomial data without prescribing the family-specific
recurrence, transfer representation or closed form in advance.

\subsection{LLM-assisted conjecture generation and held-out verification}
\label{supp:llm_discovery_protocol}

The protocol is shown in
\Figpanelref{fig:three_theta_yamada_families}{a}. A discovery set $D$ provides
invariant values for the supplied family definitions. LLM-assisted exploration
is restricted to proposing candidate recurrences, transfer structures or closed
forms. For the calculations reported here, OpenAI's GPT-5.6 Sol was used for this candidate-generation stage. A selected proposal is then converted to an explicit symbolic
expression or operator representation. Only after this form is fixed are the
additional evaluations used for testing generated. A proposed pattern is
therefore not counted as its own verification.

For the calculations reported here, the retained family constructors,
candidate expressions, explicit word lists and evaluations define the
discovery, transfer-reconstruction and verification inputs. The mixed-family
plan contains $459$ words. The pure-braid evaluation set contains $255$
distinct short words, including the Hankel input and words of lengths up to
nine, together with $20$ completed extrapolation cases of lengths $101$, $125$,
$150$ and $200$.

The computational record keeps
\[
\text{conjecture generation}
\quad\vert\quad
\text{symbolic construction}
\quad\vert\quad
\text{additional tests}
\]
separate. In particular, the words used to construct the Hankel realization
are not relabeled as holdouts; the length-$101$, $125$, $150$ and $200$ cases probe extrapolation
beyond the short-word reconstruction data. We next retain the proposed
structures and work through their algebraic consequences.

\subsection{Commuting transfer laws for homogeneous and mixed theta-derived families}
\label{supp:transfer_commuting}
\label{supp:transfer_common_notation}

The three homogeneous laws in
\Figpanelref{fig:three_theta_yamada_families}{d},
\Figpanelref{fig:three_theta_yamada_families}{g} and
\Figpanelref{fig:three_theta_yamada_families}{j}, together with the mixed law
in \Figpanelref{fig:three_theta_yamada_families}{k}, share the same
reconstructed three-channel representation. We use the factors
$a(Y),b(Y),c(Y),d(Y)$ and $\phi(Y)$ defined in
\Eqref{eq:main_family_factors}, abbreviating them as $a,b,c,d,\phi$ when no
ambiguity is possible. The common boundary vectors are
\begin{equation}
\bm\omega
=
\begin{pmatrix}
a\\
-c\\
-Y^2\phi
\end{pmatrix},
\qquad
\bm 1=
\begin{pmatrix}
1\\
1\\
1
\end{pmatrix}.
\label{supp:deriv:eq:omega}
\end{equation}

The three local cell types are denoted
$\Lambda,\downarrow,\updownarrow$, with reconstructed transfer matrices
\begin{equation}
T_{\Lambda}=\operatorname{diag}\!\left(a^2,Y^2b^2,Y^8\right),\qquad
T_{\downarrow}=\operatorname{diag}\!\left(a,Y^2b,-Y^7\right),\qquad
T_{\updownarrow}=\operatorname{diag}\!\left(ab,-Y^2(Y-1)d,Y^5(Y+1)b\right).
\label{supp:deriv:eq:transfers}
\end{equation}

\Eqsref{supp:deriv:eq:omega}{supp:deriv:eq:transfers} specify the proposed
local transfers and boundary vectors. Their products give the expressions
below by exact algebra. The normalized
homogeneous and mixed formulas in this subsection concern nonempty
constructions, $m\geq1$.

\paragraph{Cross-linked family $\Lambda_m$}
\label{supp:transfer_crosslinked}

\Figpanelref{fig:three_theta_yamada_families}{b} and
\Figpanelref{fig:three_theta_yamada_families}{c} show the cross-linked family
at $m=1$ and $m=2$, respectively, and
\Figpanelref{fig:three_theta_yamada_families}{d} displays the general-$m$
construction and closed form. A member with parameter $m$ contains $m$
identical $\Lambda$ cells; its candidate expression is
\begin{equation}
\begin{aligned}
\overline{\Upsilon}(\Lambda_m;Y)
&=\bm\omega^T T_\Lambda^m\bm 1,
\qquad
T_\Lambda^m=\operatorname{diag}\!\left(a^{2m},Y^{2m}b^{2m},Y^{8m}\right),\\
&=a\,a^{2m}-c\,Y^{2m}b^{2m}-Y^2\phi\,Y^{8m}\\
&=\boxed{a^{2m+1}-Y^{2m}c\,b^{2m}-Y^{8m+2}\phi}.
\end{aligned}
\label{supp:deriv:eq:lambda_family}
\end{equation}
The three terms of the closed form are the three transfer channels propagated
through $m$ repeated cells.

\paragraph{Lower-paired family $\Lambda_m^{\downarrow}$}
\label{supp:transfer_lower_paired}

\Figpanelref{fig:three_theta_yamada_families}{e} and
\Figpanelref{fig:three_theta_yamada_families}{f} show the lower-paired family
at $m=1$ and $m=2$, and
\Figpanelref{fig:three_theta_yamada_families}{g} displays its general-$m$ law.
For this family,
\begin{equation}
\begin{aligned}
\overline{\Upsilon}(\Lambda_m^{\downarrow};Y)
&=\bm\omega^T T_{\downarrow}^m\bm 1,
\qquad
T_{\downarrow}^m
=\operatorname{diag}\!\left(a^m,Y^{2m}b^m,(-1)^mY^{7m}\right),\\
&=a\,a^m-c\,Y^{2m}b^m-Y^2\phi\,(-1)^mY^{7m}\\
&=\boxed{a^{m+1}-Y^{2m}c\,b^m+(-1)^{m+1}Y^{7m+2}\phi}.
\end{aligned}
\label{supp:deriv:eq:down_family}
\end{equation}
The alternating sign therefore arises entirely from the third eigenvalue
$-Y^7$ of the lower-paired cell.

\paragraph{Two-sided family $\Lambda_m^{\updownarrow}$}
\label{supp:transfer_two_sided}

\Figpanelref{fig:three_theta_yamada_families}{h} and
\Figpanelref{fig:three_theta_yamada_families}{i} show the two-sided paired
family at $m=1$ and $m=2$, and
\Figpanelref{fig:three_theta_yamada_families}{j} displays its general-$m$ law.
For this family,
\begin{equation}
\begin{aligned}
\overline{\Upsilon}(\Lambda_m^{\updownarrow};Y)
&=\bm\omega^T T_{\updownarrow}^m\bm 1,\\
T_{\updownarrow}^m
&=\operatorname{diag}\!\left(
a^mb^m,\,
(-1)^mY^{2m}(Y-1)^m d^m,\,
Y^{5m}(Y+1)^m b^m
\right),\\
\overline{\Upsilon}(\Lambda_m^{\updownarrow};Y)
&=a^{m+1}b^m
-c(-1)^mY^{2m}(Y-1)^m d^m
-Y^{5m+2}\phi(Y+1)^m b^m\\
&=\boxed{
a^{m+1}b^m
+(-1)^{m+1}Y^{2m}c(Y-1)^m d^m
-Y^{5m+2}\phi(Y+1)^m b^m
}.
\end{aligned}
\label{supp:deriv:eq:updown_family}
\end{equation}

These formulas arose from LLM-assisted exploration of data generated by
\texttt{KnottedGraph}. Their common three-dimensional transfer description
uses different diagonal operators for the three local motifs. The algebraic
expansions above make their shared structure explicit.

The zero-motif boundary needs separate normalization. At $m=0$, the unshifted
matrix contraction is
$\bm\omega^T\bm1=-Y^2(1+Y+2Y^2+Y^3+Y^4)$.
Its minimum degree is two, so \Eqref{eq:main_yamada_normalization} gives
$-(1+Y+2Y^2+Y^3+Y^4)$; the unshifted expression is not the normalized result. A zero-motif
construction is not automatically the empty graph.

\paragraph{Mixed family with commuting transfer operators $\mathcal A_{\mathbf n}$}
\label{supp:transfer_abelian_mixed}

The mixed construction and its count-only family law are shown in
\Figpanelref{fig:three_theta_yamada_families}{k}. This family formula provides
an expression for an arbitrary mixture of the motifs
$\Lambda,\downarrow,\updownarrow$. Using the ordered motif word $w$ and count
vector $\mathbf n$ defined in \Eqref{eq:main_abelian_word_counts}, let $G_w$
denote the knotted graph generated by that word. The reconstructed transfer law
and its commuting reduction are
\begin{equation}
\begin{aligned}
\overline{\Upsilon}(G_w;Y)
&=\bm\omega^T\!\left(\prod_{j=1}^{m}T_{w_j}\right)\!\bm 1,\quad
[T_\Lambda,T_\downarrow]
=[T_\Lambda,T_\updownarrow]
=[T_\downarrow,T_\updownarrow]=0,\quad
\prod_{j=1}^{m}T_{w_j}
=T_\Lambda^{n_\Lambda}
T_\downarrow^{n_\downarrow}
T_\updownarrow^{n_\updownarrow}.
\end{aligned}
\label{supp:deriv:eq:mixed_commuting_transfer}
\end{equation}
Since the proposed transfer matrices commute, their products depend only on the motif counts. The resulting three transfer channels contribute
\begin{equation}
\begin{aligned}
C_1
&=a\,(a^2)^{n_\Lambda}a^{n_\downarrow}(ab)^{n_\updownarrow}
=a^{m+n_\Lambda+1}b^{n_\updownarrow},\\
C_2
&=-c\,(Y^2b^2)^{n_\Lambda}(Y^2b)^{n_\downarrow}
[-Y^2(Y-1)d]^{n_\updownarrow}\\
&=(-1)^{n_\updownarrow+1}
Y^{2m}c\,b^{2n_\Lambda+n_\downarrow}
(Y-1)^{n_\updownarrow}d^{n_\updownarrow},\\
C_3
&=-Y^2\phi\,(Y^8)^{n_\Lambda}(-Y^7)^{n_\downarrow}
[Y^5(Y+1)b]^{n_\updownarrow}\\
&=(-1)^{n_\downarrow+1}
Y^{8n_\Lambda+7n_\downarrow+5n_\updownarrow+2}
\phi\,(Y+1)^{n_\updownarrow}b^{n_\updownarrow}.
\end{aligned}
\label{supp:deriv:eq:mixed_channels}
\end{equation}
Their sum gives the candidate mixed-family invariant:
\begin{equation}
\boxed{
\begin{aligned}
\overline{\Upsilon}(\mathcal A_{\mathbf n};Y)=C_1+C_2+C_3
={}&
a^{m+n_\Lambda+1}b^{n_\updownarrow}+
(-1)^{n_\updownarrow+1}
Y^{2m}c\,
b^{2n_\Lambda+n_\downarrow}
(Y-1)^{n_\updownarrow}
d^{n_\updownarrow}
\\
&+
(-1)^{n_\downarrow+1}
Y^{8n_\Lambda+7n_\downarrow+5n_\updownarrow+2}
\phi\,
(Y+1)^{n_\updownarrow}
b^{n_\updownarrow}.
\end{aligned}}
\label{supp:deriv:eq:abelian_final}
\end{equation}
Under this transfer identification, if two words $w$ and $w'$ have the same motif-count vector,
$\mathbf n(w)=\mathbf n(w')
\Longrightarrow
\overline{\Upsilon}(G_w;Y)
=
\overline{\Upsilon}(G_{w'};Y).$
This is the count-only reduction summarized in
\Eqref{eq:main_abelianization}. Direct evaluation agrees with this predicted formula on all $459$ words in
the completed mixed-family plan. The expression in
\Figpanelref{fig:three_theta_yamada_families}{k} is invariant under motif
reordering by construction. The finite comparisons support its proposed
identification with the graph family.

\subsection{Noncommuting transfer operators retain braid-word order}
\label{supp:transfer_nonabelian_hankel}

The fixed-graph pure-braid family in
\Figpanelref{fig:three_theta_yamada_families}{l} provides the order-sensitive
motif family. Its reconstructed transfer operators satisfy the
noncommutativity condition in \Eqref{eq:main_nonabelian_commutator}, so we
treat it as an order-sensitive family with a noncommuting transfer representation. The underlying $8$-vertex,
$12$-edge cubic multigraph is fixed, and three distinguished edges carry
an ordered word in the standard Artin braid generators
\cite{kg_birman1974braids},
\begin{equation}
w=w_1\cdots w_m,\qquad
w_j\in\{A,B\},\qquad
A=\sigma_1^2,\qquad
B=\sigma_2^2.
\label{supp:deriv:eq:pure_braid_word}
\end{equation}

\begin{figure}[!t]
 \centering
 \includegraphics[width=0.62\linewidth]
 {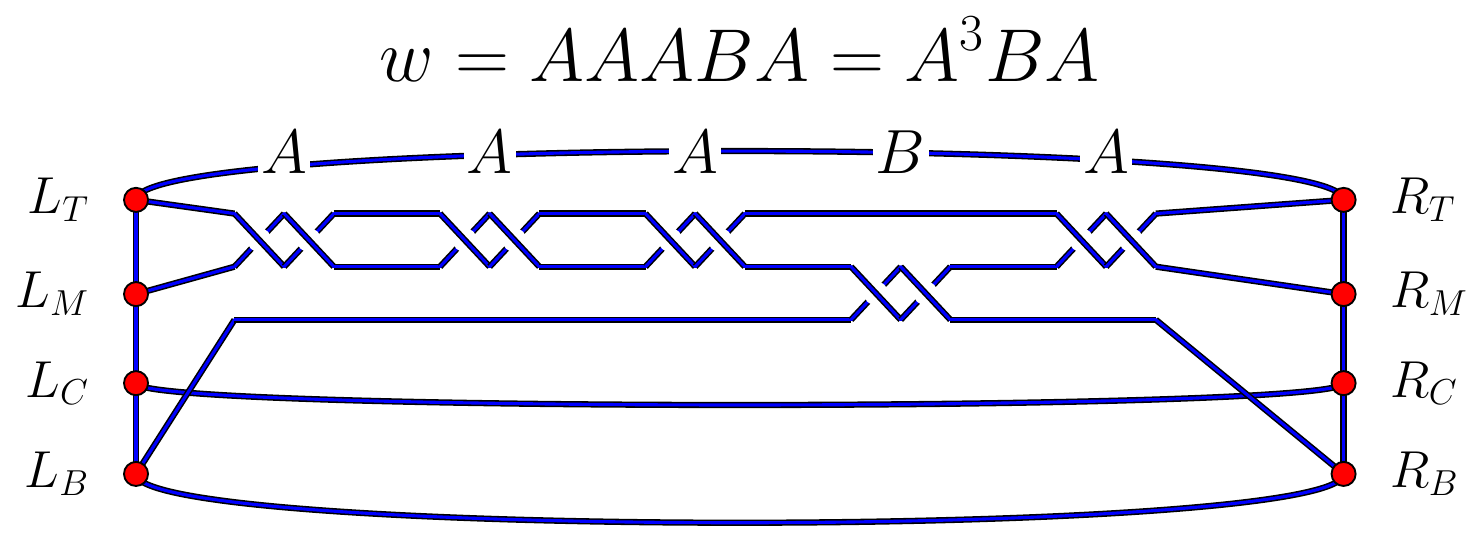}
 \caption{
 \textbf{A worked ordered braid word outside the reconstruction basis.}
 The same $8$-vertex, $12$-edge abstract multigraph used for the
 order-sensitive family is decorated here by
 $w=AAABA=A^3BA$, with $A=\sigma_1^2$ and $B=\sigma_2^2$ read from left to
 right. Each block is a pure three-braid, so the three strands return to the
 same endpoints. The word changes the spatial embedding in the highlighted
 three-strand region while preserving the surrounding cubic graph.
 The corresponding transfer product is $T_A^3T_BT_A$.
 Although this word is outside the coordinate basis, it occurs in the input
 block as $H_{AA,ABA}$ and is not an independent holdout.
 }
 \label{suppfig:nonabelian_worked_example}
\end{figure}

The geometry is shown for $w=AAABA$ in
\SuppFigref{suppfig:nonabelian_worked_example}. Here $\sigma_1$ exchanges
strands $1$ and $2$; $\sigma_2$ exchanges strands $2$ and $3$
\cite{kg_birman1974braids}. Squaring either generator restores the endpoint
permutation, so changing the
word changes the embedding of the braid region while preserving the abstract
cubic graph. The first order-sensitive example already occurs at length three. The words
$AAB$, $ABA$ and $BAA$ all have $(n_A,n_B)=(2,1)$, yet the relation in
\Eqref{eq:main_nonabelian_example} shows that the invariant retains information
beyond motif counts. 
\paragraph{A 15-dimensional Hankel realization}
\label{supp:transfer_hankel_intuition}

We encode the word dependence directly from Yamada-polynomial data. Choose
the $15$ word indices
\begin{equation}
\begin{aligned}
\mathcal B=\{&
\varepsilon,\,
A,B,\,
AA,AB,BA,BB,\,
AAB,ABA,ABB,BAA,BAB,
\\
&AABA,ABAA,ABAB
\}.
\end{aligned}
\label{supp:deriv:eq:hankel_basis}
\end{equation}
For left and right words $\xi,\eta\in\mathcal B$, define
\begin{equation}
H_{\xi,\eta}
=
\Upsilon(\mathcal N_{\xi\eta};Y).
\label{supp:deriv:eq:H_definition}
\end{equation}
Each entry is the Yamada polynomial obtained by concatenating the row word and
column word. For example,
\[
H_{A,B}=\Upsilon(\mathcal N_{AB};Y),
\qquad
H_{AA,ABA}=\Upsilon(\mathcal N_{AAABA};Y),
\]
and the second entry corresponds to the graph in
\SuppFigref{suppfig:nonabelian_worked_example}.

The symbolic $15\times15$ block in
\Eqref{supp:deriv:eq:H_definition} is nonsingular over $\mathbb Q(Y)$.
A nonzero exact determinant evaluation at $Y=2$ witnesses that its determinant
is not the zero rational function. Thus the finite block has rank fifteen:
\begin{equation}
\boxed{
\operatorname{rank}_{\mathbb Q(Y)}H=15.
}
\label{supp:deriv:eq:hankel_rank_15}
\end{equation}
This rank is consistent with the fifteen-element generating set for the
three-strand Yamada module described by Ref. \cite{kg_chbili2016}. To use it as a global upper bound for
this graph family, its local braid action and external closure must be shown
to factor through that space. Finite-rank realization theory then supplies
the connection to a minimal linear representation
\cite{kg_carlyle1971realizations,kg_balle2012spectral}.

\paragraph{Reconstructing the $A$ and $B$ actions}
\label{supp:transfer_hankel_reconstruction}

To determine how one additional braid block acts, define two shifted
matrices using the same row and column words:
\begin{equation}
(H_A)_{\xi,\eta}
=
\Upsilon(\mathcal N_{\xi A\eta};Y),
\qquad
(H_B)_{\xi,\eta}
=
\Upsilon(\mathcal N_{\xi B\eta};Y).
\label{supp:deriv:eq:hankel_matrices}
\end{equation}
For example,
\begin{equation}
H_{A,B}=\Upsilon(\mathcal N_{AB};Y),\qquad
(H_A)_{A,B}=\Upsilon(\mathcal N_{AAB};Y),\qquad
(H_B)_{A,B}=\Upsilon(\mathcal N_{ABB};Y).
\label{supp:deriv:eq:hankel_examples}
\end{equation}
$H$ joins the two words directly; $H_A$ and $H_B$ insert one additional
$A$ or $B$ between them. All entries are obtained from
\texttt{KnottedGraph}; the longest word required here has length nine.
Following the finite-rank realization construction
\cite{kg_carlyle1971realizations,kg_balle2012spectral}, the transfer matrices
are the unique operators satisfying
\begin{equation}
HT_A=H_A,\qquad
HT_B=H_B,
\qquad\Longrightarrow\qquad
\boxed{
T_A=H^{-1}H_A,\qquad
T_B=H^{-1}H_B.
}
\label{supp:deriv:eq:TA_TB}
\end{equation}
They encode the effect of adding an $A$ or $B$ block. Let $R$ select the
empty-word state and let $L^T$ be the empty-word row of $H$. With these boundary vectors, the proposed raw-word representation is
\begin{equation}
\boxed{
\Upsilon(\mathcal N_w;Y)
=
L^TT_{w_1}\cdots T_{w_m}R.
}
\label{supp:deriv:eq:raw_word_transfer}
\end{equation}

For the worked word $AAABA$, the matrices act from right to left:
\begin{equation}
\varepsilon
\xrightarrow{\;T_A\;}
A
\xrightarrow{\;T_B\;}
BA
\xrightarrow{\;T_A\;}
ABA
\xrightarrow{\;T_A\;}
AABA
\xrightarrow{\;T_A\;}
AAABA.
\label{supp:deriv:eq:AAABA_state_path}
\end{equation}
The first four nonempty words in this path,
$A,BA,ABA,AABA$, belong to $\mathcal B$. The final word $AAABA$ does not.
The last application of $T_A$ therefore leaves the set of coordinate basis
words and produces the $15$-state vector representing the continuation
behavior of $AAABA$. The final contraction with $L^T$ gives its Yamada
polynomial.

\paragraph{Three channels for repeated blocks}
\label{supp:transfer_cubic}

The reconstructed matrices satisfy the common cubic identity already stated
in \Eqref{eq:main_nonabelian_cubic}. Thus powers of either $T_A$ or $T_B$
contain only three spectral contributions. Define
\begin{equation}
P_X^{(\alpha)}
=
\prod_{\substack{\beta\in\{2,-2,-4\}\\\beta\neq\alpha}}
\frac{T_X-Y^\beta I}{Y^\alpha-Y^\beta},
\qquad
\alpha\in\{2,-2,-4\}.
\label{supp:deriv:eq:projector_general}
\end{equation}
With these projectors, the repeated-block power law is the relation already
given in \Eqref{eq:main_nonabelian_power}. Hence a repeated run $A^n$ uses the same three fixed projectors; only the
scalar powers depend on the run length. These expressions
are identities over $\mathbb Q(Y)$. At special values where the displayed
eigenvalues coincide, the projector fractions require a limit or a direct
matrix calculation before specialization.

\paragraph{Arbitrary ordered words}
\label{supp:transfer_ordered_words}

Write the word as maximal runs,
\begin{equation}
w=X_1^{n_1}\cdots X_r^{n_r},
\qquad
X_j\in\{A,B\},
\qquad
X_j\neq X_{j+1}.
\label{supp:deriv:eq:run_decomp}
\end{equation}
For example,
\[
AAABA=A^3B^1A^1,
\qquad
AAB=A^2B^1.
\]
Substituting \Eqref{eq:main_nonabelian_power} into
\Eqref{supp:deriv:eq:raw_word_transfer} gives
\begin{equation}
\boxed{
\Upsilon(\mathcal N_w;Y)
=
L^T
\prod_{j=1}^{r}
\left[
Y^{2n_j}P_{X_j}^{(2)}
+
Y^{-2n_j}P_{X_j}^{(-2)}
+
Y^{-4n_j}P_{X_j}^{(-4)}
\right]
R.
}
\label{supp:deriv:eq:raw_nonabelian_final}
\end{equation}
The run lengths enter through scalar powers, while the sequence of
$A$- and $B$-projectors remains ordered. This retained matrix order is what
allows the invariant to distinguish same-count words.

For a nonzero raw polynomial, the normalized invariant is obtained from the
general normalization rule in \Eqref{eq:main_yamada_normalization}, with
$\mu(w)=\min\deg_Y\Upsilon(\mathcal N_w;Y)$.

\paragraph{Worked prediction outside the reconstruction basis}
\label{supp:nonabelian_worked_example}

The graph in \SuppFigref{suppfig:nonabelian_worked_example} has
\[
w=AAABA=A^3BA,
\]
which is not one of the $15$ coordinate words in
$\mathcal B$. Substituting the run lengths
$(n_1,n_2,n_3)=(3,1,1)$ into
\Eqref{supp:deriv:eq:raw_nonabelian_final} gives
\begin{equation}
\begin{aligned}
\Upsilon(\mathcal N_{AAABA};Y)
={}&
L^T
\left[
Y^{6}P_A^{(2)}
+Y^{-6}P_A^{(-2)}
+Y^{-12}P_A^{(-4)}
\right]
\\
&\times
\left[
Y^{2}P_B^{(2)}
+Y^{-2}P_B^{(-2)}
+Y^{-4}P_B^{(-4)}
\right]
\\
&\times
\left[
Y^{2}P_A^{(2)}
+Y^{-2}P_A^{(-2)}
+Y^{-4}P_A^{(-4)}
\right]R
\\
={}&L^TT_A^3T_BT_AR.
\end{aligned}
\label{supp:deriv:eq:AAABA_projector_prediction}
\end{equation}

The first four transfer steps remain on coordinate states in
$\mathcal B$:
\[
R
\xrightarrow{\,T_A\,}e_A
\xrightarrow{\,T_B\,}e_{BA}
\xrightarrow{\,T_A\,}e_{ABA}
\xrightarrow{\,T_A\,}e_{AABA}.
\]
The final multiplication $T_Ae_{AABA}$ produces the state associated with
the out-of-basis word $AAABA$ as a linear combination of the $15$
reconstruction states.
Expanding \Eqref{eq:main_nonabelian_cubic} gives
\begin{equation}
T_A^3
=
\left(Y^2+Y^{-2}+Y^{-4}\right)T_A^2
-
\left(1+Y^{-2}+Y^{-6}\right)T_A
+
Y^{-4}I.
\label{supp:deriv:eq:TA3_reduction}
\end{equation}
Substitution into
\Eqref{supp:deriv:eq:AAABA_projector_prediction} yields
\begin{equation}
\begin{aligned}
\Upsilon(\mathcal N_{AAABA};Y)
={}&
\left(Y^2+Y^{-2}+Y^{-4}\right)
L^TT_A^2T_BT_AR
\\
&-
\left(1+Y^{-2}+Y^{-6}\right)
L^TT_AT_BT_AR
+
Y^{-4}L^TT_BT_AR.
\end{aligned}
\label{supp:deriv:eq:AAABA_reduced}
\end{equation}
The three remaining contractions correspond to words already contained in
$\mathcal B$:
\begin{equation}
\begin{aligned}
L^TT_A^2T_BT_AR
&=\Upsilon(\mathcal N_{AABA};Y),\\
L^TT_AT_BT_AR
&=\Upsilon(\mathcal N_{ABA};Y),\\
L^TT_BT_AR
&=\Upsilon(\mathcal N_{BA};Y).
\end{aligned}
\label{supp:deriv:eq:AAABA_basis_contractions}
\end{equation}
The polynomial of the out-of-basis word is obtained from three
basis-word polynomials:
\begin{equation}
\boxed{
\begin{aligned}
\Upsilon(\mathcal N_{AAABA};Y)
={}&
\left(Y^2+Y^{-2}+Y^{-4}\right)
\Upsilon(\mathcal N_{AABA};Y)
\\
&-
\left(1+Y^{-2}+Y^{-6}\right)
\Upsilon(\mathcal N_{ABA};Y)
+
Y^{-4}\Upsilon(\mathcal N_{BA};Y).
\end{aligned}}
\label{supp:deriv:eq:AAABA_prediction}
\end{equation}
This example shows explicitly how the finite transfer representation evaluates
a word that is not itself one of the $15$ coordinate states. Its value is
already present in the Hankel input as $H_{AA,ABA}$. The
length-$101$, $125$, $150$ and $200$ comparisons below probe the stronger extrapolation regime in
which the evaluated words are also absent from all short-word data used to
construct $H,H_A$ and $H_B$.

The matrices $H,H_A,H_B$ use only words of length at most nine. Exact checks
verify the finite-block rank, $HT_A=H_A$, $HT_B=H_B$, a nonzero commutator
and the cubic identities for both matrices. Direct evaluation agrees with
the transfer expression on $255$ distinct short words. Twenty additional
comparisons have lengths $101$, $125$, $150$ and $200$ and
agree coefficient-by-coefficient in both raw and normalized form.

\section{Software architecture and reproducibility}
\label{supp:software_architecture}

\begin{figure}[!t]
\centering
\includegraphics[width=\linewidth]{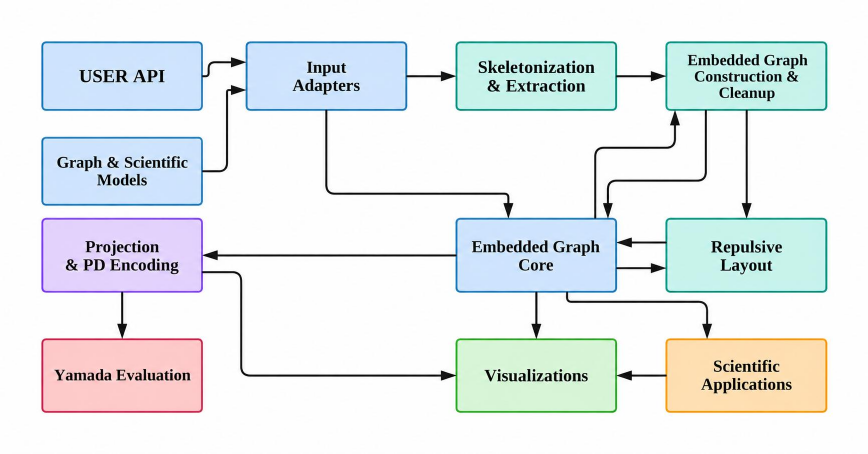}
\caption{
\textbf{Modular organization of \texttt{KnottedGraph}.}
Each box is a group of package functions, and each arrow shows the object passed from one box to the next.
Data enter at the upper left, either as user-supplied data (USER API) or as generated models (Graph \& Scientific Models), and are read by the Input Adapters. Inputs that already describe a curve or graph go directly to the Embedded Graph Core. Surfaces, volumes and fields first pass through Skeletonization \& Extraction and then Embedded Graph Construction \& Cleanup.
The resulting knotted graph in the core can be relaxed geometrically (Repulsive Layout). It can also be projected and encoded as a PD code (Projection \& PD Encoding) for Yamada Evaluation, drawn (Visualizations) and used in larger workflows (Scientific Applications). Users whose data already match the input of a later box can start at that box.
}
\label{suppfig:library_architecture}
\end{figure}

\SuppFigref{suppfig:library_architecture} shows how the source package is organized. Each box is a group of functions, and each arrow shows which object is passed from one group to the next.
Data enter at the upper left, are converted to a knotted graph along the top row, and are stored in the Embedded Graph Core at the centre of the figure. From this core object, the workflow branches to projection and Yamada evaluation on the left, geometric relaxation on the right, and visualization and scientific applications along the bottom.
The package modules follow the same division into \texttt{knotted\_graph.inputs}, \texttt{knotted\_graph.extraction}, \texttt{knotted\_graph.core}, \texttt{knotted\_graph.layout}, \texttt{knotted\_graph.projection}, \texttt{knotted\_graph.invariants}, \texttt{knotted\_graph.visualization} and \texttt{knotted\_graph.applications}. The modules are divided by the type of object they receive and return, not by scientific application, so the same code serves every example in this work. The boxes are described below in the order in which data pass through them.

\paragraph{Entry points: USER API and Graph \& Scientific Models}
A calculation can start in two ways. Through the USER API box, the user supplies their own data: a structure file, a coordinate array, a node--edge table or a surface mesh. Through the Graph \& Scientific Models box, the user instead asks the package to generate an input from a definition, such as a named knot or link, a braid word, a mathematical spatial graph or a Hamiltonian model (\SuppTabref{supptab:input_interfaces}). Both routes lead to the Input Adapters.

\paragraph{Input Adapters}
The Input Adapters (\texttt{knotted\_graph.inputs}) read each input and apply the checks and conversions listed in \SuppTabref{supptab:input_interfaces}. Examples include selecting a protein chain, ordering polymer atoms and checking node coordinates.
Where the output goes depends on what the input already contains. Inputs that already describe a curve or graph are passed directly to the Embedded Graph Core (downward arrow). These include backbones, polymer chains, coordinate curves and node--edge networks. Surfaces, volumes and fields do not yet contain a graph, so they are sent to Skeletonization \& Extraction (rightward arrow).

\paragraph{Skeletonization \& Extraction}
This box (\texttt{knotted\_graph.extraction}) receives a surface or volume and returns a first, uncleaned knotted graph. It fills the surface to a voxel volume and thins this volume to a one-voxel-wide skeleton. It then converts the skeleton to graph vertices and edges using the multiscale junction selection of \SuppSubsecref{supp:extraction} (\SuppFigpanelref{suppfig:skeletonization_steps}{a--d}).
For large skeletons, a compiled C++ implementation performs this step. It applies the same shortcut-removal and graph-selection rules as the Python implementation, so it changes the running time but not the selected graph.

\paragraph{Embedded Graph Construction \& Cleanup}
The raw graph can still contain artifacts of voxelization. Typical artifacts are short spurious branches, single junctions split into several nearby vertices, and jagged edges. This box removes them with the operations shown in \SuppFigpanelref{suppfig:skeletonization_steps}{e--h}: leaf removal, degree-two chain collapse, short-edge contraction and polyline smoothing.
Leaf removal and short-edge contraction are applied only when suitable for the target object, because they can delete real features, such as the terminal branches of a biological tree. The cleaned graph is stored in the Embedded Graph Core, or it can be passed directly to Repulsive Layout. The two opposite arrows between this box and the core indicate that the same cleanup operations can also be applied to a graph already stored in the core.

\paragraph{Embedded Graph Core}
The core (\texttt{knotted\_graph.core}) holds the central object of the package: a knotted graph compatible with \texttt{networkx.MultiGraph} \cite{kg_hagberg2008networkx}. Each vertex stores its three-dimensional position in the attribute \texttt{pos}. Each edge stores its ordered three-dimensional polyline in \texttt{pts}, and the two ends of this polyline coincide with the positions of the edge's end vertices. Parallel edges and self-loops are kept as separate edges, so closed curves and multigraphs are represented without loss.
The downstream projection, layout, visualization and application boxes can use this same graph. A graph built from any input can therefore be projected, relaxed, drawn or analysed without further input conversion; Yamada Evaluation receives the PD representation produced by projection.

\paragraph{Repulsive Layout}
This optional box (\texttt{knotted\_graph.layout}) changes only the coordinates of a graph, not its connectivity. It is useful when an embedding is valid but hard to project, for example when separate edges pass very close to each other. It relaxes the curves by tangent-point-energy repulsion (\SuppNoteref{supp:repulsive_layout}).
Graph vertices are kept fixed by default, and each coordinate update can be checked to confirm that no edge passes through another. The relaxed coordinates are written back to the same graph. That graph returns to the core (leftward arrow) and continues to projection like any other input.

\paragraph{Projection \& PD Encoding}
This box (\texttt{knotted\_graph.projection}) converts the three-dimensional graph into a planar diagram, the form required by the Yamada calculation (\SuppNoteref{supp:projection}). It samples viewing directions and rejects views in which a crossing cannot be resolved reliably. It records which edge passes over at each crossing and writes the diagram as a PD code (\SuppFigref{suppfig:pd_code_generation}).
The crossings and their positions along both edges are found in a single geometric search, so they do not have to be located again when the PD code is written. The output is a \texttt{ProjectionResult} object, which stores the selected view together with its diagram and PD code. This object is passed to Yamada Evaluation and can also be drawn by Visualizations.

\paragraph{Yamada Evaluation}
This box (\texttt{knotted\_graph.invariants}) receives a PD code and computes the exact Yamada polynomial, optionally in normalized form (\SuppNoteref{supp:yamada_engine}). A user who already has a diagram or PD code can start here directly, without any three-dimensional geometry (\SuppTabref{supptab:input_interfaces}).
Coefficients are first computed with $64$-bit integers. If any coefficient exceeds that range, the calculation is automatically repeated with arbitrary-size integers, so the result remains exact. The output is a \texttt{YamadaComputationResult}, which stores the polynomial together with the projection it was computed from. Each value can therefore be traced back to its diagram. Reproducing the value from the original data also requires the source data, the conversion choices and the software revision (see below).

\paragraph{Visualizations and Scientific Applications}
The two boxes at the bottom right use the objects produced above rather than creating new representations. The Visualizations box (\texttt{knotted\_graph.visualization}) draws graphs from the core, diagrams from projection and results from applications. Scientific Applications (\texttt{knotted\_graph.applications}) contains the workflows used in this work: material-surface utilities (\texttt{materials.py}); mathematical examples (\texttt{mathematical.py}); knot-deformation workflows (\texttt{knot\_deformation.py}); nodal models (\texttt{nodal/}); and parameter sweeps (\texttt{phase\_maps.py}).
These modules prepare scientific inputs and then call the same extraction, projection and invariant functions described above. For example, the Hamiltonian scans in \SuppNoteref{supp:knot_fields} and the family calculations in \SuppNoteref{supp:transfer_derivation} use the same graph-construction and Yamada code. Application code is kept outside the core, so adding a new application does not change the topology implementation.
The \texttt{phase\_map\_examples} package provides three commands. \texttt{inspect} and \texttt{plot} read and display saved scan records, whereas \texttt{scan} runs a new scan. Its quick scan profile is a small demonstration only; its output is not a convergence study and does not replace the research calculations reported here.

\paragraph{Installation and dependencies}
The base installation contains only what is needed to build a knotted graph, project it and evaluate its invariants. This includes NumPy for numerical arrays \cite{kg_harris2020numpy}, SymPy for symbolic expressions \cite{kg_meurer2017sympy}, and the required graph-processing and projection libraries. Workflow-specific functionality is available as optional extras: \texttt{knot-fields}, \texttt{nodal}, \texttt{surface}, \texttt{viz}, \texttt{repulsion}, \texttt{notebook} and \texttt{benchmark}.
For example, a user who only needs graph or PD-code input and Yamada evaluation does not have to install the surface-processing, interactive-visualization or benchmark packages. Development, testing and documentation dependencies are kept separate from the runtime package. The native \texttt{Repulsor} solver used by Repulsive Layout is not built by installing the Python \texttt{repulsion} extra. It requires its own source setup and compilation, which are described in the repulsive-layout documentation.

The package requires Python 3.11 or later. Installation and usage are documented in the repository \texttt{README} and in the \texttt{doc/} directory. This directory contains a quick-start guide, application tutorials, API reference and developer documentation.
Example scripts and notebooks are included in the source repository but are not part of the installed Python package. Likewise, an installed wheel should not be assumed to include research-reference CSV files, geometry bundles or large interactive HTML results; the phase-map tools can instead read records supplied by the user. Reproducing the final input figures additionally requires the figure-composition (compositor) source code and its accepted bundle of external panels.

\paragraph{Reproducibility}
Reproducibility follows the same module boundaries. Unit and regression tests check the reusable package interfaces. The application notebooks in \texttt{User\_guide/applications} write their outputs in machine-readable form. The repository listed under Code and data availability provides the public interfaces and a \texttt{uv.lock} file that records the exact dependency versions.
The package alone does not reproduce a specific calculation. Each calculation also needs its input data, preprocessing choices, parameters, implementation choices and verification records. For example, the finite mathematical checks in \SuppNoteref{supp:transfer_derivation} retain their constructors, matrices and explicit word lists. The software configuration of a calculation is fixed by three records: the source revision, the lockfile and the parameter record. Hardware and implementation details are stored with the corresponding benchmark records.

\FloatBarrier

\end{document}